\documentclass{emulateapj}
\usepackage[]{}
\pdfoutput=1

\newcommand{\kms}{km~s$^{-1}$}
\newcommand{\br}{\ensuremath{B\!-\!R}}
\newcommand{\ha}{H\hbox{$\alpha$}}
\newcommand{\hi}{\ion{H}{1}}

\newcommand{\amax}{\ensuremath{a_{\epsilon}}}
\newcommand{\lbar}{\ensuremath{L_{\mathrm{bar}}}}

\newcommand{\sigsky}{\ensuremath{\sigma_{\rm sky}}}
\newcommand{\mulim}{\ensuremath{\mu_{\rm crit}}}

\slugcomment{To appear in The Astronomical Journal}

\shorttitle{Outer Disks of Barred Galaxies}
\shortauthors{Erwin, Pohlen, \& Beckman}

\begin{document}

\title{The Outer Disks of Early-Type Galaxies.  I. Surface-Brightness
Profiles of Barred Galaxies}

\author{Peter Erwin\altaffilmark{1,2}}
\affil{Max-Planck-Insitut f\"{u}r extraterrestrische Physik, 
Giessenbachstrasse, 85748 Garching, Germany}
\email{erwin@mpe.mpg.de}
\and
\author{Michael Pohlen}
\affil{Astronomical Institute, University of Groningen, PO Box 800,
NL-9700 AV Groningen, The Netherlands}
\email{pohlen@astro.rug.nl}
\and
\author{John E. Beckman\altaffilmark{3}}
\affil{Instituto de Astrofisica de Canarias, C/ Via L\'{a}ctea s/n, 
38200 La Laguna, Tenerife, Spain}
\email{jeb@iac.es}

\altaffiltext{1}{Universit\"{a}ts-Sternwarte M\"{u}nchen, Scheinerstrasse 1,
D-81679 M\"{u}nchen, Germany}
\altaffiltext{2}{Guest investigator of the UK Astronomy Data Centre}
\altaffiltext{3}{Consejo Superior de Investigaciones Cientificas, 
Spain}

\begin{abstract}
  
We present a study of 66 barred, early-type (S0--Sb) disk galaxies, focused on
the disk surface brightness profile outside the bar region, with the aim of
throwing light on the nature of Freeman Type I and II profiles, their origins,
and their possible relation to disk truncations.  This paper discusses the
data and their reduction, outlines our classification system, and presents
$R$-band profiles for all galaxies in the sample, along with their
classifications.  In subsequent papers, we will explore the structure of outer
disks as revealed by these profiles, and investigate their possible origins.

The profiles are derived from a variety of different sources, including the
Sloan Digital Sky Survey (Data Release 5).  For about half of the galaxies, we
have profiles derived from more than one telescope; this allows us to check
the stability and repeatability of our profile extraction and classification.
The vast majority of the profiles are reliable down to levels of $\mu_{R}
\approx 27$ mag arcsec$^{-2}$; in exceptional cases, we can trace
profiles down to $\mu_{R} > 28$.  We can typically follow disk profiles out to
at least 1.5 times the traditional optical radius $R_{25}$; for some galaxies,
we find light extending to $\sim 3 \times R_{25}$.  For Type~I
(single-exponential) profiles, this means that we can trace the exponential
disk out to 6--7 scale lengths.

We classify the profiles into three main groups: Type I (single-exponential),
Type II (down-bending), and Type III (up-bending).  The frequencies of these
types are approximately 27\%, 42\%, and 24\%, respectively, plus another 6\%
which are combinations of Types II and III. We further classify Type II
profiles by where the break falls in relation to the bar length, and in terms of the postulated mechanisms for breaks at large radii (``classical trunction'' of star formation versus the influence of the Outer Lindblad Resonance of the bar).  We also classify the Type
III profiles by the probable morphology of the outer light (disk or spheroid).
Illustrations are given for all cases.
\end{abstract}


\section{Introduction}

The radial surface brightness profiles of galaxy stellar disks are usually
assumed to be exponential in nature, though it is by no means obvious that a
galaxy disk \textit{should} have an exponential profile.  One of the more
successful early attempts to show that an exponential disk could form
naturally was by \citet{yoshii89}, who found that an exponential disk forms if
the time scales for viscosity and the star formation are comparable; more
recent studies along these lines include \citet{ferguson01} and
\citet{slyz02}.  Although early attempts to derive exponential disks from
first principles in cosmological simulations yielded disks with excess
brightness in the inner profiles \citep[e.g.,][]{navarro94}, more recent
experiments in which star formation was taken at least partially into account
have been more successful \citep{robertson04,governato07}.  Even so, the
exponentiality of the radial brightness profile is perhaps best treated as an
empirical datum.

In his pioneering paper on galaxy disks, \citet{freeman70} pointed out that
not all disks are simple exponentials.  In particular, he identified two basic
types of disk profiles: Type I, in which the disk \textit{does} in fact show a
simple exponential form; and Type II, where the outer part of the disk shows a
purely exponential fall-off, but where the inner part of the profile falls
below the inward projection of the outer exponential.  (In both cases, the
profile usually rises more steeply in the innermost part of the galaxy; this
is typically assumed to represent the contribution of the central bulge.)
Although it has been argued that Type II profiles are simply an illusion
generated by excessive dust extinction at certain radii
\citep{adamson87,phillips91}, so that the actual \textit{stellar} profile is
still Type I, the very existence of Type II profiles in S0 galaxies tends to
discount this explanation.  \citet{macarthur03} combined optical and
near-IR imaging to show that dust extinction is responsible for only a subset
of Type II profiles in intermediate and late-type spirals.

Another feature which is sometimes taken to be a general (or even universal)
property of disks is that of a \textit{truncation} of the stellar population
at large radii, typically 2--4 exponential scale lengths \citep[see, e.g., the
review by][]{pohlen04}.  Van der Kruit (1979)\nocite{vdk79} and
\citet{vdk81a,vdk81b} first drew attention to this phenomenon, which they
inferred primarily from the major-axis profiles of edge-on, late-type spirals.
The term ``truncation'' is perhaps misleading, since even the original studies
did not argue for a complete absence of stars beyond the truncation radius.
More recently, \citet{pohlen02} used deep images of three face-on spirals to
show that the truncation actually takes the form of a change in slope, from
the shallow exponential of the main disk to a steeper exponential at larger
radii \citep[see also][]{degrijs01}.  From this perspective, truncations can
be seen as another form of Freeman's Type II profile, with breaks at fainter
surface brightness levels than was typical of Freeman's original sample.

Theoretical models have ascribed truncations to a feature of the initial
collapsing cloud which formed the disk \citep{vdk87}, or to the effects of a
star formation threshold due to changes in the gas density or phase at large
radii \citep{kennicutt89,schaye04,elmegreen06}.  \citet{zhang00} and
\citet{ferguson01} have put forward viscous evolutionary disk models which
produce truncation-like features, though the latter authors argue that their
truncations would tend to smooth away during the evolution of the galaxy.  The
influence of magnetic fields has also been proposed to account for the
truncated form of disk edges \citep{battaner02}, and recent $N$-body
simulations by \citet{debattista06} suggest that purely stellar-dynamical
effects could be a plausible mechanism.

Most of those working on explantations for truncations have implicitly assumed
that all disks, or at least the great majority, are truncated.  However, this
appears not to be the case.  At least some spirals show a single exponential
brightness profile traceable out to eight or even ten scale lengths from the
center, with no sign of truncation \citep[see,
e.g.][]{barton97,weiner01,bland-hawthorn05}.

Evidence for a \textit{third} general class of disk profiles has been
presented by \citet{erwin05-type3} for early-type disks and by
\citet{hunter06} and \citet{pohlen-trujillo} for late-type disks; an earlier
identification of this phenomenon in extreme late-type spirals is that of
\citet{matthews97}.  In this class, dubbed ``Type III'' (or
``antitruncation'') by Erwin et al., the inner profile is a relatively steep
exponential, which gives way to a \textit{shallower} surface brightness
profile (which may or may not be exponential itself) beyond the break radius.
This profile is thus something like the inverse of a Type II profile, bending
``up'' instead of ``down'' beyond the break radius.  \citet{elmegreen06}
showed that something like a Type~III profile could result from star formation
if the initial gas disk has the right (ad-hoc) radial density profile; more
recently, \citet{younger07} argued that minor mergers can produce Type~III
profiles.

Stellar disks thus seem to be a mixed and somewhat confusing bag: some disks
are exponential out to very large radii, some are apparently truncated, some
display classical Freeman Type II profiles (if these are indeed really
distinct from truncations), and some have shallower profiles beyond a certain
radius.  This diversity is probably telling us something important about
galaxy formation and evolution --- for example, the outer part of the disk may
record useful information about past accretion and interactions or the lack
thereof \citep[e.g.,][]{ibata05,younger07}.  It would clearly be useful,
however, to have a better understanding of just how often, where, and in what
fashion disks deviate from the simple exponential model.

This paper is part of a larger study focused on the outer disks of S0 and
early-type spiral galaxies, complementing the study of late-type disks by
\citet{pohlen-trujillo}.  Our aim is to map out some of the actual complexity
in galaxy disk profiles, look for patterns and order within this complexity,
establish a general taxonomy for disk profiles, and ultimately attempt to
understand \textit{why} disk profiles behave the way they do.  Along the way,
we hope to test some recent models of star formation in galaxy disks, and lay
the groundwork for more general testing of disk galaxy formation models.

Here, we present surface brightness profiles and classifications for a sample
of 66 S0--Sb galaxies.  These galaxies are the \textit{barred} subset of our
early-type sample (a total of 118 galaxies); the data and analysis for the
unbarred galaxies will be presented in a subsequent paper \citep{gutierrez07}.
We concentrate first on the barred galaxies --- which are the majority of the sample
--- because they have a unique and useful characteristic: the bar can be
used as a measuring rod.  As we will show below and in \citet{erwin07}, the
bar size provides a useful and informative way to analyze Type II profiles in
particular, and there are hints of strong connections between the bars and the
disk profiles.

The outline of this paper is as follows.  In Section 2 we describe the galaxy
sample, the imaging observations made, a brief description of the data taken
from archives, and the reduction and photometric calibration of the images.
In Section 3 we explain how we obtained the radial surface-brightness profiles
of the galaxies, the comparison of the profiles made with data from different
telescopes, and the quality of the profiles taken from Sloan Digital Sky
Survey images.  Section 4 presents the detailed classification scheme for the
surface brightness profiles, with illustrative examples.  Finally, in Section
5 we show individual profiles for all the galaxies and supply explanatory
notes for individual galaxies.

\section{Sample and Observations}

\subsection{The Sample} 

Our sample is essentially the same as that presented in \citet{erwin05}, which
in turn was an expansion of a sample originally studied by \citet{erwin03}.
The expanded sample consists of all galaxies from the Uppsala General Catalog
\citep{ugc} which are nearby (heliocentric redshift $\leq 2000$ \kms),
northern ($\delta > -10\arcdeg$), and large (diameters $\geq 2.0$\arcmin),
with Hubble types S0--Sb, and strong or weak bars \citep[SB or SAB bar types
from][hereafter RC3]{rc3}.  Because of uncertainties about how consistent
Hubble types for spirals are between the Virgo Cluster and the field
\citep[e.g.,][]{vandb76,koopmann98}, we excluded Virgo Cluster spirals (but
retained the S0 galaxies).  We also removed nine galaxies which did not appear
to have bars, despite their SB or SAB classification, or which were involved
in strong interactions; see \citet{erwin05}.

The trimmed sample has a total of 66 galaxies.  This is one more than the
sample presented in \citet{erwin05} because inspection of an $H$-band image
from the Galaxy On Line Database Milano Network
\citep{gavazzi03}\footnote{available online at
http://goldmine.mib.infn.it/} shows that NGC~4531 \textit{is} in fact barred,
though Erwin had described it as unbarred based on available optical images.
The galaxies and their global properties are listed in Table~\ref{tab:basic}.

As mentioned above, the sample is restricted to barred galaxies.  Because we
include galaxies with both strong and weak bars, we cover at least two-thirds
of local disk galaxies
\citep[e.g.,][]{eskridge00,laurikainen04,menendez-delmestre07,erwin07-rc3}.
We also have images for the corresponding \textit{unbarred} galaxy sample,
which are currently being analyzed \citep{aladro07}.  Details of their
surface-brightness profiles and similarities with (or differences from) the
barred galaxies will be discussed in future papers \citep{gutierrez07}.

\subsection{Observations}\label{sec:obs} 

The field S0--Sa galaxies in our sample were previously studied by
\nocite{erwin03}Erwin \& Sparke (2002, 2003), who imaged almost all of them in
$B$ and $R$ with the 3.5m WIYN Telescope.  Subsequently most of the Sab and Sb
galaxies, as well as the Virgo S0 galaxies, were imaged in $B$ and $R$ with
the Andalucia Faint Object Spectrograph and Camera (ALFOSC) on the Nordic
Optical Telescope (NOT) in 2001 and 2002.

Unfortunately, the majority of these images turned out not to be useful for
the present study, for two reasons.  First, the small fields of view of the
imagers ($6.8\arcmin \times 6.8\arcmin$ for WIYN and $6.4\arcmin \times
6.4\arcmin$ for NOT/ALFOSC) meant that the outer disks of many of our galaxies
filled most of the CCDs.  We have found that light from the stellar disks can
often be traced to at least \textit{twice} the RC3 $D_{25}$ diameter.  Since
all of our galaxies have, due to the way the sample is defined, $D_{25} \geq
2\arcmin$, each galaxy's disk thus extends to at least 4\arcmin{} in at least
one dimension; galaxies with $D_{25} > 3\arcmin$ will essentially fill the CCD
in at least one dimension (in both dimensions if the galaxy is close to
face-on).  This means that there is often little or no area on the image where
the sky background can be measured.

Second, many of the images proved to have significant scattered light
problems, which manifest as backgrounds which are not flat at large distances
from the galaxy.  From comparison with images having a larger field of view
(see below), we suspect that the problematic WIYN images are usually affected
by scattered light from nearby bright stars; images where this was not the
case (and where the galaxy was relatively small) were still usable.  On the
other hand, the NOT/ALFOSC images routinely showed large-scale variations out
to the borders of the images, even when there were no bright stars in the
vicinity.  In this case, we suspect that scattered light from the galaxy
itself is the culprit, in part because the pattern of excess light seems to 
match the orientation of the more elongated galaxies, and because the effect 
seemed to be independent of lunar phase.

To remedy this problem, we turned to images obtained with the Wide Field Camera
(WFC) of the 2.5m Isaac Newton Telescope (INT).  These were taken during two
observing runs: 2003 September 19--21 and 2004 March 14--17.  Conditions were
photometric on the first night of the 2003 run and on all four nights of the
2004 run.  Seeing varied from 1.0--1.5\arcsec{} in the 2003 run and from
0.7--3.4\arcsec{} during the 2004 run; poor seeing is \textit{not} a problem
for our analysis, since we are interested primarily in the outer disks, where
we must average the light over large spatial areas.

For four more galaxies, we were able to retrieve usable images from the Isaac
Newton Group (ING) archives; these were taken either with the INT-WFC, the
earlier Prime Focus Camera Unit on the INT, or with the 1m Jacobus Kapteyn
Telescope (JKT).  Additional image sources included the BARS Project
\citep{lourenso01} for NGC~4151 and 4596, and the catalog of \citet{frei96}
for NGC~5701.  For NGC~4612, we made use of $R$-band images from the 2.4m MDM
Telescope at Kitt Peak (1996 March, courtesy Paul Schechter).

Finally, we found images for about three quarters of our sample in the Sloan
Digital Sky Survey \citep{york00}, including Data Release 5 \citep{sdss-dr5}.
This provided an additional source of comparison for galaxies already
observed, as well as images of \textit{better} quality for 22 galaxies (mainly
because the sky background in the SDSS images tended to be very uniform).  The
reduced images were retrieved from the SDSS archive; in cases where a galaxy
extended off the top or bottom of a given field, we retrieved the adjacent
field(s) from the same scan and merged them to create larger images including
as much of the galaxy as possible.  Adjacent SDSS fields are sequential
products of a given observing run and so can be merged without fear of
significant changes in orientation or observing conditions, though secular
changes in the sky background can show up as vertical gradients.

Table~\ref{tab:obs} lists the primary and secondary images used for each
galaxy.  The first (and sometimes only) image is what we used to extract the
surface-brightness profile; if another image was useful (e.g., for photometric
calibration or validation of the surface-brightness profile shape), it is
listed afterwards.

\subsection{Reduction} 

The reduction of images taken with the WIYN Telescope has already been 
discussed by \citet{erwin03}.  Here, we focus on those images taken with the 
INT-WFC, MDM, NOT, and the archival ING images.

All images were reduced using standard tools and techniques within
\textsc{iraf}, including bias subtraction and flat fielding.  During our first
set of INT-WFC observations (2003 September), we obtained both dome flats and
twilight flats.  Test reductions using both sets of flat fields showed that
the twilight flats produced significantly flatter backgrounds, so we used
twilight flats for those and all subsequent INT-WFC reductions.

For all of the galaxies observed with the INT-WFC, the central CCD (Chip 4)
was large enough to include both the galaxy and a significant amount of sky
around it, so we reduced the Chip 4 images in isolation and did not attempt to
construct full image mosaics.  Although the WFC suffers from strong optical
distortions near the edges of its field, the center of Chip 4 is close to the
optical axis, so galaxy isophotes on that chip are not significantly affected.

Multiple exposures were typically offset by only 10--20\arcsec, so we aligned
them prior to coadding with simple linear shifts; the shifts were calculated
using positions of stars near each galaxy.

\subsubsection{Sky Subtraction} 
\label{sec:sky-sub}

Proper sky subtraction is essential for deriving accurate surface brightness 
profiles, since under- or over-subtraction can introduce false curvature or 
obscure existing curvature.

We inspected all images for the existence of 2D structure in the sky
background by generating median-smoothed copies of each image.  Isolated,
small-scale structures (e.g., dust-grain halos not removed by flat-fielding)
were masked out, but large-scale gradients, including occasional vertical
gradients in SDSS images, were removed using the \textsc{iraf}
\texttt{imsurfit} routine.  Regions for the fit were selected to be both far
from the galaxy and free of contamination by bright stars.  We usually fit the
background with a linear function, but occasionally higher-order polynomials
were needed.

The main stage of sky subtraction was the determination of the mean sky
background level, including the level in the residual images which resulted
from \texttt{imsurfit} processing.  The method we used involved measuring the
median level in $10 \times 10$ pixel boxes in regions devoid of bright stars
and scattered light, located well outside the galaxy.  Since, as we show
below, stellar light can often be traced out to at least twice $R_{25}$, we
were careful to measure the sky background only in regions farther away from
the galaxy.  The number of regions sampled depended on the image size and the
area that was free of bright stars, neighboring galaxies, etc., but ranged
between 40 and 100.  The final sky level was the mean of these 40--100 median
values.

The advantage of measuring the sky background this way is twofold.  First, we
were able to avoid sampling regions that contain galaxy light or scattered
light from bright stars (the latter were identified by inspection of
median-filtered copies of the images).  Second, it enabled us to compute an
estimate of our uncertainty by bootstrap resampling: we re-computed the mean
from a resampled set of the median values 500 times, and took the standard
deviation of these 500 estimates as our uncertainty \sigsky.  A possible
disadvantage of our technique is that we may occasionally understimate the sky
background for the galaxy itself if there are patches of scattered light
within the galaxy-dominated part of the image, though we do mask bright stars
within the galaxy when we do our ellipse fitting.

A more ideal method for both estimating the background level and the
uncertainty thereof might be that used by \citet{barton97} and
\citet{pohlen-trujillo}.  In this approach, concentric fixed ellipses are used
to determine the intensity level as a function of radius from the galaxy
center.  By extending this to well outside the galaxy, the intensity will
asymptotically approach the background level; variations from one annulus to
the next can be used to determine the uncertainty on the background estimate.
Unfortunately, this requires an image which is both very large (with the
galaxy well centered) and very uniform, whereas a significant number of our
images are either not large enough in both dimensions or contain regions
affected by scattered light, or both.  But a comparison of background and
\sigsky{} estimates for 8 SDSS images measured using both methods shows that
the background measurements typically differ by less than 1.5\sigsky, and the
\sigsky{} values themselves agree to within 10\%.

Following \citet{pohlen-trujillo}, we use the sky uncertainty \sigsky{} to
define a ``critical'' surface brightness level below which we consider our
profiles to be uncertain.  We do this by setting $\mulim = 4.94 \, \sigsky$,
which corresponds to the level at which a 1-$\sigma$ error in the sky
subtraction would shift the profile by 0.2 mag arcsec$^{-2}$ (see the bottom 
panel of Fig.~2 in Pohlen \& Trujillo).  This level is indicated on our 
profile plots by a horizontal dashed line.  For the majority of our profiles, 
$\mulim \sim 26.5$--27 in $R$, but fainter levels exist: there are seven 
galaxies with $\mulim \gtrsim 28$ mag arcsec$^{-2}$.

\subsection{Photometric Calibration}\label{sec:calib}

With three exceptions, all of our images and surface brightness profiles are
calibrated to Cousins $R$, either directly via observations of standard stars
or the use of published aperture photometry, or indirectly via conversion of
SDSS zero points from $r$ to $R$.  The exceptions --- NGC~1022, NGC~4319, and
UGC~11920 --- were cases where none of our images were obtained during
photometric conditions, and for which we could find no literature
calibrations.  Although we used the SDSS $r$ filter for many of our own
observations (to reduce the possibility of fringing in the sky background), we
opted to calibrate all images to Cousins $R$.

Our images come from a variety of sources, and we used a variety of methods to
calibrate them.  The largest set was the direct calibration of our 2003
September and 2004 March INT-WFC observations via standard-star observations
made during the same night.  For other images, we did the calibrations using
short exposures from the photometric INT-WFC runs, aperture photometry from
the literature, or SDSS images, as explained below.

\subsubsection{INT-WFC Photometric Observations}

The first night of our 2003 September run at the INT, and all four nights of
our 2004 March run, were photometric.  Standard stars from \citet{landolt92}
fields were observed in both $B$ and $r$ filters throughout each photometric
night.\footnote{The exception is the first night of the 2004 March run, when
no $B$-band observations were made.} The photometric calibration was then done
by fitting following equations using the \texttt{fitparams} task from the
\textsc{iraf} photcal package:
\begin{eqnarray}
	B_{\rm inst} & = & B - Z_{B} + k_{1,B} X + k_{2,B} (\br) \\
	R_{\rm inst} & = & R - Z_{R} + k_{1,R} X + k_{2,R} (\br)
\end{eqnarray}
where the instrumental magnitudes are $B_{\rm inst}$ and $R_{\rm inst}$, the
catalog magnitudes from \citet{landolt92} are $B$ and $R$, the zero points are
$Z_{B}$ and $Z_{R}$, and $X$ is the airmass.

For the $R$-band calibrations we used a fixed extinction term $k_{1,R}$ taken
from the Carlsberg Meridian Telescope's nightly extinction
measurements.\footnote{\url{http://www.ast.cam.ac.uk/\~{}dwe/SRF/camc\_extinction.htm}}
We proceeded by first determining color terms ($k_{2}$) for each night.  We
then derived mean color terms for the entire run and then re-determined the
zero points ($Z_{B}$ and $Z_{R}$) using the fixed color terms.  The final
values are given in Table~\ref{tab:calib}.

Proper calibration of individual galaxy images using the coefficients derived
above requires knowing a galaxy's \br{} color.  For some of the galaxies, we
used large-aperture colors from the compilation of \citet{ph98}; for others,
we determined \br{} colors from large-aperture photometry on our $B$ and $r$
images, iterating until the color values converged.  However, for the majority
of galaxies observed with the INT-WFC, we were unable to find or derive
individual \br{} colors, so we assumed the following default colors based on
Hubble type: $\br = 1.5$ for S0, $\br = 1.4$ for S0/a and Sa, and $\br = 1.3$
for Sb.  These values are based on the calibrated colors of other galaxies in
our sample, as well as additional unbarred galaxies observed during the same
INT-WFC observing runs.  Since an error of 0.1 in \br{} translates to a
zero-point change of less than 0.01 mag, our calibrations are not strongly
affected by uncertainties in galaxy colors.

For four galaxies observed with the INT-WFC under photometric conditions (plus
three additional galaxies not in this barred-galaxy sample), we found aperture
magnitudes in the literature \citep{ph98}.  We used these to test the accuracy
of our standard-star calibrations; the results indicate that our calibrations
agree with literature photometry to within 0.1 magnitudes, and usually to
better than 0.05 magnitudes.

Because most of the previous observations with the WIYN Telescope were made
under non-photometric conditions, we also observed several of these S0--Sa
galaxies with the INT-WFC during the 2004 March run, usually with very short
(30--120s) exposures.  These were used to calibrate the pre-existing deeper
exposures from the WIYN Telescope, by matching surface brightness profiles
derived from the WIYN and INT images using the same fixed-ellipse-fit
parameters.  (While it is also possible to calibrate the images by performing 
matching aperture photometry, several of the deep WIYN images were saturated 
in the galaxy center.)  We also used this approach to calibrate a deep image 
of NGC 6654 from a non-photometric night of our 2003 September INT-WFC run.

For NGC 2787, we used archival JKT observations (from 2001 January 29) of
NGC~2787 to calibrate our WIYN image of the galaxy.  The JKT observations,
originally made by Edo Noordermeer, were obtained under photometric conditions,
and were accompanied by observations of Landolt standards.  Because the
standard star observations covered only a limited range of airmass, we used
the $r$-band extinction measurement from the Carlsberg Meridian Telescope
observations.  The WIYN image was calibrated by matching aperture photometry
performed on the WIYN image to that performed on the calibrated JKT images.

\subsubsection{Calibration of Images Using Aperture Photometry}

Other galaxy images were calibrated using aperture photometry from the
compilation of \citet{ph98}; these are marked with ``PH98'' in column 6 of
Table~\ref{tab:obs}.  We used the \texttt{apphot} task from the
\texttt{digiphot} package of \textsc{iraf} to perform aperture photometry on
our sky-subtracted images, using apertures of the same sizes as in the
literature (we did not use data from apertures with diameters smaller than
20\arcsec{}, in order to minimize possible problems from differences in seeing
or centering between our measurements and those in the literature).  The
instrumental magnitudes were compared with the literature values to derive
appropriate zero points for the images.  For NGC~4699 and NGC~7743, the only
red photometry in \citet{ph98} is the aperture photometry of \citet{sv78},
which uses their own $r$ filter (\textit{not} the same as the SDSS $r$
filter); we used their conversion of $V - r$ to $V - R$.  Since it is not
clear if the ``$R$'' in this conversion corresponds to Johnson $R$ or Cousins
$R$, and because there has apparently been little or no follow-up calibration
of this filter system, the calibrations for NGC 4699 and NGC 7743 should be
considered more uncertain than those of other galaxies.

\subsubsection{Calibration using SDSS Images}

For five galaxies, we used profiles from non-SDSS images but calibrated them
using SDSS images, as follows.  We measured $g - r$ colors in large apertures
centered on the galaxies in the SDSS images (large enough to encompass most of
the galaxy while still staying within the image boundaries), and used these
colors to convert the SDSS $r$-band zero points (derived from the headers of
the tsField files accompanying each field) to Cousins $R$.  The conversion
used Table~7 of \citet{smith02}, so that the Cousins $R$ magnitude is
\begin{equation}
	R = r - 0.14(g - r) - 0.14.
\end{equation}
The non-SDSS images were then calibrated using matching aperture photometry.

Finally, there were 22 galaxies for which we used profiles derived directly 
from SDSS images.  The SDSS $r$-band zero points were converted to Cousins 
$R$ as we have just described.

\section{The Profiles}

In this study we are primarily interested in the question \textit{What
are the radial surface-brightness profiles of stellar disks?}.  There
are several related questions, such as: \textit{Where and how often
are stellar disks truncated?} \textit{What form does a truncation
take?} \textit{Are features such as bars, rings, and spiral arms
simply azimuthal redistributions of the underlying exponential disk?}
For all of these questions, a key first step is to determine the mean
surface brightness as a function of radius.

Another reason that the mean (i.e., azimuthally averaged) surface-brightness
profiles are important is their potential for testing models of disk formation
and evolution.  Current models for the formation of exponential disks
\citep[e.g.,][]{ferguson01,slyz02}, as well as for such features as disk
truncations or antitruncations
\citep[e.g.,][]{battaner02,schaye04,elmegreen06,debattista06,younger07}, are
almost all strictly one-dimensional.

To obtain our profiles, we used of the \textsc{iraf} task \texttt{ellipse}.
It is important to distinguish between two related uses of this and other
routines which fit ellipses to galaxy isophotes.  The more general fitting is
with ``free'' ellipses, where the ellipse position angle (PA) and ellipticity
are allowed to vary (as well as, optionally, the ellipse centers).  In
contrast, one can also fit using ``fixed'' ellipses, where the ellipse center,
PA, and ellipticity are held fixed, and only the semi-major axis and intensity
are allowed to vary.  If the ellipse shape and orientation are matched to that
of the projected galaxy disk, then this is (almost) equivalent to averaging on
concentric circles for a face-on galaxy.  (There will be differences if parts
of the galaxy, such as its bulge, are not flat, or if the disk is very thick
and the galaxy is highly inclined.)

\begin{figure*}
\vspace{0.15cm}
\centerline{\includegraphics[width=18cm]{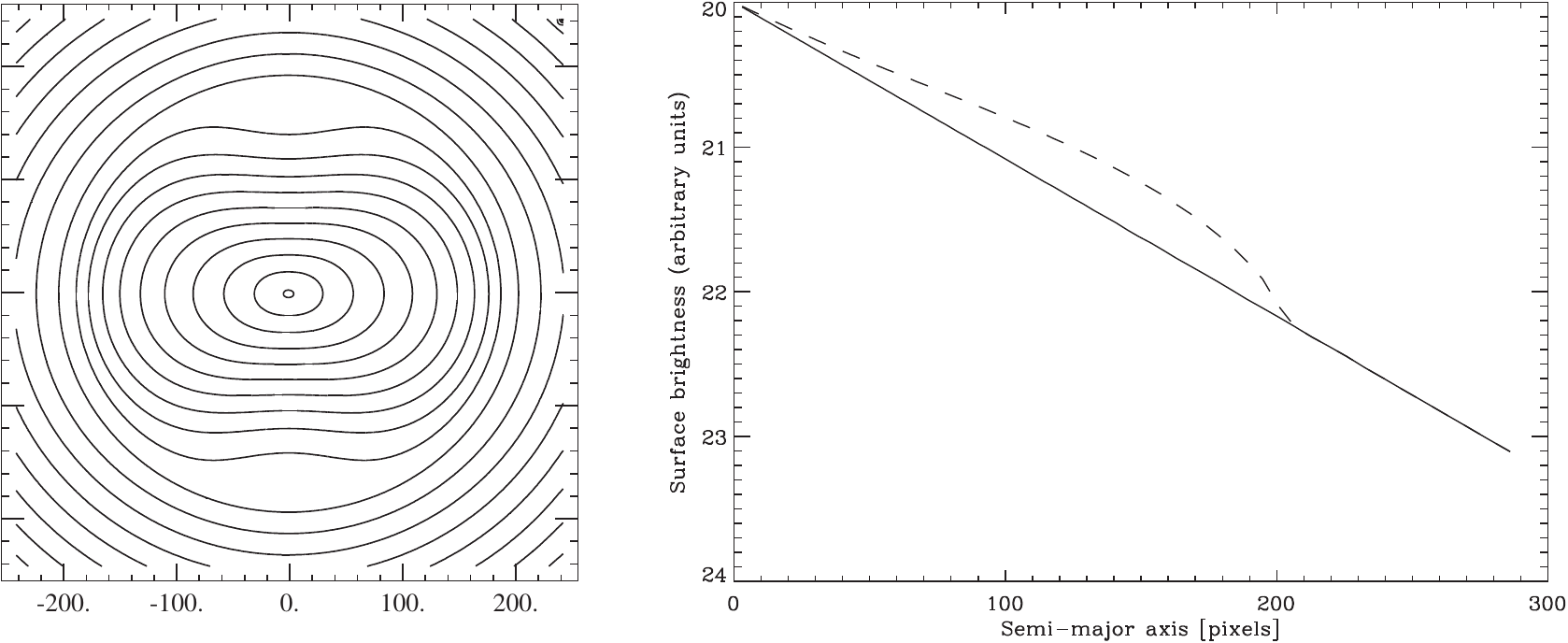}}

\vspace{0.2cm}
\caption{A model galaxy image (isophotes in left panel) and
corresponding surface-brightness profiles (right panel) derived using
fixed ellipse fits (solid line, matching the ellipticity and position
angle of the outer disk) and free ellipse fits (dashed line, with
ellipse position angle and ellipticity allowed to vary).  The model
galaxy has a bar which is a purely azimuthal ($\cos 2 \theta$)
perturbation (for $r < 200$ pixels) of an exponential disk.  The
fixed-ellipse profile recovers the true mean profile; the free-ellipse
profile deviates at $r < 200$ pixels where the ellipses alter shape to
trace the bar.\label{fig:azavg-model}}

\end{figure*}

\begin{figure*}
\vspace{0.5cm}
\centerline{\includegraphics[width=18cm]{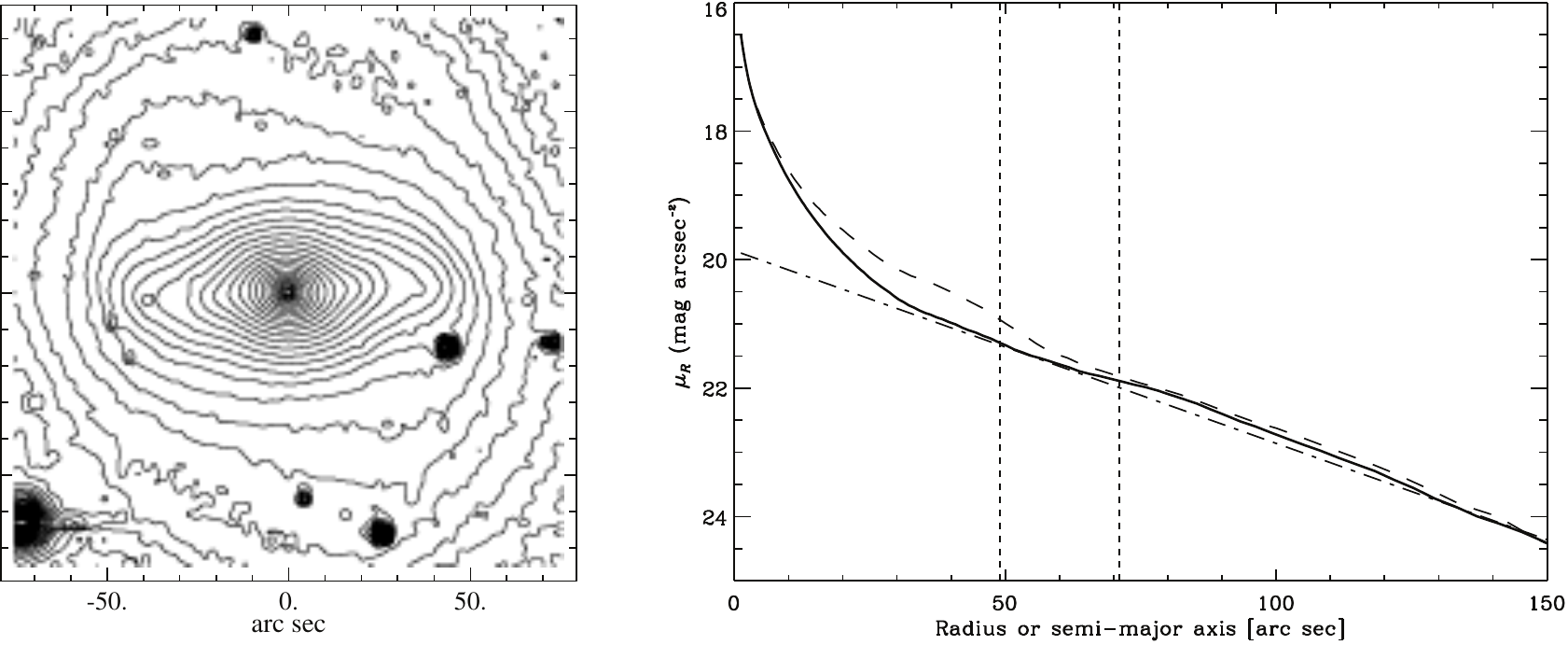}}

\vspace{0.2cm}
\caption{As for Figure~\ref{fig:azavg-model}, but now showing
surface-brightness profiles from an $r$-band image of the SB0/a galaxy
NGC~4665.  The vertical dotted lines indicate lower and upper limits on the
bar size (\amax{} and \lbar{} from Table~\ref{tab:basic}); the diagonal dot-dashed line is an exponential fit to the
fixed-ellipse profile at $r = 125$--179\arcsec (see
Figure~\ref{fig:profiles}).  The free-ellipse-fit profile (dashed line) is
significantly brighter in the region $a \sim 20$--70\arcsec, due to the fact
that the ellipses trace the bar.\label{fig:n4665}}

\vspace{0.5cm}
\end{figure*}

Even though free-ellipse fits are often used to generate surface brightness
profiles for galaxies, there \textit{can} be significant differences between
profiles generated with free ellipses and profiles generated with fixed
ellipses.  Figure~\ref{fig:azavg-model} shows the different profiles produced
by free-ellipse and fixed-ellipse fits for the same image, in this case an
artificial galaxy consisting of an exponential disk with, at $r < 200$ pixels,
a bar which is a purely azimuthal redistribution of the underlying disk:
\begin{equation}
	I(r, \theta) = I_{0} \exp(-r/h) \cos( 2 \theta).
\end{equation}
The fixed-ellipse fit (solid line) recovers the underlying exponential
profile, as we would expect.  However, the free-ellipse-fit profile
(dashed line) differs in the bar region, because the ellipses alter
shape to track the bar-distorted isophotes.  A possible real-world
example of this phenomenon is shown in Figure~\ref{fig:n4665} for the galaxy 
NGC 4665.

An additional reason for using fixed-ellipse fits is the fact that at large
radii the S/N becomes so low that the free-ellipse fitting algorithm fails to
converge, and the \texttt{ellipse} program automatically switches to fixed
ellipses for larger radii.  In this case \texttt{ellipse} uses the outermost
successful ellipse fit as a template for larger radii, and this may not always
be a good match to the true disk orientation (especially if the latter can be
determined from, e.g., \hi{} kinematics).

In order to know what ellipticity and position angle to use for the profile
extraction, we of course need to know the orientation of the disk.  We do this
using a variety of techniques (see \citealt{erwin03}, \citealt{erwin05}, and
Section~\ref{sec:notes} for discussions of the determinations for individual
galaxies), including kinematic determinations if they are available.  For most
galaxies, we rely on free-ellipse fits and the assumption that the outer disk
(specifically, the disk \textit{outside} any outer ring) is intrinsically
circular.  Note that because most of our profiles can be traced significantly
further out than $R_{25}$, we sometimes derive outer-disk orientations which
are different from those in RC3 or LEDA, which may be based on intermediate
structures such as outer rings.

The final set of fixed-ellipse profiles is presented in
Section~\ref{sec:profile-plots} (Figure~\ref{fig:profiles}), along with
comments on the individual galaxies.  These profiles were generated using
logarithmic radial spacing (the semi-major axis of each successive annulus is
1.03 times larger), with the actual intensity being the median of those pixels
in the annulus after a sigma-clipping algorithm was applied.  In all cases,
background or neighboring galaxies and the halos of bright stars were masked
out; a typical example of this process can be seen in the upper panel of
Figure~2 of \citet{pohlen-trujillo}, although for this study we used circular
and elliptical masks to better match the shapes of galaxies and stellar halos.

\subsection{Comparison of Profiles from Different Telescopes}

\begin{figure*}
\vspace{0.15cm}
\centerline{\includegraphics[scale=0.7]{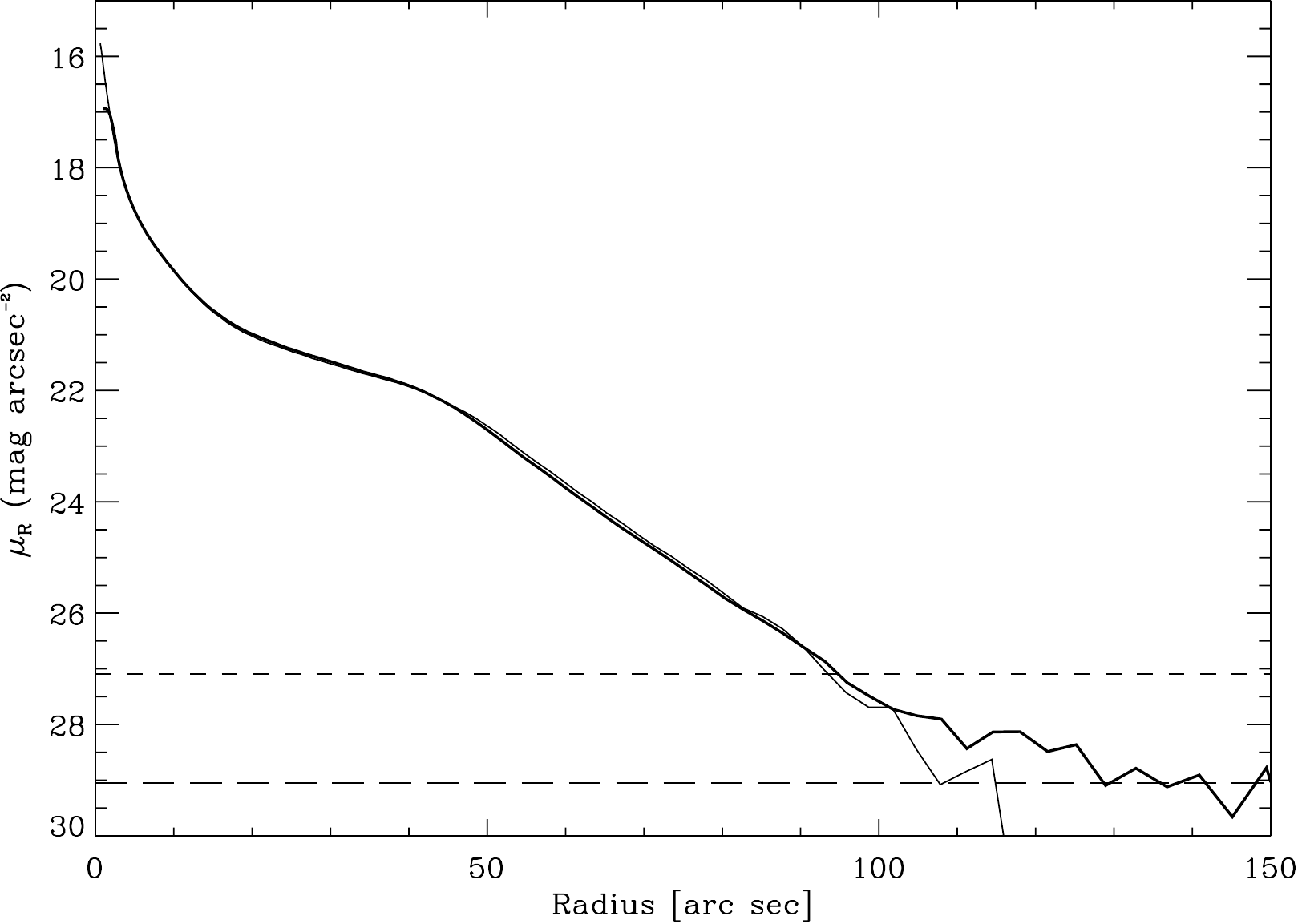}}

\vspace{0.2cm}
\caption{Comparison of surface-brightness profiles for NGC~7280 derived from
different observations.  The thick profile is from a 10200s observation with
the 2.5m Isaac Newton Telescope's Wide Field Camera, while the thin line is
the profile from a 300s observation with the 3.5m WIYN Telescope.  The two
profiles have been scaled to match in the region $r = 5$--80\arcsec.  The
horizontal dashed lines mark the limiting surface brightness levels \mulim{}
for the two profiles (thick for INT-WFC profile, thin for WIYN
profile).\label{fig:comp}}

\vspace{0.5cm}
\end{figure*}

We have usable images from at least two different telescopes for 31 of the 66
galaxies in our sample.  (By ``usable,'' we mean that the images are large
enough to have measurable sky background outside the galaxy, and the
background is reasonably flat and free of complex scattered light.)  This
allows us to check the validity of our profiles and our profile
classifications: how ``repeatable'' and reliable are our profiles?
Figure~\ref{fig:comp} shows one such comparison, for the galaxy NGC~7280;
Figure~\ref{fig:sdss} shows more comparisons, concentrating on those involving
SDSS images.  Although small variations in the profiles at large radii are
present, in almost all cases the basic shape of the profile is consistent.

\subsubsection{The Quality of Profiles from SDSS Images}

As noted in Section~\ref{sec:obs}, we use profiles derived from SDSS
$r$-band images for 22 galaxies, either because images from other
telescopes suffered from small fields of view or scattered light
problems, or because no other images were available.  At first glance,
SDSS images might not seem deep enough for reliable surface-brightness
profiles of the faint outer disk ($\mu_{R} \gtrsim 24$); after all,
the images are from a 2.5m telescope with effective exposure times of
only 54 seconds.

In practice, we have found the SDSS images to be surprisingly useful: it is
possible to measure reliable (azimuthally averaged!)  surface brightness
profiles down to at least $\mu_{R} \sim 26$, and often down to $\mu_{R} \sim
27$.  \citet{pohlen-trujillo} reached a similar conclusion in their analysis
of a large sample of SDSS-derived profiles.  This appears to be due to a
combination of three factors: SDSS images are always taken during dark time;
the images are taken in drift-scan mode, which allows for very accurate
flat-fielding; and excellent telescope and camera design which significantly
reduces scattered light problems \nocite{gunn98,gunn06}(Gunn et al.\ 1998,
2006).

To show how reliable the profiles from SDSS images are, Figure~\ref{fig:sdss}
compares profiles from SDSS images with profiles from our own deeper
observations (i.e., longer exposure times with similar-sized or larger
telescopes).  In some cases, the other images are only marginally deeper
(e.g., 120s on the 2.5m INT), but in others we have combined images with
cumulative exposures times of 10, 20, or even 40 minutes.  As can be seen from
the figure, in almost all cases the SDSS-based profiles agree very well with
the profiles from deeper images.  Even in cases where the profiles start to
diverge at faint surface-brightness levels (e.g., NGC~3941, NGC~4037, and
NGC~5338), the basic nature of the profiles are unchanged: for example, both
profiles for NGC~4037 show a downward break at $r \sim 80\arcsec$, indicating
a Type~II profile.  A similar comparison using very deep images of two
galaxies from \citet{pohlen02} is presented in \citet[][their
Fig.~3]{pohlen-trujillo}.

\begin{figure*}
\centerline{\includegraphics[scale=0.9]{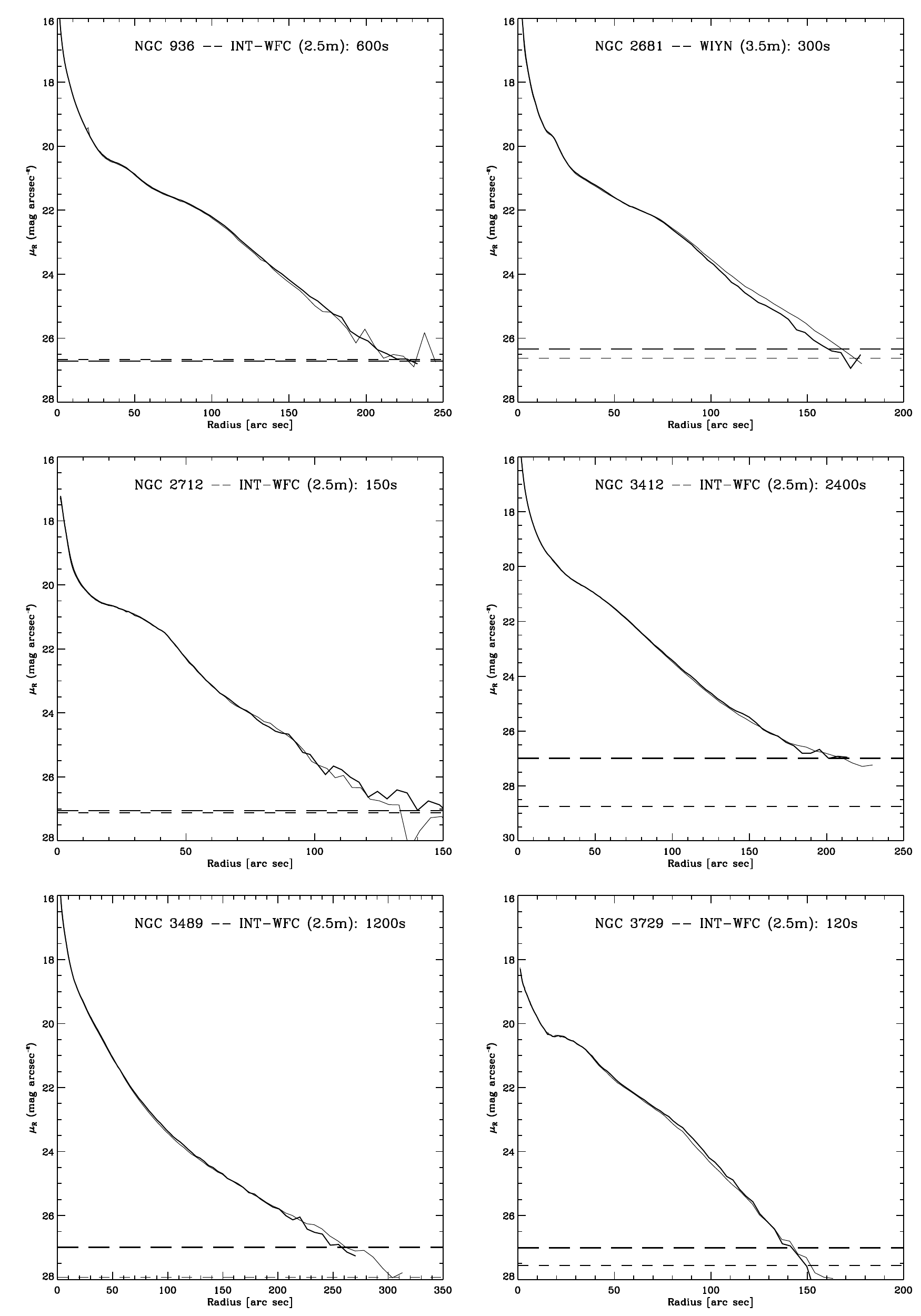}}

\vspace{0.5cm}
\caption{Comparison of profiles from SDSS images (thick lines) with profiles
from deeper images of the same galaxy obtained with other telescopes (thin
lines).  The telescope and exposure time of the deeper image is indicated
within each plot.  The horizontal dashed lines mark the sky uncertainty limits
\mulim.  Even though the SDSS images are short exposures, in most cases they
match the profiles from deeper exposures quite well.\label{fig:sdss}}

\end{figure*}

\begin{figure*}
\centerline{\includegraphics[scale=0.9]{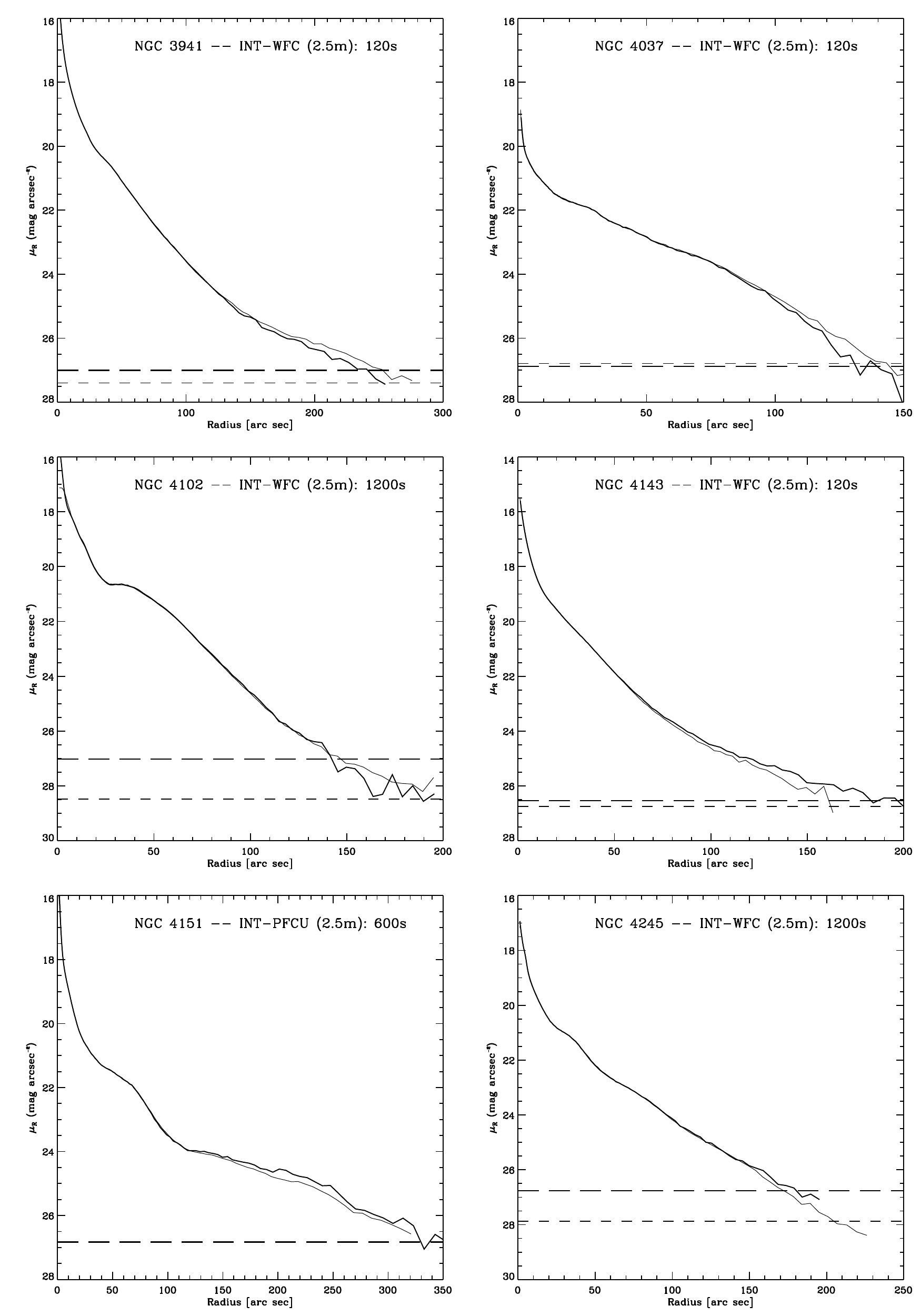}}

\addtocounter{figure}{-1}
\vspace{0.2cm}
\caption{continued.}

\end{figure*}


\begin{figure*}
\centerline{\includegraphics[scale=0.9]{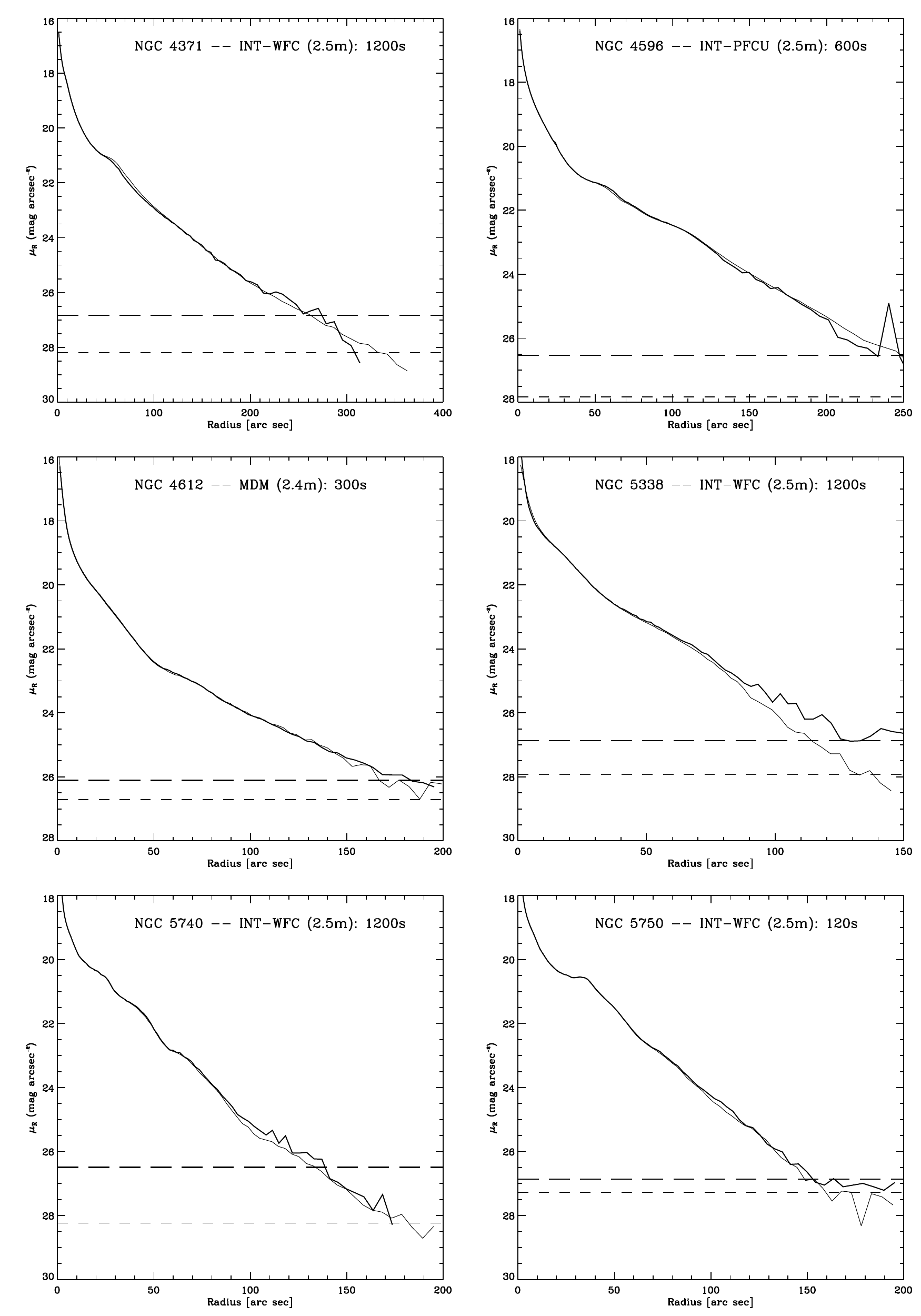}}

\addtocounter{figure}{-1}
\vspace{0.2cm}
\caption{continued.}

\end{figure*}

\section{Classification of Surface-Brightness Profiles}
\label{sec:profile-types}

\begin{figure*}
\centerline{\includegraphics[scale=1.0]{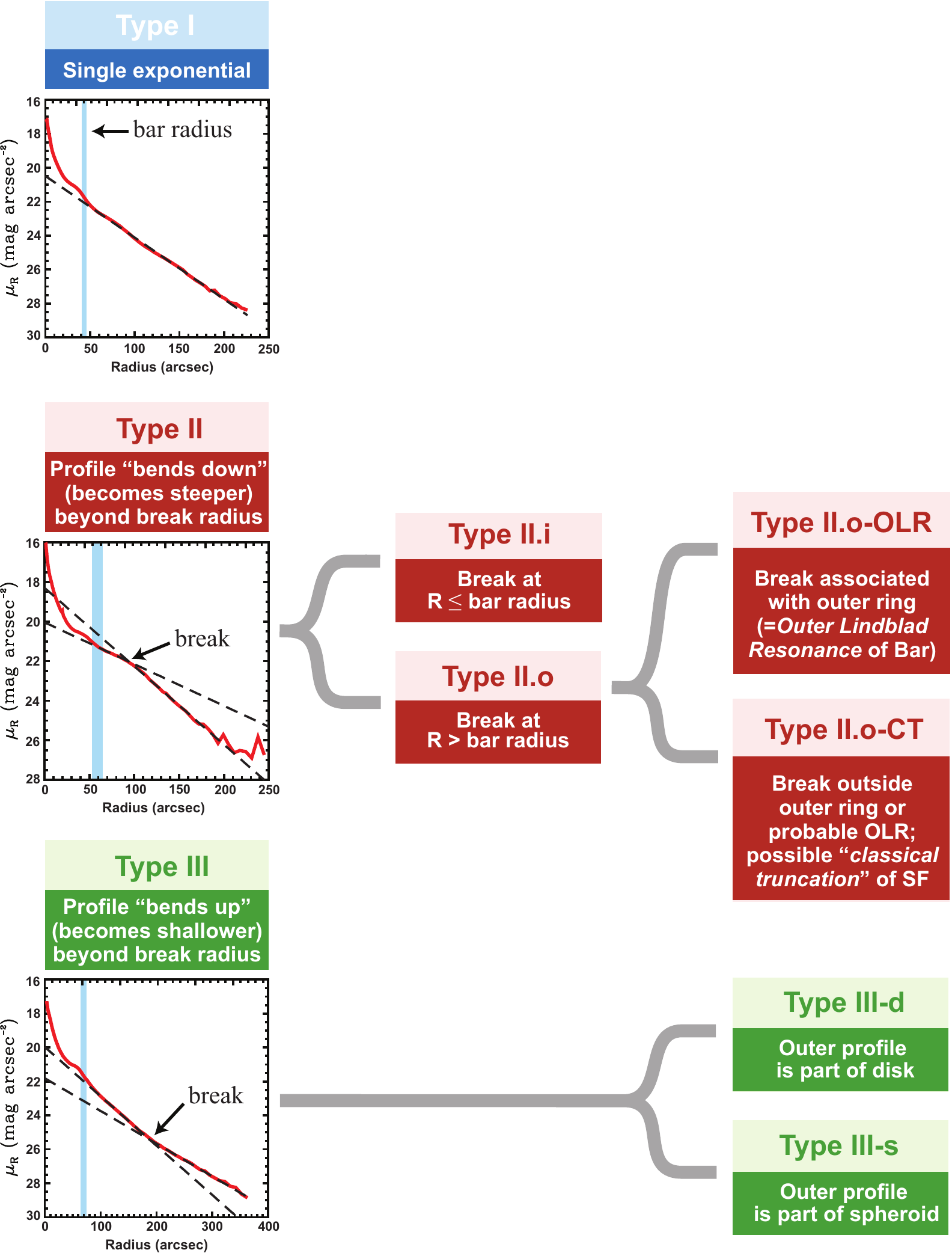}}

\vspace{0.5cm}

\caption{An overview of our scheme for classifying surface-brightness
profiles.  The basic level recognizes Types~I, II, and III, based on their
overall shape (ignoring the central excess associated with the bar/bulge).
Type II profiles can be further subdivided into II.i and II.o, based on where
the break in the profile is located.  Finally, Type II.o and Type III profiles
can be further classified based on the probable nature of the break (Type
II.o-OLR vs.\ Type II.o-CT) or the disk vs.\ spheroid nature of the outer
profile (Type III-d vs.\ Type III-s).  See text and subsequent figures for more
details.\label{fig:scheme}}

\end{figure*}

In this section we lay out the classification scheme we have worked out for
galaxy surface brightness profiles.  This scheme, summarized in
Figure~\ref{fig:scheme}, can be thought of as having three levels: a purely
descriptive classification consisting of three basic types (I, II, and III); a
refinement for Type~II profiles based on the location of the break relative to
the bar size; and a final interpretive level applied to Type~II and Type~III
profiles.  Some of this has already been discussed briefly in
\citet{erwin05-type3}, and in more detail by \citet{pohlen-trujillo}.

The profile classifications for individual galaxies are given in
Table~\ref{tab:results}.

\subsection{Basic Classifications: Types I, II and III}

\begin{figure*}
\vspace{0.15cm}
\centerline{\includegraphics[scale=0.9]{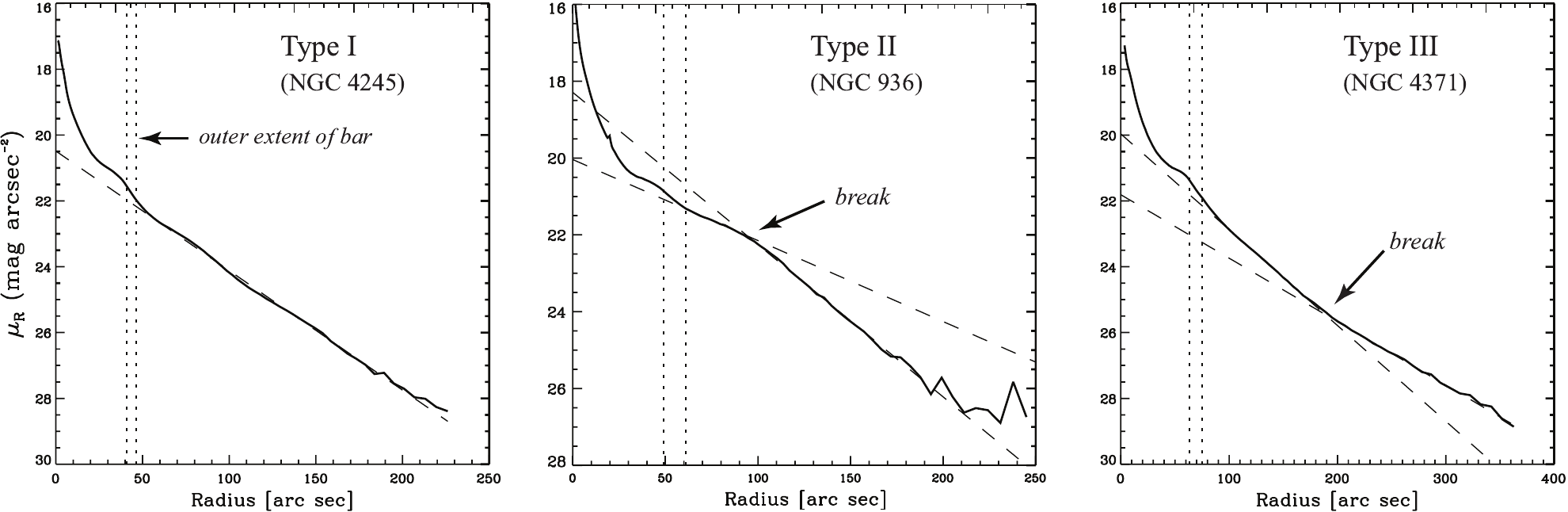}}

\vspace{0.2cm}
\caption{The three basic classes of surface-brightness profiles.  Type~I
(left) is the simple, single-exponential profile; note that we focus on the
profile \textit{outside} the bar region (the vertical dotted lines mark lower-
and upper-limit estimates of the bar size, from Table~\ref{tab:basic}).
Type~II profiles (center) have a break at which the profile changes slope from
shallow to steep.  Type~III profiles (``antitruncations,'' right) have the
reverse behavior: the profile slope changes from steep to shallow at the
break.\label{fig:type123}}

\end{figure*}

The first level --- \textbf{Type~I} vs.\ \textbf{Type~II} vs.\
\textbf{Type~III} --- is a very general, empirical description of the profile
shape outside the bar radius.  This is an extension of the scheme introduced
by \citet{freeman70}, who divided disk profiles into Types~I and II. Our
innovation --- made possible by profiles which extend to fainter
surface-brightness levels than Freeman had access to --- is the identification
of \textit{Type~III} profiles as a third major class \citep{erwin05-type3}.  
Examples of these three basic types are given in Figure~\ref{fig:type123}.

\textbf{Type I profiles} are single-exponential profiles, with the profile 
continuing out to the limit set by our sky-background uncertainty \mulim.

\textbf{Type II profiles} contain a break, where the profile bends ``down'' --
that is, the profile becomes steeper outside the break.  In most cases, the
profile is exponential both inside and outside the break, with two different
slopes; in a few extreme cases, the profile inside the break is \textit{not}
exponential (e.g., NGC~2859).  In some galaxies the break is quite sharp; in 
others it can be an extended region of gradual curvature.  This class 
includes so-called ``truncations.''

\textbf{Type III profiles} are similar to Type~II profiles, except that the
profile bends ``up'' beyond the break --- that is, it changes from a steep
exponential to something shallower at large radii.  The outer part of the
profile (beyond the break) is often exponential, but is sometimes curved.
Again, the break is sometimes sharp and sometimes gradual.  We also refer to 
these profiles as ``antitruncations.''

\subsection{Subdivisions of Type II: Type II.i and Type II.o}

\begin{figure*}
\vspace{0.5cm}
\centerline{\includegraphics[scale=0.9]{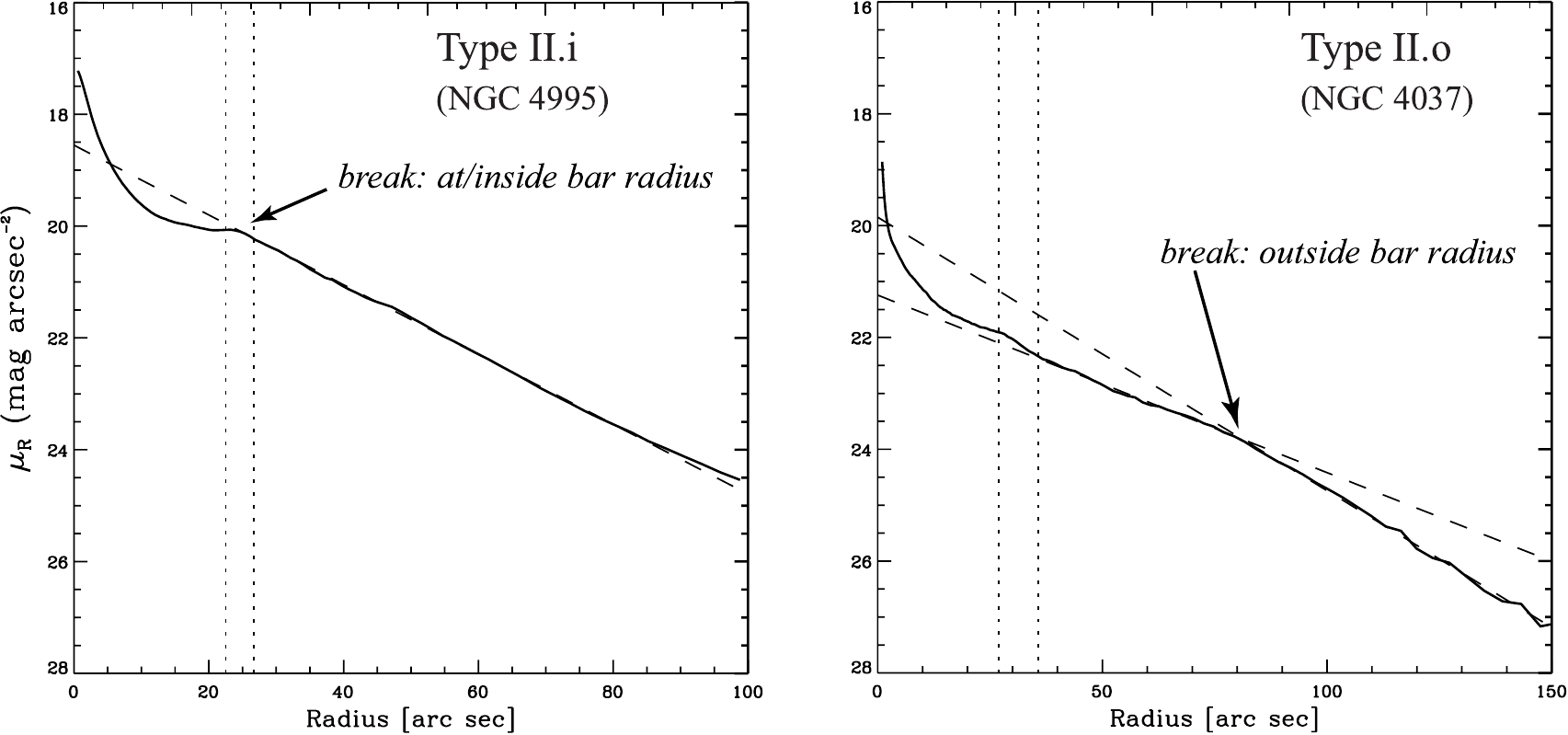}}

\vspace{0.2cm}
\caption{Examples of our Type II profile subdivisions.  On the left is a Type
II.i profile, in which the break occurs near the end of the bar (so that the
deficit is fully \textit{inside} the bar radius).  On the right is a Type II.o
profile, with the break occurring well \textit{outside} the bar.
\label{fig:type2io}}

\vspace{0.5cm}

\end{figure*}

The next level is a subdivision of the Type~II class, in which we note whether
the break is an ``inner'' break or an ``outer'' break, based on where the
break takes place \textit{relative to the bar radius} (see
Figure~\ref{fig:type2io}).  A break which is located near or at the bar radius
is an ``inner'' break, which we call \textbf{Type~II.i}.  A break which
happens outside the bar is an ``outer'' break and is called
\textbf{Type~II.o}.  This subclassification obviously depends on the galaxy
having a bar whose length we can measure!  In the absence of a bar, we are
left with a plain ``Type~II'' profile, as used for unbarred galaxies in
\citet{pohlen-trujillo}.

\subsection{Interpretive Levels for Type II and III}

The final classification stage is an interpretive one, in which we attempt to
say something about the \textit{nature}, and possibly the cause, of the breaks
in Type~II and III profiles.

Note that the use of ``(?)'' after a profile classification in
Table~\ref{tab:results} means that the \textit{last} part of the
classification is uncertain.  E.g., ``III-s(?)'' means that the profile is
definitely Type III, but that the spheroid identification for the morphology
of the outer light (Section~\ref{sec:type3}) is uncertain.

\subsubsection{Type II.o-OLR versus Type II.o-CT}
\label{sec:type2}

\begin{figure*}
\vspace{0.15cm}
\centerline{\includegraphics[scale=0.9]{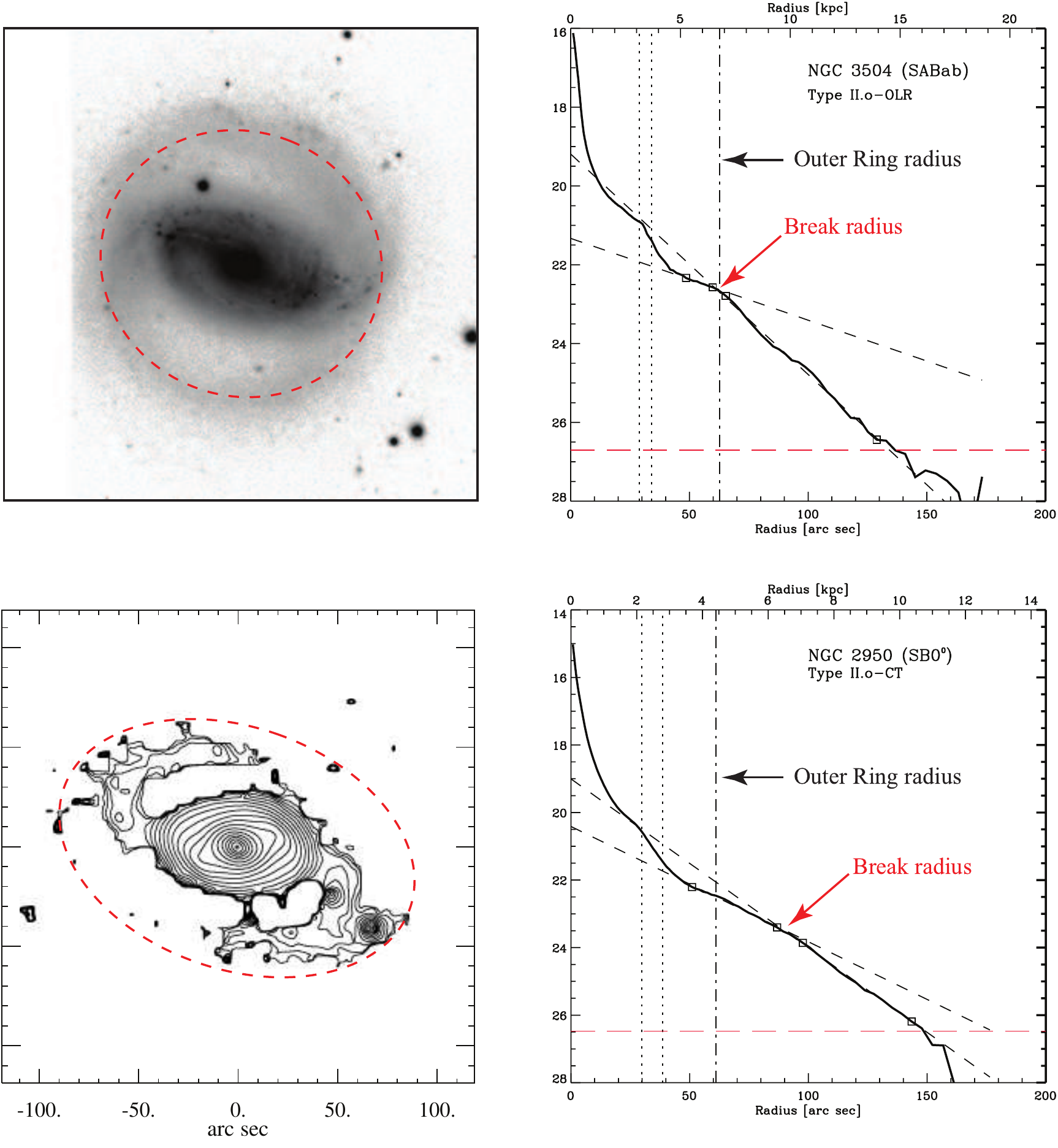}}

\vspace{0.2cm}
\caption{Comparison of Type II.o-OLR and Type II.o-CT profiles.  NGC~3504 (top
panels) has a Type II.o profile where the break coincides with an outer ring;
NGC~2950 (bottom panels) has a Type II.o profile where the break is well
\textit{outside} the outer ring.  In both figures, the break radius is
indicated by the dashed red ellipse (left-hand panels) and the red arrow
(right-hand panels).  For NGC~3504, we display the SDSS $r$-band image; for
NGC~2950, we have subtracted a model of the outer disk from the SDSS image in
order to bring out the (faint) outer pseudo-ring.
\label{fig:type2olr-vs-ct}}

\vspace{0.5cm}
\end{figure*}

First, we observe that many Type~II.o profiles have a break which coincides
with an outer ring (see Figure~\ref{fig:type2olr-vs-ct}, top panels, for an
example).  Even for those cases where no outer ring is visible (as in many S0
galaxies), the break occurs at $\sim 2$--3 times the bar radius, which is
where outer rings are usually found \citep[e.g.,][]{bc93}.  Since outer rings
are generally understood to be linked to the Outer Lindblad Resonance (OLR) of
bars \citep[e.g.,][and references therein]{buta96}, we call these \textbf{Type
II.o-OLR} profiles (or ``OLR breaks'').  If the OLR link is more
circumstantial --- that is, if there is no outer ring seen in the galaxy, but
the break is at $\sim 2$--3 times the bar radius, --- then we refer to them as
\textbf{Type II.o-OLR(?)} profiles.  More detailed arguments for this
classification will be presented in \citet{erwin07}.

In two galaxies with Type~II.o profiles (NGC~2273 and NGC~2950), the break
occurs well \textit{outside} what appears to be the outer ring; see
Figure~\ref{fig:type2olr-vs-ct}, bottom panels.  In these cases, the break is
apparently \textit{not} related to the bar's OLR. We therefore consider these
to be examples of ``classical truncations,'' so named because they seem
similar to the ``truncations'' first seen in edge-on, late-type galaxies
\citep[e.g.,][and references therein]{pohlen04}.  Such breaks are usually
supposed to be related to star-formation thresholds
\citep[e.g.,][]{kennicutt89,schaye04,elmegreen06}.  We call these profiles
\textbf{Type II.o-CT}.

In NGC 5338, the break is at 3--4 times the bar radius\footnote{Depending on
which of the two bar-radius measurements one uses.}.  Although no outer ring
is visible in this galaxy, we conclude that this break is outside the probable
radius of the OLR, and so we classify the profile as Type~II.o-CT(?).  In the
case of NGC 4037, the break is at 2.4--3.1 time the bar radius, and again
there is no visible outer ring.  Because the break is at a low surface
brightness level --- as is the case for the breaks in NGC 2273, 2950, and 5338
--- it is tempting to consider this another classical truncation.  However,
since the break radius is not clearly beyond the limits of plausible OLR
breaks, the situation is ambiguous; we leave this galaxy with an unmodified
II.o classification.

Finally, we note that we do \textit{not} see any examples of the
``Type~II-AB'' (``apparent/asymmetric break'') profiles that
\citet{pohlen-trujillo} found in 13\% of their galaxies.  They noted that
these seemed to occur only in Sc--Sd galaxies and not in any of the earlier
(Sb--Sbc) galaxies in their sample; it is apparently due to strong
lopsidedness in the outer isophotes.  Since all galaxies discussed in
\textit{this} paper are Sb or earlier, this would appear to reinforce their
suggestion that the Type~II-AB profile is a late-type phenomenon.

\subsubsection{Type III-s versus Type III-d}
\label{sec:type3}

We also define an interpretive subdivision for Type~III profiles.  This is
based on whether the evidence indicates that the outer part of the profile,
beyond the break, is still part of the disk (\textbf{Type III-d}) or whether
it is due to a more spheroidal component (\textbf{Type III-s}).  The
distinction between these two subdivisions is explained in somewhat more
detail, with illustrative examples, in \citet{erwin05-type3}.

\begin{figure*}
\vspace{0.15cm}
\centerline{\includegraphics[scale=0.9]{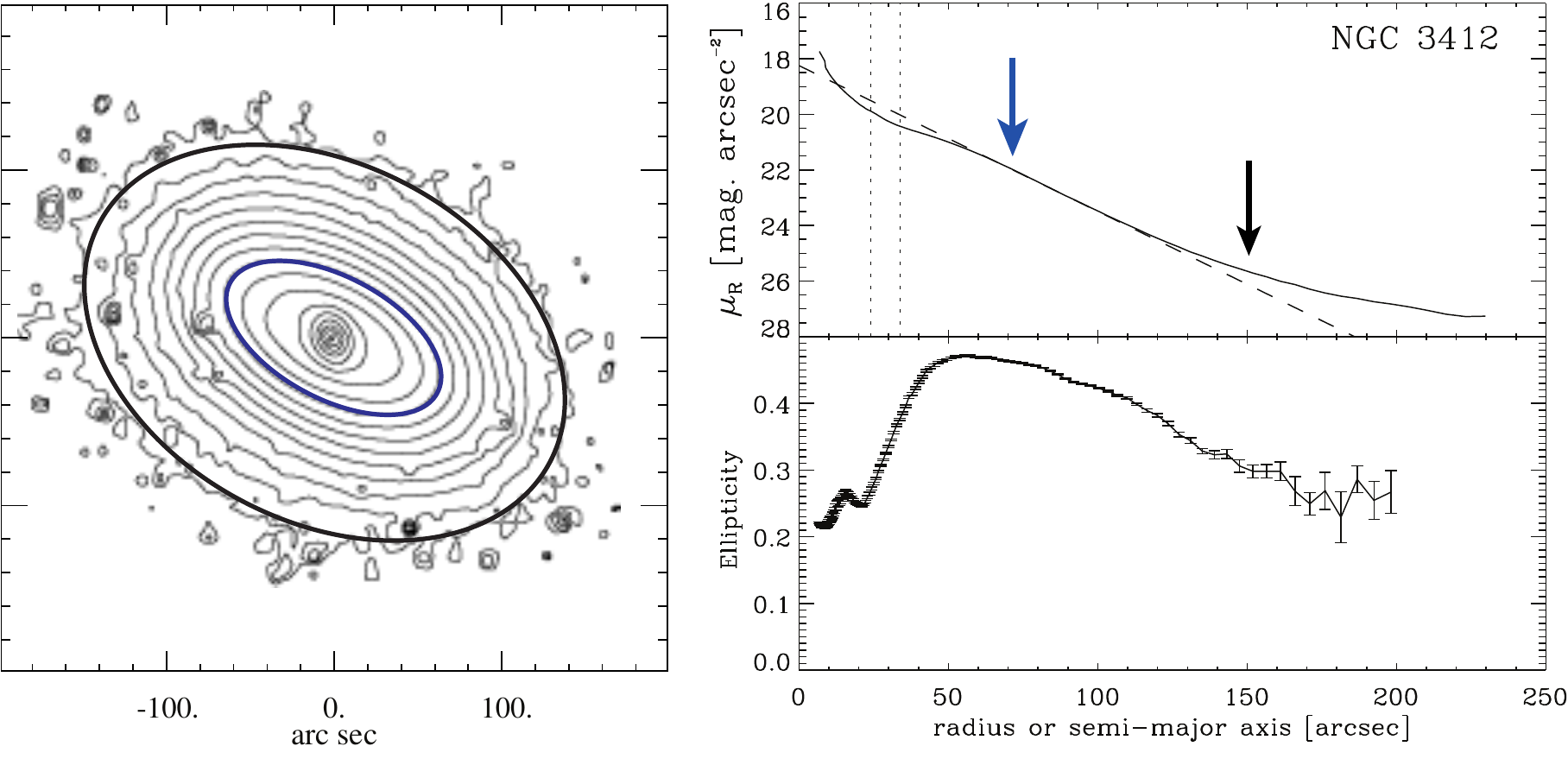}}

\vspace{0.1cm}
\caption{Identifying a Type III-s (``Type III-spheroid'') profile in an
inclined galaxy: The outer isophotes of NGC~3412 (e.g., large black ellipse in
left panel, black arrow in upper-right panel) are clearly rounder than the
inner isophotes (e.g., blue ellipse and arrow), and become even
rounder at larger radii (see ellipse fits in lower-right panel).  This
suggests that the outer part of the profile ($r \gtrsim 130\arcsec$) is due to
an intrinsically rounder component --- presumably the outer part of this S0
galaxy's bulge.
\label{fig:type3s}}

\vspace{0.45cm}
\end{figure*}

The clearest signature of a spheroidal component (Type III-s) is when the
isophotes for an inclined galaxy become progressively rounder at larger radii,
and the transition between inner and outer slopes is smooth, not abrupt
(Figure~\ref{fig:type3s}).  This is what one would expect if the light at
large radii is coming from a rounder structure, in which the inclined disk is
embedded.

\begin{figure*}
\centerline{\includegraphics[scale=0.9]{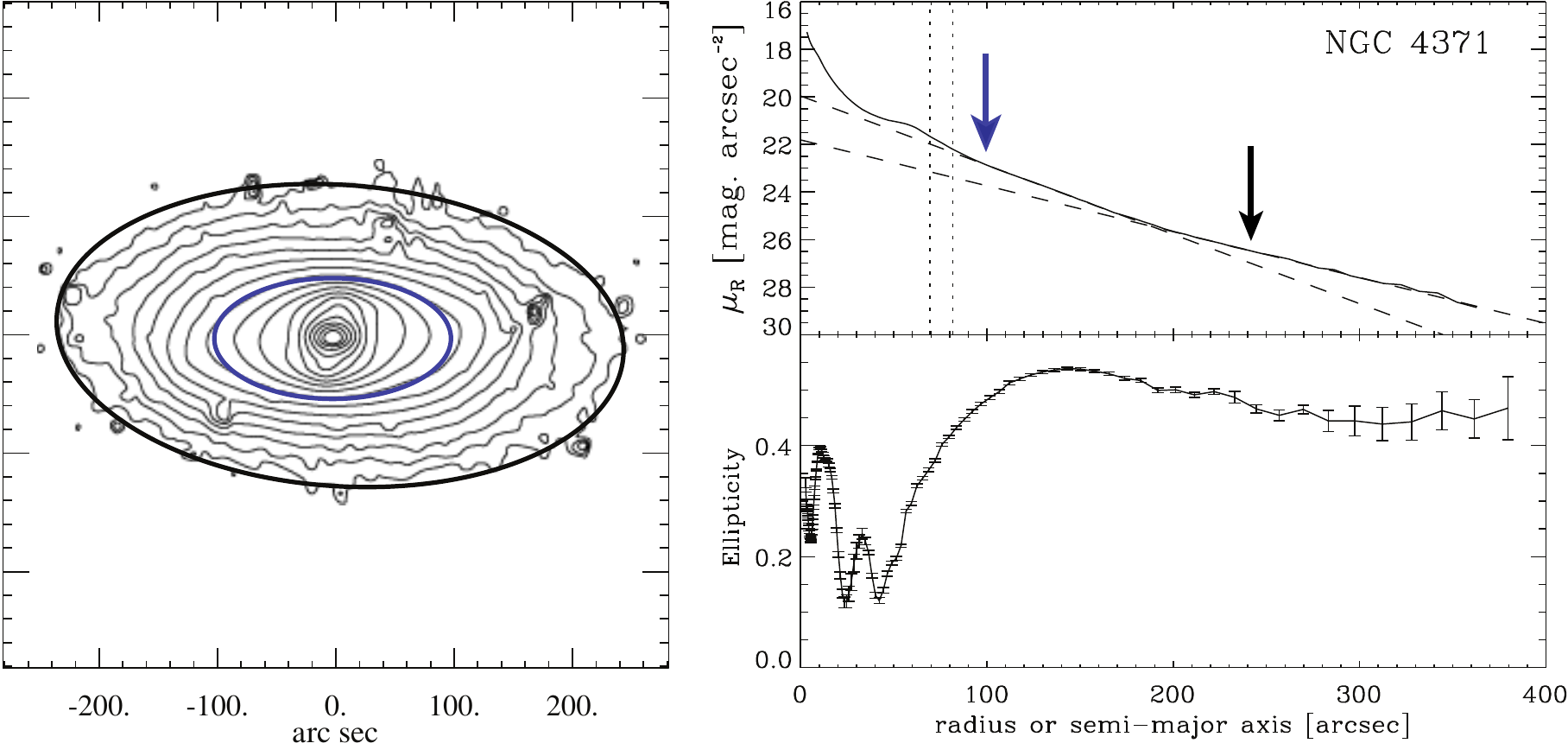}}

\vspace{0.2cm}
\caption{Identifying a Type III-d (``Type III-disk'') profile in an inclined
galaxy: The outer isophotes of NGC~4371 (e.g., large black ellipse in left
panel and black arrow in upper-right panel) are approximately as elliptical as
the inner isophotes (e.g., blue ellipse and arrow) and show no signs of
becoming rounder at the largest radii, suggesting that the outer part of the
profile ($r > 200\arcsec$) is still part of the disk.  The ellipticity peak at
$r \sim 150\arcsec$ is associated with twisted and partly boxy isophotes, and
is probably due to an outer ring.\label{fig:type3d}}

\vspace{0.5cm}
\end{figure*}

\begin{figure*}
\vspace{0.15cm}
\centerline{\includegraphics[scale=0.9]{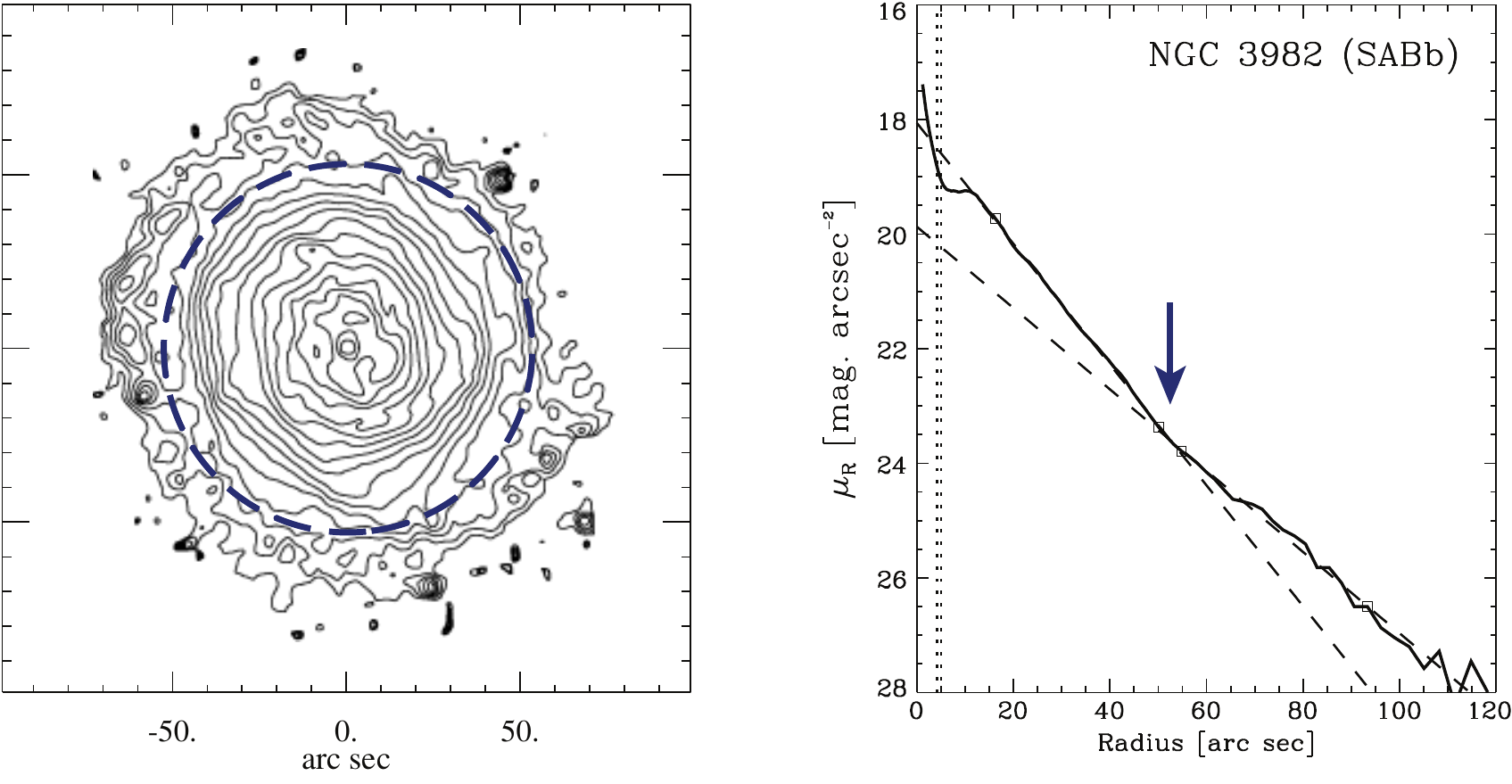}}

\vspace{0.2cm}
\caption{Identifying a Type III-d profile in a face-on galaxy: Although NGC
3982 is almost face-on, we can still identify the outer part of the profile
($r > 53$\arcsec) as being part of the disk because the light is clearly
dominated by spiral arms.  The dashed ellipse in the left panel (SDSS $g$-band
isophotes) marks the break, also indicated by the arrow in the right panel
($R$-band profile).\label{fig:type3d-spirals}}

\vspace{0.5cm}
\end{figure*}

Cases where the outer light is  part of the disk (Type III-d) can be
identified in two ways.  When the galaxy is inclined, the outer light appears
to have the same ellipticity as the inner light, suggesting it is still coming
from the disk (Figure~\ref{fig:type3d}).  In other cases (e.g., face-on
galaxies, where both disks and spheroids will produce roughly circular
isophotes), we can sometimes see clear spiral arms in the outer region, which
again are signs that the outer light is still coming from a disk
(Figure~\ref{fig:type3d-spirals}).

One could argue that the Type III-s classification doesn't really represent
the \textit{disk} profile, so that we might just as easily refer to these as,
e.g., ``Type I + spheroid'' and reserve ``Type III'' strictly for the disky
cases (what we currently call Type III-d).  For consistency, however, and
because in some cases we cannot be certain the outer light \textit{is} from a
spheroid, we keep the Type III-s term.

\subsection{Profiles with Multiple Classifications}

\begin{figure*}
\vspace{0.15cm}
\centerline{\includegraphics[scale=0.9]{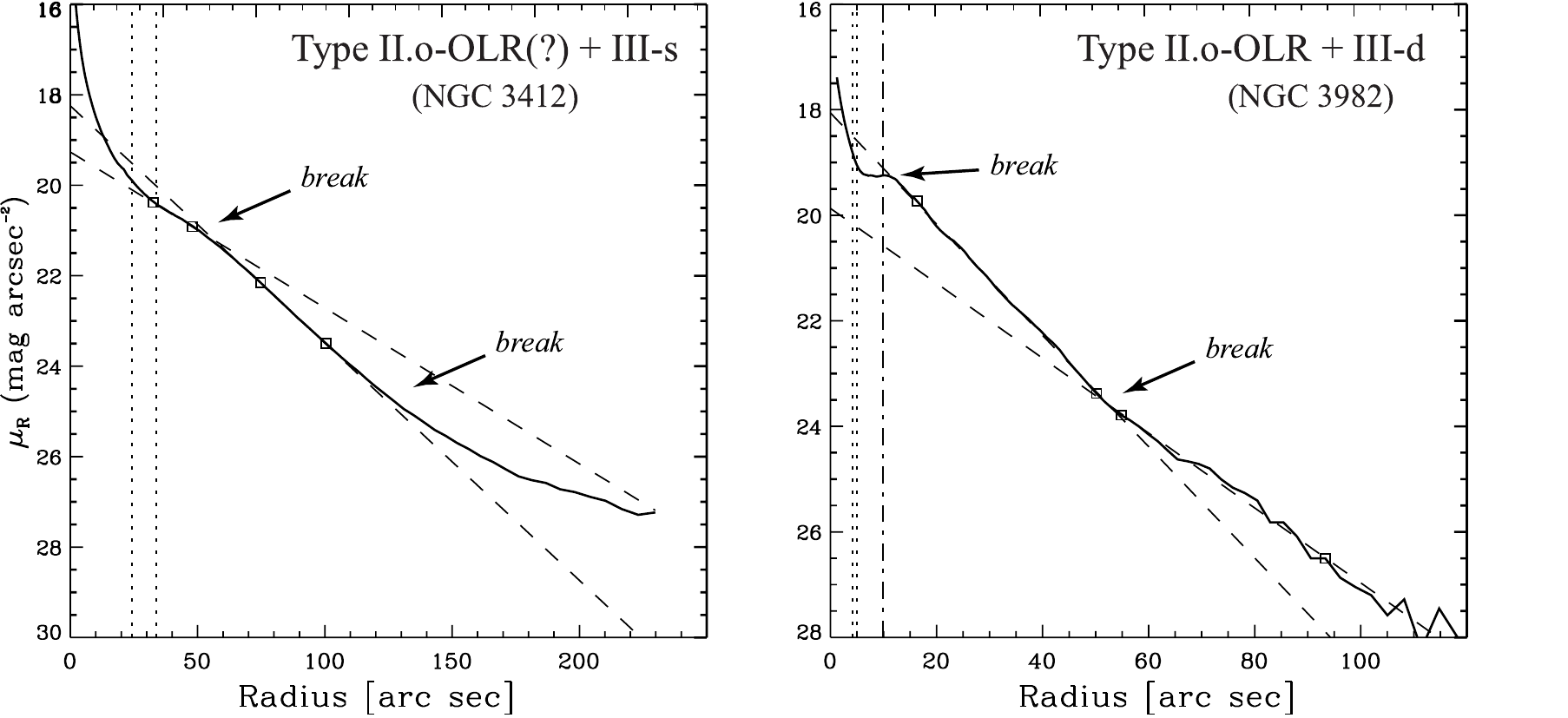}}

\vspace{0.2cm}
\caption{Examples of composite profiles.  In both galaxies, the inner profile
is Type~II.o, with the break at $\sim$ 2 $\times$ the bar radius; further
outside is a second break, with a shallower profile beyond.  The outermost
part of NGC 3412's profile (left) corresponds to rounder isophotes (see
Fig.~\ref{fig:type3s}); this plus the smooth nature of the outer break
suggests additional light from a spheroid.  In contrast, the outer break in
NGC 3982's profile (right) is quite sharp, and the light beyond that point is
still part of the disk (Fig~\ref{fig:type3d-spirals}).
\label{fig:composite}}

\vspace{0.5cm}
\end{figure*}

The Type~I/II/III classification scheme, with the associated subtypes, does a
good job of capturing the main variations we see in the disk profiles.
Nevertheless, nature is nothing if not perverse, and there are at least four
galaxies whose profiles are more complicated, combining elements of more than
one type.  (\nocite{pohlen-trujillo}Pohlen \& Trujillo 2006 also found
examples of composite profiles in their late-type sample.)  These are all
cases where the inner part of the profile (that is, outside the central
photometric ``bulge'' and any excess or ``shoulder'' associated with the bar)
has a Type~II character, but at larger radii the profile appears to be
Type~III; see Figure~\ref{fig:composite} for two examples.  We do \textit{not}
see any profiles with \textit{two} downward-bending breaks, nor do we see any
cases of Type~III profiles with truncations.\footnote{\citet{pohlen-trujillo}
did, however, find one case of an apparent double-downward-break profile; see
their Fig.~5.}

In some galaxies (e.g., NGC~3412, Figure~\ref{fig:type3s}), the complex
profile appears to be a simple case of a Type~II \textit{disk} plus light from
a spheroid which dominates at large radii to produce the Type~III-s profile.
There are other galaxies where the outer excess light is still part of the
disk.  For example, the outer excess light in the profile of NGC~3982 ($r >
53\arcsec$) comes from a region dominated by two blue spiral arms
(Figure~\ref{fig:type3d-spirals}).  In all cases, we indicate such composite
profiles with a plus sign, e.g., ``Type II.o-OLR + III-d,'' where the first
type is the innermost.

There are also some profiles where we have hints of excess light at large
radii (e.g., NGC~3507), but the S/N at those radii is too low for us to be
absolutely certain.  To deal with these cases, we use a criterion based on the
magnitude difference between the point where the apparent outer excess begins
($\mu_{R} = 25.5$ in the case of NGC~3507) and the limiting magnitude from the
sky-background uncertainty ($\mulim = 26.6$ for NGC~3507).  If this difference
is $> 1.5$ mag, we consider the Type~III classification secure and list it in
Table~\ref{tab:results}; if it is between 0.5 and 1.5 mag, we include a
tentative note on the plot (``[+ III?]'' in the case of NGC~3507) and in the
``Notes'' column of Table~\ref{tab:results}; if the difference is smaller
than 0.5 mag (e.g., NGC~3368 or NGC~4267), then we do not consider it a
significant detection.

\subsection{Exponential Fits and Measuring the Break Radius} 

We fit exponentials to portions of the surface-brightness profile which appear
approximately linear on the plots.  In almost all cases, we look for a
reasonably linear zone outside the bar region; in many galaxies, the bar is
marked by an excess ``shoulder'' over the exponential profile outside (see,
e.g., NGC~4245 and 936 in Fig.~\ref{fig:type123}).  We also exclude clear,
extended bumps due to outer rings, on the principle that such features can be
treated as excess flux on top of an underlying exponential.  When fitting
regions of the profile that extend down to our limiting surface-brightness
level \mulim, we set the outer limit of the fit at the radius where $\mu(r) =
\mulim$, except in cases where the profile becomes visibly noisy near that
limit, in which case we stop the fit just outside the beginning of the noisy
part of the profile.

After isolating the region or regions to fit, we regrid the data to a linear
radial spacing using cubic-spline interpolation, and then perform a simple
linear, least-squares fit.  We regrid because the logarithmic spacing in the
original profile means more data points at smaller radii; this can bias the
fit so that it does not match the outer part of the profile well.  This is not
an ideal solution, by any means; in doing this, we ignore the fact that the
data points at smaller radii have higher S/N, and could provide better
constraint on the fit if they were properly weighted --- at least if the
underlying profile were known to be intrinsically exponential.  However, local
non-exponentiality (e.g., due to spiral arms or dust) at small radii could
then bias the fit.  We note that a limited comparison was made by
\citet{erwin05}, who found that exponential slopes for Type~I profiles
obtained this way generally agreed with slopes determined by \citet{bba98} for
the same galaxies, even though the latter authors used Poisson-noise weighting
in their fitting.

For the case of Type II.o profiles, we attempt to fit the parts of the profile
both inside and outside of the break in a piecewise fashion, as done by, e.g.,
\citet{pohlen02} and \citet{pohlen-trujillo}.  \nocite{pohlen02}Pohlen et al.\
demonstrated that it makes sense to try doing this even when one is dealing
with classical truncations in late-type spirals, since the profile outside the
break is clearly exponential.  One might, however, ask why we bother fitting
two exponentials to some of the more extreme Type II.o profiles, where the
inner profile is ``exponential'' only over a relatively short range
\citep[e.g., NGC~2962 and NGC 3945; see also the discussion of NGC~6654
(``VII Zw 793'') in][]{kormendy77}.  We do this because we want to start
from a position of agnosticism on the question of when a profile is
``truncated'' or not.  Discussions of disk truncation usually assume that
``the'' exponential disk is the part of the profile \textit{inside} of the
break.  In contrast, the traditional picture of Type II profiles is that the
exponential disk is the \textit{outer} part of the profile, outside of the
break --- even if the inner profile is also exponential.  If we resolutely fit
\textit{both} parts of the profile, inside and outside the break radius, we
can use the results as part of a general analysis of Type II.o profiles
\citep{erwin07}.  In addition, there are a small number of Type II.o profiles
which are genuinely ambiguous: the break is located at about the right radius
for an OLR, but the inner profile is fairly steep and extended and so the
break is \textit{also} plausible as a classical truncation.  Future analysis
--- and better development of models which predict the characteristics of
broken profiles --- may allow us to more cleanly classify such profiles, but
in the meantime we adopt a pluralistic approach.  We note that there are only
two Type II.o profiles where the profile inside the break (i.e., between the
bar and the break) is clearly non-exponential: NGC~2859 and NGC
3982\footnote{In the case of NGC 3982, we are referring to the Type II.o break
at $r = 12\arcsec$, not the Type~III break further out.}; for these galaxies,
we do not report an inner scale length.

Note that we do \textit{not} fit the outer part of a Type~III profile if we
can identify it as Type~III-s, since in such cases the outer part of the
profile is often non-exponential (e.g., NGC 3941), and in any case we are
assuming that it is part of a surrounding spheroid, not an exponential
subsection of the disk.\footnote{We do list approximate ``break radii'' for
such profiles in Table~\ref{tab:results}; these are the radii where the outer
profiles begin to dominate over the inner exponentials.} We also do not fit the
region inside the break of Type~II.i profiles, since this is within the bar
region and is often non-exponential.  In this sense, we treat Type~II.i
profiles as being similar to Type~I profiles: they have a single exponential
disk outside the bar.

The results of these fits --- the central surface brightness and exponential
scale length for each of the exponential fits --- are given in
Table~\ref{tab:results}.  The inner-profile fits (or the single fit for a
Type~I profile) are denoted by $\mu_{0,i}$ and $h_{i}$ for the central surface
brightness and the exponential scale length, respectively; the outer-profile
fits are denoted by $\mu_{0,o}$ and $h_{o}$.  These are \textit{observed}
values; we have not applied any corrections for inclination or dust extinction
(intrinsic or Galactic).  We also do not apply any redshift corrections, but
since these galaxies are all at redshifts of 2000 \kms{} or less, any such
corrections would be negligible.  The measurements of the break radius and the
surface brightness of the break for Type~II and III profiles, also listed in
Table~\ref{tab:results}, is explained below.

\subsubsection{Determining the Break Radius} 

The break radius in Type II.o and Type~III-d profiles can be estimated by eye
from the profiles, or by determining the point where the exponentials fitted
to the inner and outer slopes (``piecewise'' fits) intersect.  A somewhat more
precise and rigorous approach is to fit the whole profile (excluding the bulge
and bar region) with a function containing a parameterized break.  We do this
using a ``broken-exponential'' function, inspired in part by the broken
power-law function known as the ``Nuker law'' \citep{lauer95,byun96} used to
model HST profiles of galaxy centers.  Our broken-exponential function is
described in more detail in \citet{erwin07-n1543}, but in brief it consists of
two exponential pieces joined by a transition region of variable
``sharpness'':
\begin{equation}
	I(r) \; = \; S \, I_{0} \, e^{\frac{-r}{\gamma}} [1 + e^{\alpha(r \, - \,
	R_{b})}]^{\frac{1}{\alpha} (\frac{1}{\gamma} \, - \, \frac{1}{\beta})},
\end{equation}
where $I_{0}$ is the central intensity of the inner exponential, $\gamma$ and
$\beta$ are the inner and outer exponential scale lengths (corresponding to
$h_{i}$ and $h_{o}$ in our piecewise fits), $R_{b}$ is the break radius, and
$\alpha$ parameterizes the sharpness of the break.  Low values of $\alpha$
mean very smooth, gradual breaks, while high values correspond to abrupt
transitions; typical values of $\alpha$ for our profiles range between 0.1 and
1.  $S$ is a scaling factor, given by
\begin{equation}
  S \; = \; (1 + e^{-\alpha R_{b}})^{\frac{1}{\alpha} (\frac{1}{\gamma} \, - 
  \, \frac{1}{\beta})}.
\end{equation}

\begin{figure*}
\vspace{0.15cm}
\centerline{\includegraphics[scale=0.9]{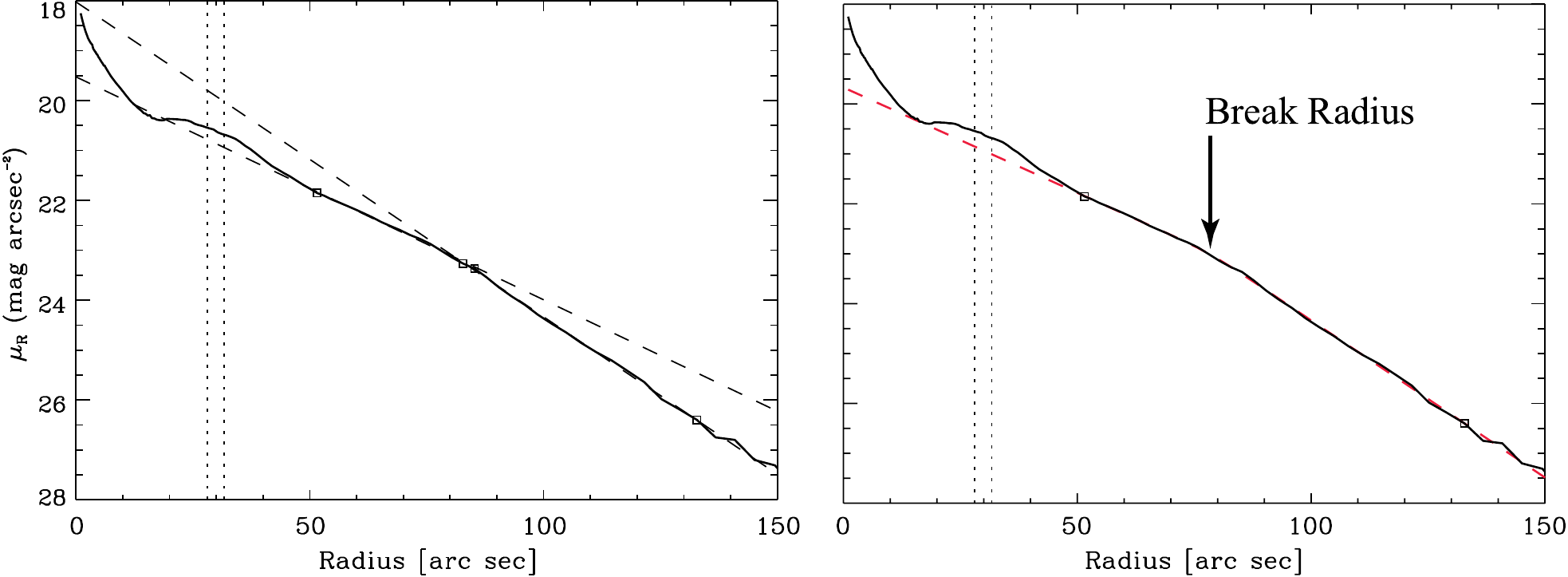}}

\vspace{0.2cm}
\caption{Example of a broken-exponential fit to the Type II.o profile of NGC
3729.  \textit{Left:} Piecewise fit to zones (delimited by small squares)
inside and outside apparent break.  \textit{Right:} Broken-exponential fit
(dashed red line) to the same profile; small squares mark the inner and outer
limits of the fitted region.  The break radius of the fit is $R_{b} =
78.5\arcsec$.\label{fig:broken-exp}}

\vspace{0.5cm}
\end{figure*}

The inner and outer scale lengths from the broken-exponential fits generally 
agree quite well with the piecewise fits: the median difference is 1--5\%, 
depending on whether one compares inner or outer scale lengths, and whether 
one is considering Type~II.o or Type~III profiles.

An example of a broken-exponential fit to one of our profiles is shown in 
Figure~\ref{fig:broken-exp}.  The only profiles for which we do not use this 
approach are those for which the inner region is non-exponential --- 
specifically, the Type~II.o profiles of NGC 2859 and NGC 3982, where the 
break radius and surface brightness is measured by eye.

For Type I and Type II.i profiles, we provide \textit{lower limits} on
potential break radii in Table~\ref{tab:results}, based on where the profile
reaches \mulim.  This is indicated by the ``$>$'' symbols in
Table~\ref{tab:results}.  Since the Type II.i profiles \textit{do} have breaks
at much smaller radii (i.e., near the end of the bar), we list these break
radii separately in Table~\ref{tab:2i}.  Break radii for Type III-s profiles
are listed in Table~\ref{tab:results}; these are approximate values measured
by eye, and indicate the radius where the outer excess light (presumed to come
from a spheroid rather than the disk) begins to deviate from the inner
exponential.

\section{Profiles and Notes for Individual Galaxies}\label{sec:profile-plots}

Figure~\ref{fig:profiles} presents the individual surface-brightness profiles
for all the galaxies in our sample.  For each galaxy, we plot the azimuthally
averaged surface brightness profile from our fixed-ellipse fits, exponential
fits to different regions of each profile, bar sizes (\amax{} and \lbar, from
Table~\ref{tab:basic}), and sizes of outer rings, if any (see
Table~\ref{tab:basic}).  We do not apply any extinction corrections, intrinsic
or Galactic, to these profiles.

\subsection{Note on Individual Galaxies}\label{sec:notes}

Here, we include notes for individual galaxies.  If a galaxy is \textit{not}
listed here, then the reader is directed to \citet{erwin03} and
\citet{erwin05} for explanations of how the bar sizes and disk orientations
were determined.  For the reader's convenience, we include the profile type 
immediately after the galaxy's name.

\textbf{NGC 718 (II.o-OLR).} See \citet{erwin03}.  The break in the profile at
$\sim 41\arcsec$ matches the outer pseudoring \citep[studied
by][]{kennicutt86}.  Note that there is an additional, faint ring at larger
radii, visible as a bump in the profile at $r \sim 70\arcsec$.  The scale
length (24\arcsec) of the major-axis fit by \citet{bba98} is intermediate
between our $h_{i}$ and $h_{o}$ values; this is not surprising, since their
fit extends somewhat beyond the break radius and thus includes light from both
parts of the profile.

\textbf{NGC 936 (II.o-OLR(?)).} See \citet{erwin03}.  This is one of the
stronger cases of a probable OLR system: there is no visible outer ring in
this S0 galaxy, but the break is at about twice the bar radius.  The Type II
(broken-exponential) nature of the profile is not always visible in shallower
exposures \citep[e.g.,][]{kent89,laurikainen05}, but has previously been noted
by \citet{kormendy84} and \citet{wozniak91}.  \citet{bba98} fit their
major-axis profile using a disk with a large hole, another indication of a
Type~II profile; their disk scale length (23.8\arcsec) is a reasonably good
match to our $h_{o} = 27.4\arcsec$.  The break -- and the relatively shallow
inner slope -- also agree with the ``outer lens'' classification by
\citet{kormendy79}.

\textbf{NGC 1022 (I).} See \citet{erwin03}.  Note that the outer ring is
associated with a slight excess in the profile at $r \sim 50$--60\arcsec; the
profile decreases beyond this point, but recovers to an exponential for $r
\gtrsim 100\arcsec$.  This is one of the few galaxies in our sample for which
we could find no $R$-band photometric calibration.

\textbf{NGC 2273 (II.o-CT).} See \citet{erwin03}.  This galaxy is notable for
having \textit{two} distinct outer rings (as well as an inner pseudoring just
outside the bar and a nuclear ring inside the bar); both outer rings are
associated with \hi{} \citep{vandriel91}.  As discussed in \citet{erwin07}, we
identify the smaller of the two outer rings with the bar's OLR, so the break
in the profile (just beyond the outermost ring) is then judged to be a
classical truncation.

\textbf{NGC 2681 (I).} See \citet{erwin99} and \citet{erwin03}.  Even
though we treat the ``outer disk'' as the region outside the largest
of this galaxy's three bars ($r \gtrsim 110\arcsec$), the exponential
fit to the outer disk matches the disk at $r \sim 30$--50\arcsec{}
(outside the middle bar) as well.

\textbf{NGC 2712 (I).} See \citet{erwin05} for notes on the bar measurements;
the disk orientation is from the \hi{} map of \citet{krumm82}.  There is a
small bump in the profile associated with the outer ring, whose existence and
size we report here for the first time; there is also a very extended excess
associated with spiral arms outside the bar ($r \sim 30$--60\arcsec).  Note
that in principle it might be possible to interpret this excess as a Type
II.o-OLR profile outside the bar, with the exponential at $r > 50\arcsec$
making the whole profile Type II.o-OLR + III like that of NGC 3982.

\textbf{NGC 2787 (I).} See \citet{erwin03} and \citet{erwin03-id}.
Scattered light problems prevent tracing the disk to fainter light
levels, and make it difficult to be certain about the possible excess
light at $r \gtrsim 130\arcsec$.  (Similar problems are present in
images taken with the Jacobus Kapteyn Telescope in 2001 January 29, obtained
from the ING Archive, which we used for the photometric calibration.)

\textbf{NGC 2859 (II.o-OLR).} See \citet{erwin03}.  This is one of three
galaxies with ``extreme outer-ring'' profiles; the profile between the end of
the bar and the outer ring is in this case clearly \textit{not} exponential,
and so we do not attempt to fit it.  The disk orientation parameters presented
here are updated slightly from those in \citet{erwin03} and \citet{erwin05},
and are based on the SDSS image.  \citet{laurikainen05} suggest a somewhat
higher inclination (outer-disk ellipticity = 0.24 instead of 0.13) than we
use, based on their deep $B$-band image; however, the profile derived using
their orientation parameters is very similar to the profile we present here.

\textbf{NGC 2880 (III-s).} See \citet{erwin03}.  This is the most dramatic
case of a Type III-s profile, with most of the light probably coming from a
spheroidal component; an E/S0 classification might be more appropriate for
this galaxy.  The central surface brightness of our exponential disk fit is
undoubtedly an overestimate, since it neglects any contribution from the
spheroid.

\textbf{NGC 2950 (II.o-CT).} See \citet{erwin03}.  There is a faint outer ring
\citep[e.g.,][]{hubble-atlas} associated with a very slight excess in the
profile at $r \sim 60\arcsec$.  The break in the profile is well outside this
ring, making this a classical truncation (II.o-CT) rather than an OLR break
(see Figure~\ref{fig:type2olr-vs-ct}).

\textbf{NGC 3049 (II.o-OLR(?)).} This profile is slightly ambiguous, since the
break is fairly close to the end of the bar, and the profile between the end
of the bar and the break suggests a steeper profile (with another break
inside).  This is the only profile for which we include a region
\textit{inside} the bar radius in the fit.  There is a slight hint of excess
light at $r > 90\arcsec$, hence we include ``[+ III?]'' in the label on the
plot.

The bar size measurements have been slightly updated from those published in
\citet{erwin05}, using a Spitzer 3.5$\mu$m (IRAC1) image from the Spitzer 
Infrared Nearby Galaxy Survey \citep{kennicutt03}.

\textbf{NGC 3412 (II.o-OLR(?)  + III-s).} See \citet{erwin03}.  As shown in
\citet{erwin05-type3} and Figure~\ref{fig:type3s}, this is a clear case of an
outer spheroid dominating the light at large radii; at smaller radii, the
profile is Type II.o-OLR(?).

\textbf{NGC 3485 (I).} An alternate interpretation of the profile might be Type 
II.i + III; however, the part of the profile from the end of the bar out to 
$r \sim 50\arcsec$ is perhaps more easily understood as an extended bar 
``shoulder'' (as for, e.g., NGC 2712 or NGC 3507).

\textbf{NGC 3489 (III-d).} See \citet{erwin03}.  This is a somewhat peculiar
Type III profile: the break is very smooth and gradual, which suggests that
the outer light could be from an additive, additional component (like a bulge
or halo); but the isophotes at $r > 150\arcsec$ are more elliptical than any
part of the galaxy except the outer ring; see Figure~21 of Erwin \& Sparke.

\textbf{NGC 3945 (II.o-OLR).} See \citet{erwin99} and \citet{erwin03}.  This
is another galaxy with a very strong outer ring, similar to NGC~2859 and
NGC~5701, thought the break is relatively sharp.

\textbf{NGC 3982 (II.o-OLR + III-d).} This is perhaps the most complicated
profile in our sample.  The profile out to $r \sim 50\arcsec$ is a good
example of the II.o form, with a ring right at the break.  This is one of
three galaxies with ``extreme outer-ring'' profiles; in this case, the profile
between the end of the bar and the outer ring is \textit{not} exponential.
(Due to the small size of this ring, \citet{rc1} described it as an ``inner
ring;'' but since it lies at almost exactly twice the bar radius, it is far
more likely to be an OLR-associated outer ring.)  Beyond 50\arcsec, the
profile changes to a shallower exponential, forming a Type III profile.
Although the galaxy is face-on, making determination of the outer geometry
difficult, there is a clear pair of spiral arms at $r > 50\arcsec$, indicating
that the outermost light is still from the disk
(Figure~\ref{fig:type3d-spirals}).

\citet{pohlen-trujillo} note the inner break associated with the ring but
exclude it from their analysis; consequently, they consider this a Type~III
profile only.  Our analysis of the Type III part of the profile matches
extremely well with theirs; the inner scale length and break radius
measurements agree to within 2\%, and the outer scale length measurements
(more affected by sky subtraction errors) differ by only 6\%.

We adopt the Cepheid-based distance estimate of \citet{riess05}, which is
based on a re-analysis of the data and estimates of \citet{saha01} and
\citet{stetson01}.

\textbf{NGC 4037 (II.o).} This galaxy is unusually low surface brightness for our
sample, with a central surface brightness of $\mu_{R} > 18$ mag arcsec$^{-2}$.
It is also somewhat unusual in having a break well outside the bar radius:
$R_{\rm brk} = 3.1 \, \amax$ and 2.4 \lbar).  As pointed out in
Section~\ref{sec:type2}, this is large enough to be potentially unrelated to
the bar's OLR, but not convincingly so; we leave it with a bare Type~II.o
classification.

\textbf{NGC 4102 (II.o-OLR + III-s(?)).} Our classification of the profile is
identical to that of \citet{pohlen-trujillo}, except that our deeper INT-WFC
image allows us to examine the shape of the isophotes outside the second break
radius better.  Since these show some sign of becoming rounder at larger
radii, as might be expected from a spheroid rather than an extension of the
(inclined) disk, we tentatively classify the it as III-s(?), and do not
attempt to fit an exponential to the outermost part of the profile.

We do, however, differ from Pohlen \& Trujillo in our analysis of the Type
II.o part of the profile.  We fit the short, nearly flat region from $r
\approx 27$--35\arcsec, which as it happens is the region just inside the
radius of the outer ring; Pohlen \& Trujillo fit a steeper region
\textit{outside} the outer ring (which we consider part of the transition
region).  Consequently, we have rather different values for $h_{i}$ and the
break radius.  Since our break radius is based on fitting a larger range of
the profile, on both sides of the outer ring, and since the resulting break
radius (42.9\arcsec) is closer to the radius of the outer ring (36\arcsec), we
consider our analysis more correct.

This is one of our deepest profiles, with $\mulim \approx 28.5$ $R$-mag
arcsec$^{-2}$.  The apparent turnover at $r < 5\arcsec$ is due to saturation
of the galaxy center in the INT-WFC images.

\textbf{NGC 4151 (I).} The best $R$-band image available is not large enough
to guarantee good sky subtraction, so we truncate our profile at a level well
above the nominal uncertainty (as determined from the corners of the image).
The resulting profile agrees very well with the profile from an SDSS image
(Fig.~\ref{fig:sdss}); unfortunately, the faintness of the outer disk means we
cannot use the SDSS image to probe any further out in radius.  The outer ring
size is measured directly from the image, since the value from \citet{rc1}
appears to be an underestimate.

\citet{laurikainen04} reported a disk scale length of 27.1\arcsec{} from a 2D
decomposition of a $B$-band image from the Ohio State University Bright Spiral
Galaxy Survey.  This is about one-third the value we find (78.6\arcsec; we
find a very similar scale length using the SDSS $r$-band image).

\textbf{NGC 4203 (II.o-OLR(?)  + III-s(?)).} This is a clear Type II.o profile
with a weak but significant excess over the outer exponential at $r \gtrsim
140\arcsec$.  The galaxy is close to face-on, which makes determining the
shape of the outer isophotes difficult.  However, there is evidence that the
isophotes become systematically rounder for $r > 100\arcsec$, so our tentative
conclusion is that the outer light may be coming from a spheroid.

The nature of the inner (Type II.o) break is also somewhat ambiguous, since
the bar size is uncertain; as noted by \citet{erwin03} and \citet{erwin05},
the large value of $\lbar = 46\arcsec$ is almost certainly an overestimate of
the bar size.  So it is difficult to judge whether the break is too far out
from the bar to be associated with its OLR. Nonetheless, the high surface
brightness of the break suggests it is \textit{not} due to a
(star-formation-related) classical truncation \citep[see][]{erwin07}, so we
tentatively classify it as an OLR break.

\textbf{NGC 4267 (I).}  For this galaxy, we adopt the distance reported by
\citet{mei07}, based on surface-brightness fluctuation measurements.

\textbf{NGC 4319 (III-d).} This galaxy possesses a spectacular outer 1-arm
spiral, which could also be interpreted as a tidal tail.  This structure
produces the dramatic bump in the profile at $r \sim 70\arcsec$, obscuring the
transition between the inner disk and the outer disk.  The III-d
classification --- and the break radius measurement -- is therefore somewhat
dubious.  A combination of bright stars and strong residual structure in the
image (including flat-fielding artifacts) prevents us from reliably tracing
the profile down to our nominal \mulim{} level.

\textbf{NGC 4340 (I).} The outer-disk orientation parameters for this galaxy
have been updated from those published in \citet{erwin05}; the new parameters
are based on our analysis of the SDSS image from DR5.  The adopted distance is
the mean distance to the Virgo Cluster from \citet{mei07}.

\textbf{NGC 4371 (III-d).} See \citet{erwin99} and \citet{erwin05}.  As shown
in Figure~\ref{fig:type3d}, the outer isophotes have roughly constant
ellipticity consistent with that of the inner disk, so this is a III-d
profile.  For this galaxy, we adopt the distance reported by \citet{mei07},
based on surface-brightness fluctuation measurements.  This is one of our
deepest profiles, with $\mulim \approx 28.2$ $R$-mag arcsec$^{-2}$.

\textbf{NGC 4477 (I).} The adopted distance is the mean Virgo Cluster distance
of \citet{mei07}.

\textbf{NGC 4531 (III-s).} This galaxy was \textit{not} included in the
bar-size study of \citet{erwin05} because no bar was visible in the optical
images.  However, analysis of an $H$-band image from the GOLDMine database
\citep{gavazzi03} indicates that there \textit{is} a bar in this galaxy.
Using the same approach as \citet{erwin05}, we find bar sizes of $\amax =
8.5\arcsec$ and $\lbar = 14\arcsec$ for this galaxy, with the bar having a
position angle of 127\arcdeg.  The distance is from \citet{tonry01}.

\textbf{NGC 4596 (I).} The adopted distance is the mean Virgo Cluster distance
of \citet{mei07}.  Because of the small size of the image, the outer reaches
of the galaxy along the semi-major axis lie outside the image, preventing us
from tracing the profile as far out as we might (even though the sky
background in the far corners of the image gives a very faint \mulim{} value).

\textbf{NGC 4608 (II.i).} The adopted distance is the mean Virgo Cluster
distance of \citet{mei07}.

\textbf{NGC 4612 (III-d).} For this galaxy, we adopt the
surface-brightness-fluctuation distance reported by \citet{mei07}.

\textbf{NGC 4691 (III-d).} See \citet{erwin03}.  The inner part of the profile
($r < 125\arcsec$ could conceivably be interpreted as a weak Type~II profile,
in which case this galaxy would be similar to NGC~3982.

\textbf{NGC 4699 (III-d).} The only $R$-band image we were able to obtain of
this galaxy was with the NOT. Due to the small field of view and strong
scattered light problems, we are not able to trace the outer disk profile out
as far as we might otherwise.  We note, however, that the basic Type~III
nature of the profile is present in a profile profile derived from the
publically available $B$-band image of the Ohio State University Bright Spiral
Galaxy Survey \citep{eskridge02}; unfortunately, the latter image also has a
small field of view, and so the precise nature of the profile beyond $r \sim
150\arcsec$ remains unknown.

\textbf{NGC 4995 (II.i).} An extremely bright star $\sim 4\arcmin$ north of
the galaxy produced a halo covering most of the central chip of the INT-WFC
array; consequently, our sky subtraction in the vicinity of the galaxy is
unreliable, and we stop tracing the profile at $\sim 100\arcsec$.  (The sky
background and uncertainty were measured in the extreme edges of the image,
and do not necessarily reflect the background nearer the galaxy.)

\textbf{NGC 5338 (II.o-CT(?)).} See \citet{erwin03}.  As explained in
Section~\ref{sec:type2}, the break is sufficiently far outside the bar
($R_{\rm brk} = 3.8 \, \amax$ and $2.8 \, \lbar$) that we consider an OLR
connection unlikely, although no outer ring is visible.  Consequently, we
consider this an (uncertain) example of a classical truncation (Type II.o-CT).

\textbf{NGC 5701 (II.o-OLR).} See \citet{erwin03}.  This is one of three
galaxies with ``extreme outer-ring'' profiles; the profile between the end of
the bar and the outer ring is only barely exponential, and the break is quite
gradual.  From visual inspection of the image \citep[e.g.,][]{erwin03}, it is
clear that the break is due to the luminous outer ring, as for NGC 2859 and
NGC 3945.

\textbf{NGC 5740 (III-d).} The isophotes outside the break radius show
signatures of lopsided, asymmetric spirals, and the overall ellipticity is
slightly higher than that of the isophotes inside the break radius.  Thus, we
consider the the light beyond the break radius to be part of the disk, and
classify this as a Type III-d profile.  This is one of our deepest profiles,
with $\mulim \approx 28.2$ $R$-mag arcsec$^{-2}$.

\textbf{NGC 5806 (III-d).} We find evidence for irregular and asymmetric
isophotes beyond the break radius in our INT-WFC image; in addition, the
isophotes retain a high ellipticity out to at least $r \sim 170\arcsec$.
Consequently, we consider this a III-d profile.  Our classification is
otherwise similar to that of \citet{pohlen-trujillo}, including very similar
values for the inner scale length and the break radius.

\textbf{NGC 5832 (II.i).} Our values for the disk orientation, which are
derived from the outer isophotes of our INT-WFC $r$-band image, agree well
with those derived from a Fabry-Perot \ha{} velocity field by \citet[][PA $=
45\arcdeg\pm3\arcdeg$, $i = 54\arcdeg\pm6\arcdeg$]{garrido05}.  This is one of
our deepest profiles, with $\mulim \approx 28.3$ $R$-mag arcsec$^{-2}$.

\textbf{NGC 5957 (II.o).} This galaxy was also observed with the NOT; the
resulting profile, like all of our NOT images, has significant problems with
sky subtraction (Section~\ref{sec:obs}) but clearly shows the same overall
Type~II.o shape as the SDSS profile.

\textbf{NGC 6012 (III-d).} Our images of this galaxy are strongly affected by
bright stars near the galaxy, which necessitated large-scale masking.  Our
profile should be considered somewhat uncertain, especially when it comes to
the parameters of the outer part of the Type III profile.

\textbf{NGC 6654 (II.o-OLR).} Our profile is very similar to that of
\citet{kormendy77}, who commented on the extreme shallowness of the inner
profile and the relatively sharp break; our deeper profile shows that the
outer slope is definitely exponential.

\textbf{NGC 7177 (III-d).} The shape of the outer isophotes is somewhat hard
to determine due to the low signal-to-noise, so our III-d classification is
uncertain.  The broad, smooth transition between the inner and outer
exponentials is similar to that of NGC 3489.

\textbf{NGC 7280 (II.o-OLR(?)).} The profile is derived from our combined deep
INT-WFC images, supplemented by a profile from our WIYN image at $r <
5\arcsec$, since the latter was higher resolution and was less saturated in
the center.  Figure~\ref{fig:comp} compares the individual profiles from both
images.  The composite profile is the deepest one in our sample, with $\mulim
\approx 29.0$ $R$-mag arcsec$^{-2}$.  The excess flux at $r > 100\arcsec$ in
the INT-WFC profile appears to be real, but this is a region where we have
heavily masked extended halos from nearby bright stars, so it is very
difficult to derive accurate isophote shapes.  Since the deviation from the
outer exponential starts at only $\sim 1$ mag arcsec$^{-2}$ above \mulim, we
consider this a tentative III profile only.

The inner part of the profile ($r < 100\arcsec$) is clearly Type II.o. Whether
this is likely an OLR break or a classical truncation is somewhat harder to
determine (there is no visible outer ring to help us), because the two
estimates of the bar length vary so much.  The break radius is at $\sim 4.4$
times \amax, but only 1.9 times \lbar.  Since the latter value is perfectly
consistent with an OLR interpretation, and since the break occurs at a high
surface brightness level inconsistent with classical-truncation predictions
\citep[see][]{erwin07}, we tentatively classify this as an OLR break.

\textbf{NGC 7743 (I).} Due to the relatively small size of the WIYN image, we
are unable to trace the profile of this galaxy very far out.  There is a
suggestion of a break, but nothing definite, so we consider this an
(uncertain) Type I profile.

\textbf{IC 676 (I).} See \citet{erwin03}.  There is an extended excess on top
of the underlying exponential between $r \approx 30$ and 55\arcsec, probably
due to the outer ring.

\textbf{IC 1067 (II.o-OLR(?)).} The outer-disk orientation parameters for this
galaxy have been updated from those published in \citet{erwin05}, based on our
analysis of the SDSS image and our INT-WFC images (the latter have higher S/N
than the SDSS image, but suffer from strong scattered light from nearby bright
stars).

Our classification of this galaxy as II.o-OLR(?) is essentially identical to
that of \citet{pohlen-trujillo}, though they were unable to measure the inner
scale length and break radius.

\textbf{UGC 3685 (I).} There is a suggestion of a downward break in the
profile at $r \sim 195\arcsec$, but beyond this point the profile appears to
recover and approach a continuation of the exponential fit, so we consider
this a Type~I profile.

\textbf{UGC 11920 (II.o-OLR(?)).} See \citet{erwin03}.  The profusion of
foreground stars in our image of this galaxy makes it impossible to trace the
profile as far out as our \mulim{} measurement would suggest.

\section{Discussion and Summary}

We have presented surface-brightness profiles for a sample 66 barred, S0 and
early-type spiral galaxies.  These profiles, derived from $R$-band images,
have been classified into several categories based on their overall shape.
Our basic classification is an extension of \nocite{freeman70}Freeman's (1970)
Type I and II system:
\begin{itemize}
  \item Type I: single exponential profiles, making up 27\% of the sample;
  \item Type II: profiles with a downward break, with a steep exponential 
  outside the break and (usually) a shallow exponential inside, making up 
  48\% of the sample;
  \item Type III (antitruncations): profiles with an upward break, with a steep 
  exponential inside the break and a shallow exponential (or sometimes non-exponential) 
  profile outside, making up at least 30\% of the sample.
\end{itemize}
There are four galaxies (6\% of the sample) which combine Type II and III
characteristics --- that is, they have a shallow profile beginning just
outside the bar, followed by a break to a steeper profile, followed by a break
to a shallower profile at the largest radii (e.g.,
Figure~\ref{fig:composite}).

We recognize several subdivisions within these general categories.  At a
descriptive level, we note that a few Type II profiles have ``inner'' breaks,
near or at the end of the bar (Type II.i), while the majority have ``outer'' 
breaks, beyond the end of the bar (Type II.o).

The Type II.i profiles are rare (only 6\% of the sample).  These appear
similar to profiles seen in $N$-body simulations of bar formation
\citep[e.g.,][]{athan02,valenzuela03}, and thus we suspect they are
stellar-dynamical side effects of the bar-formation process.

Type II.o profiles, on the other hand, are common (42\% of the sample), and
appear to fall into two subtypes.  Type II.o-OLR profiles (35\% of the sample)
are those where the break is close to an outer ring, or to where an outer ring
would be expected (2--3 times the bar radius).  Since outer rings are
understood as being due to the Outer Lindblad Resonance (OLR) of the bar
\citep[e.g.,][]{buta96}, we suggest that these breaks are associated with the
OLRs as well \citep[see][]{erwin07}.  Profiles where the break is located well
outside the outer ring (or the likely radius of the bar's OLR, if there is no
visible outer ring) are assumed to be similar to the ``classical truncations''
of late-type spirals and are called Type II.o-CT; these make up only 5\% of
the sample.  (One galaxy, NGC 4037, is ambiguous and could fall into either of
these subtypes.)  More detailed arguments in favor of these respective
classifications will be presented in \citet{erwin07}.

We also subdivide the Type III profiles on an interpretative basis
\citep[see][]{erwin05-type3}.  In some cases, we can use the shape of the
outer isophotes, or the presence of clear disk features such as spiral arms,
to argue that the outer part of the profile is still part of the disk (Type
III-d, 18\% of the sample).  In other cases, the morphology suggests that the
outer part of the profile is actually due to light from a separate, more
spheroidal component, added to that of the disk (Type III-s, 12\% of the
sample); these may be cases of extended bulges or luminous stellar halos.

There are four galaxies with limited evidence for Type III profiles at very
large radii; these are indicated in the Notes column of
Table~\ref{tab:results}.  Taking these into account would raise the Type III
frequency to 36\%, and in principle it could be higher still, since we may be
missing light at large radii in some of our shallower images.

\citet{pohlen-trujillo}, who used essentially the same classification system
as presented here, studied a sample of late-type spirals (Sb--Sm).  Although
the sample selection was not identical, it is similar enough to make for an
interesting comparison, particularly if we focus on the barred galaxies (that
is, those galaxies with SB or SAB classifications from \nocite{rc3}RC3) in
their sample.  In what follows, we will use ``late types'' to refer to the
barred, Sbc--Sm subset of the Pohlen \& Trujillo sample (47 of the 85 galaxies
in their sample with disk-profile classifications).

The comparison shows some striking differences between early and late types.
In particular, Type~I profiles --- the paradigmatic single-exponential disk
profiles --- are much rarer in the late types ($11 \pm 5$\% vs.\ $27 \pm
5$\%), while Type~II profiles are significantly \textit{more} common in the
late types ($77 \pm 6$\% vs.\ $48 \pm 6$\%).  Type~III profiles are slightly
rarer in the late types, but not significantly so ($23 \pm 6$\% vs.\ $30 \pm
6$\%).  When we look at subclasses, other differences appear.  By far the most
common subclass of Type~II.o profiles in the early types are the OLR breaks
(Type II.o-OLR), which makes up $35 \pm 6$\% of our sample.  But in the late
types, OLR breaks are only $11 \pm 5$\% of the sample, while classical
truncations (found in only $5 \pm 3$\% of the early types) are found in almost
half ($45 \pm 7$\%) of the late types; late types also have
apparent/asymmetric breaks (Type II-AB, $19 \pm 6$\% of the sample), which we
do not find in the early types.  Curiously, Type II.i profiles seem to be
equally rare in both early and late types ($6 \pm$ 3\% and $4 \pm 3$\%,
respectively).

A similar analysis of the corresponding \textit{unbarred} S0 and early-type
spirals will be presented and discussed in a subsequent paper \citep{gutierrez07}.  Other
papers will examine the nature of the Type II and III profiles, their
connection to bar properties and galaxy environments, and will use them to
test star-formation and galaxy-formation models.

\acknowledgments 

We would like to thank, among others, Andrew Cardwell, Ignacio Trujillo,
Alister Graham, Bo Milvang-Jensen, and Bruce Elmegreen for helpful and
stimulating conversations.  We are also grateful to Juan Carlos Vega
Beltr\'{a}n for his help with the Nordic Optical Telescope observations, and
to Paul Schechter for the MDM observations of NGC~4612.  Gabriel P\'{e}rez
helped prepare Figure~\ref{fig:scheme}.  Finally, an anonymous referee made 
several helpful suggestions.

P.E. was supported by DFG Priority Program 1177.  M.P. was (partly)
supported by a Marie Curie Intra-European Fellowship within the 6th European
Community Framework Programme.  This work was also supported by grants No.\
AYA2004-08251-CO2-01 from the Spanish Ministry of Education and Science and
P3/86 of the Instituto de Astrofisica de Canarias.  P.E. and M.P. acknowledge the hospitality of the IAC, where much of the initial research was carried out.

This research is based in part on data from the ING Archive, and on
observations made with both the Isaac Newton Group of Telescopes, operated on
behalf of the UK Particle Physics and Astronomy Research Council (PPARC) and
the Nederlandse Organisatie voor Wetenschappelijk Onderzoek (NWO) on the
island of La Palma, and the Nordic Optical Telescope, operated on the island
of La Palma jointly by Denmark, Finland, Iceland, Norway, and Sweden.  Both
the ING and the NOT are part of the Spanish Observatorio del Roque de los
Muchachos of the Instituto de Astrof\'{\i}sica de Canarias.  We also used
images from the Barred and Ringed Spirals (BARS) database, for which time was
awarded by the Comit\'e Cient\'{\i}fico Internacional of the Canary Islands
Observatories.

The data from the NOT presented here have been taken using ALFOSC, which is
owned by the Instituto de Astrofisica de Andalucia (IAA) and operated at the
Nordic Optical Telescope under agreement between IAA and the NBIfAFG of the
Astronomical Observatory of Copenhagen.

Funding for the creation and distribution of the SDSS Archive has been
provided by the Alfred P. Sloan Foundation, the Participating
Institutions, the National Aeronautics and Space Administration, the
National Science Foundation, the U.S. Department of Energy, the
Japanese Monbukagakusho, and the Max Planck Society.  The SDSS Web
site is http://www.sdss.org/.

The SDSS is managed by the Astrophysical Research Consortium (ARC) for
the Participating Institutions.  The Participating Institutions are
The University of Chicago, Fermilab, the Institute for Advanced Study,
the Japan Participation Group, The Johns Hopkins University, the
Korean Scientist Group, Los Alamos National Laboratory, the
Max-Planck-Institute for Astronomy (MPIA), the Max-Planck-Institute
for Astrophysics (MPA), New Mexico State University, University of
Pittsburgh, University of Portsmouth, Princeton University, the United
States Naval Observatory, and the University of Washington.

Finally, this research has made extensive use of the Lyon-Meudon Extragalactic
Database (LEDA; http://leda.univ-lyon1.fr) and of the NASA/IPAC Extragalactic
Database (NED); the latter is operated by the Jet Propulsion Laboratory,
California Institute of Technology, under contract with the National
Aeronautics and Space Administration.



\begin{deluxetable}{llrrrrrrrrrr}
\tablewidth{0pt}
\tablecaption{Basic Galaxy Data\label{tab:basic}} 
\tablecolumns{12}
\tablehead{
\colhead{Name} & \colhead{Type (RC3)} & \colhead{Distance} & 
\colhead{Source} & \colhead{$M_{B}$} & \colhead{$R_{25}$} &
\multicolumn{2}{l}{Outer Disk} & \multicolumn{2}{l}{Bar Size} & 
\colhead{$R_{\rm OR}$} & \colhead{Source} \\
 & & & & & & PA & $i$ & \amax{} & \lbar{} & & \\
 & & (Mpc) & & & ($\arcsec$) & ($\arcdeg$) & ($\arcdeg$) & ($\arcsec$) & 
 ($\arcsec$) & ($\arcsec$) & \\
\colhead{(1)} & \colhead{(2)} & \colhead{(3)} & \colhead{(4)} & \colhead{(5)} &
\colhead{(6)} & \colhead{(7)} & \colhead{(8)} & \colhead{(9)} & \colhead{(10)} &
\colhead{(11)} & \colhead{(12)}}

\startdata
 NGC 718 &                  SAB(s)a &  22.6 &   & $-19.43$ &   71 &    5 &   30 &   21 &   31 &   42 & 1 \\
 NGC 936 &            SB(rs)$0^{+}$ &  23.0 &   & $-20.86$ &  140 &  130 &   41 &   49 &   61 &      &   \\
NGC 1022 &     (R$^{\prime}$)SB(s)a &  18.1 &   & $-19.46$ &   72 &  174 &   24 &   20 &   24 &   56 & 1 \\
NGC 2273 &                  SB(r)a: &  27.3 &   & $-20.11$ &   97 &   50 &   50 &   21 &   25 &   66 & 1 \\
NGC 2681 & (R$^{\prime}$)SAB(rs)0/a &  17.2 &   & $-20.20$ &  109 &  140 &   18 &   52 &   63 &   68 & 2 \\
NGC 2712 &                  SB(r)b: &  26.5 &   & $-19.88$ &   87 &   10 &   59 &   26 &   28 &   85 & 3 \\
NGC 2787 &             SB(r)$0^{+}$ &   7.5 &   & $-18.20$ &   95 &  109 &   55 &   43 &   54 &      &   \\
NGC 2859 &          (R)SB(r)$0^{+}$ &  24.2 &   & $-20.21$ &  128 &   85 &   32 &   40 &   50 &  102 & 1 \\
NGC 2880 &                SB$0^{-}$ &  21.9 &   & $-19.38$ &   62 &  144 &   52 &   12 &   14 &      &   \\
NGC 2950 &          (R)SB(r)$0^{0}$ &  14.9 &   & $-19.14$ &   80 &  125 &   48 &   29 &   37 &   61 & 1 \\
NGC 2962 &        (R)SAB(rs)$0^{+}$ &  30.0 & 1 & $-19.71$ &   79 &    7 &   53 &   32 &   47 &   66 & 1 \\
NGC 3049 &                 SB(rs)ab &  20.2 &   & $-18.65$ &   66 &   26 &   51 &   35 &   38 &      &   \\
NGC 3185 &                (R)SB(r)a &  17.5 &   & $-18.61$ &   71 &  140 &   49 &   35 &   36 &   75 & 1 \\
NGC 3351 &                   SB(r)b &  10.0 & 2 & $-19.94$ &  222 &   10 &   46 &   74 &   83 &  165 & 2 \\
NGC 3368 &                SAB(rs)ab &  10.5 & 2 & $-20.37$ &  228 &  172 &   50 &   86 &  106 &  185 & 1 \\
NGC 3412 &             SB(s)$0^{0}$ &  11.3 &   & $-18.98$ &  109 &  152 &   58 &   24 &   34 &      &   \\
NGC 3485 &                  SB(r)b: &  20.0 &   & $-19.03$ &   69 &    5 &   26 &   21 &   25 &      &   \\
NGC 3489 &           SAB(rs)$0^{+}$ &  12.1 &   & $-19.45$ &  106 &   71 &   58 &   22 &   29 &   51 & 4 \\
NGC 3504 &              (R)SAB(s)ab &  22.3 &   & $-20.29$ &   81 &  149 &   22 &   29 &   34 &   63 & 2 \\
NGC 3507 &                   SB(s)b &  14.2 &   & $-19.21$ &  102 &   90 &   27 &   26 &   30 &      &   \\
NGC 3729 &                   SB(r)a &  16.8 &   & $-19.35$ &   85 &  174 &   53 &   28 &   32 &      &   \\
NGC 3941 &             SB(s)$0^{0}$ &  12.2 & 3 & $-19.31$ &  104 &    8 &   52 &   23 &   35 &      &   \\
NGC 3945 &         (R)SB(rs)$0^{+}$ &  19.8 &   & $-19.94$ &  158 &  158 &   55 &   56 &   68 &  135 & 1 \\
NGC 3982 &                 SAB(r)b: &  20.9 & 4 & $-19.95$ &   70 &   17 &   29 &    4 &    5 &   10 & 1 \\
NGC 4037 &                 SB(rs)b: &  13.5 &   & $-17.79$ &   75 &  150 &   32 &   29 &   36 &      &   \\
NGC 4045 &                  SAB(r)a &  26.8 &   & $-19.70$ &   81 &   90 &   48 &   26 &   29 &   66 & 4 \\
NGC 4102 &                 SAB(s)b? &  14.4 &   & $-19.22$ &   91 &   38 &   55 &   12 &   18 &   36 & 1 \\
NGC 4143 &            SAB(s)$0^{0}$ &  15.9 &   & $-19.40$ &   68 &  144 &   59 &   19 &   32 &      &   \\
NGC 4151 & (R$^{\prime}$)SAB(rs)ab: &  15.9 &   & $-20.70$ &  189 &   22 &   20 &   69 &   95 &  230 & 3 \\
NGC 4203 &               SAB$0^{-}$ &  15.1 &   & $-19.21$ &  102 &   11 &   28 &   13 &   46 &      &   \\
NGC 4245 &                 SB(r)0/a &  12.0 & 5 & $-18.28$ &   87 &  173 &   38 &   41 &   46 &      &   \\
NGC 4267 &            SB(s)$0^{-}$? &  15.8 & 6 & $-19.32$ &   97 &  127 &   25 &   20 &   29 &      &   \\
NGC 4314 &                  SB(rs)a &  12.0 & 5 & $-19.12$ &  125 &   65 &   25 &   74 &   88 &  109 & 2 \\
NGC 4319 &                  SB(r)ab &  23.5 &   & $-19.26$ &   89 &  135 &   42 &   16 &   18 &   38 & 1 \\
NGC 4340 &             SB(r)$0^{+}$ &  16.5 & 7 & $-19.07$ &  105 &  101 &   56 &   67 &   82 &      &   \\
NGC 4371 &             SB(r)$0^{+}$ &  17.0 & 6 & $-19.55$ &  119 &   86 &   58 &   64 &   75 &      &   \\
NGC 4386 &              SAB$0^{0}$: &  27.0 &   & $-19.68$ &   74 &  140 &   48 &   25 &   36 &      &   \\
NGC 4477 &           SB(s)$0^{0}$:? &  16.5 & 7 & $-19.86$ &  114 &   80 &   33 &   29 &   43 &      &   \\
NGC 4531 &               SB$0^{+}$: &  15.2 & 3 & $-18.67$ &   93 &  154 &   49 &   10 &   16 &      &   \\
NGC 4596 &             SB(r)$0^{+}$ &  16.5 & 7 & $-19.80$ &  119 &  120 &   42 &   62 &   68 &      &   \\
NGC 4608 &             SB(r)$0^{0}$ &  16.5 & 7 & $-19.19$ &   97 &  100 &   36 &   54 &   60 &      &   \\
NGC 4612 &            (R)SAB$0^{0}$ &  16.6 & 6 & $-19.19$ &   74 &  143 &   44 &   22 &   26 &   67 & 1 \\
NGC 4643 &                SB(rs)0/a &  18.3 &   & $-19.85$ &   93 &   53 &   38 &   63 &   78 &      &   \\
NGC 4665 &                 SB(s)0/a &  10.9 &   & $-18.87$ &  114 &  120 &   26 &   49 &   71 &   72 & 1 \\
NGC 4691 &              (R)SB(s)0/a &  15.1 &   & $-19.43$ &   85 &   30 &   38 &   35 &   53 &   81 & 1 \\
NGC 4699 &                 SAB(rs)b &  18.9 &   & $-21.37$ &  114 &   37 &   42 &   13 &   16 &   81 & 1 \\
NGC 4725 &                 SAB(r)ab &  12.4 & 2 & $-20.69$ &  321 &   40 &   42 &  119 &  127 &  300 & 1 \\
NGC 4754 &            SB(r)$0^{-}$: &  16.8 & 3 & $-19.78$ &  137 &   22 &   62 &   44 &   52 &      &   \\
NGC 4995 &                  SAB(r)b &  23.6 &   & $-20.41$ &   74 &   93 &   47 &   22 &   27 &      &   \\
NGC 5338 &               SB$0^{0}$: &  12.8 &   & $-16.70$ &   76 &   95 &   66 &   17 &   23 &      &   \\
NGC 5377 &                (R)SB(s)a &  27.1 &   & $-20.29$ &  111 &   25 &   59 &   67 &   77 &  110 & 1 \\
NGC 5701 &             (R)SB(rs)0/a &  21.3 &   & $-19.97$ &  128 &   45 &   20 &   41 &   60 &  102 & 1 \\
NGC 5740 &                 SAB(rs)b &  22.0 &   & $-19.67$ &   89 &  161 &   60 &   18 &   20 &      &   \\
NGC 5750 &                 SB(r)0/a &  26.6 &   & $-19.94$ &   91 &   65 &   62 &   37 &   41 &      &   \\
NGC 5806 &                  SAB(s)b &  19.2 &   & $-19.67$ &   93 &  166 &   58 &   38 &   39 &      &   \\
NGC 5832 &                 SB(rs)b? &   9.9 &   & $-17.15$ &  111 &   45 &   55 &   43 &   49 &      &   \\
NGC 5957 &    (R$^{\prime}$)SAB(r)b &  26.2 &   & $-19.36$ &   85 &  100 &   15 &   24 &   27 &      &   \\
NGC 6012 &              (R)SB(r)ab: &  26.7 &   & $-19.78$ &   63 &   45 &   33 &   34 &   41 &      &   \\
NGC 6654 &   (R$^{\prime}$)SB(s)0/a &  28.3 &   & $-19.65$ &   79 &    0 &   44 &   27 &   39 &   56 & 1 \\
NGC 7177 &                  SAB(r)b &  16.8 &   & $-19.79$ &   93 &   83 &   48 &   14 &   16 &      &   \\
NGC 7280 &            SAB(r)$0^{+}$ &  24.3 &   & $-19.16$ &   66 &   74 &   50 &   10 &   23 &      &   \\
NGC 7743 &          (R)SB(s)$0^{+}$ &  20.7 &   & $-19.49$ &   91 &  105 &   28 &   31 &   37 &  117 & 1 \\
  IC 676 &          (R)SB(r)$0^{+}$ &  19.4 &   & $-18.42$ &   74 &   15 &   47 &   15 &   21 &   55 & 4 \\
 IC 1067 &                   SB(s)b &  22.2 &   & $-18.82$ &   64 &  120 &   44 &   21 &   21 &      &   \\
UGC 3685 &                  SB(rs)b &  26.8 &   & $-19.51$ &   99 &  119 &   31 &   25 &   27 &      &   \\
UGC 11920 &                    SB0/a &  18.0 &   & $-19.71$ &   72 &   50 &   52 &   26 &   39 &      &   \\
\enddata

\tablecomments{Col.\ (1) Galaxy name; (2) Hubble type from RC3; (3) distance
in Mpc, from LEDA unless otherwise specified; (4) source for distance if not
from LEDA: 1 = \citet{ajhar01}, 2 = \citet{freedman01}, 3 = \citet{tonry01}, 4
= \citet{riess05}, 5 = \citet{forbes96}, 6 = \citet{mei07}, 7 = mean Virgo
Cluster distance from Mei et al.; (5) absolute blue magnitude, based on
$B_{\rm tc}$ in LEDA; (6) one-half of the corrected $\mu_{B} = 25$ magnitude
diameter $D_{0}$ from RC3; (7) and (8) position angle and inclination of the
outer disk, from \citet{erwin05}; (9) and (10) lower and upper limits on the
bar length, from \citet{erwin05}, deprojected; (11) semi-major axis of the
outer ring, if any; (12) source for outer-ring measurement: 1 = \citet{rc1}, 2
= \citet{bc93}, 3 = this paper, 4 = \citet{erwin03}.}

\end{deluxetable}

\begin{deluxetable}{lrrrr}
\tablewidth{0pt}
\tablecaption{INT-WFC Photometric Calibrations\label{tab:calib}} 
\tablecolumns{5}
\tablehead{
\colhead{Date} & \colhead{Filter} & \colhead{$Z$} & \colhead{$k_{1}$} & 
\colhead{$k_{2}$}}

\startdata
2003 Sept.\ 19 & $B$ & 24.683 & $-0.165$ & 0.033 \\
2003 Sept.\ 19 & $R$ & 24.760 & $-0.090$ & $-0.110$ \\

2004 March 14  & $B$ & \nodata & \nodata & \nodata \\
2004 March 14  & $R$ & 24.682 & $-0.116$ & 0.006 \\

2004 March 15  & $B$ & 24.469 & $-0.085$ & 0.132 \\
2004 March 15  & $R$ & 24.614 & $-0.122$ & 0.071 \\

2004 March 16  & $B$ & 24.833 & $-0.263$ & 0.059 \\
2004 March 16  & $R$ & 24.776 & $-0.115$ & $-0.057$ \\

2004 March 17  & $B$ & 24.767 & $-0.255$ & 0.074 \\
2004 March 17  & $R$ & 26.649 & $-0.117$ & 0.027 \\
\enddata

\tablecomments{Photometric parameters for our INT-WFC observations; the zero
point for a given night is $Z + k_{1}X + k_{2}(B - R)$, where $X$ is the
airmass and $B - R$ is the color of the object.  See Section~\ref{sec:calib}
for more details.}

\end{deluxetable}

\begin{deluxetable}{lllrrrr}
\tablewidth{0pt}
\tablecaption{Observations and Calibrations\label{tab:obs}} 
\tablecolumns{7}
\tablehead{
\colhead{Galaxy} & \colhead{Telescope/Instrument} & \colhead{Date} & \colhead{$t_{exp}$ (s)} &
\colhead{Filter} & \colhead{Calibration} & \colhead{Notes}}

\startdata
 NGC 718 &   INT-WFC  &  2003 Sep 19  & 10200 &  $r$ & standards & \\
 NGC 936 &   INT-WFC  &  2003 Sep 21  &   600 &  $r$ & PH98      & \\
         &   SDSS     &  \nodata      &    54 &  $r$ & SDSS      & \\
NGC 1022 &   WIYN     &  1997 Mar 2   &   300 &  $R$ & \nodata   & 1 \\
NGC 2273 &   WIYN     &  1995 Dec 27  &   300 &  $R$ & INT-WFC   & \\
         &   INT-WFC  &  2004 Mar 15  &   120 &  $r$ & standards & \\
NGC 2681 &   WIYN     &  1997 Mar  2  &   300 &  $R$ & SDSS      & \\
         &   SDSS     &  \nodata      &    54 &  $r$ & SDSS      & \\
NGC 2712 &   INT-WFC  &  2000 Mar 29  &   150 &  $R$ & SDSS      & 2 \\
         &   SDSS     &  \nodata      &    54 &  $r$ & SDSS      & \\
NGC 2787 &   WIYN     &  1995 Dec 26  &   240 &  $R$ & JKT       & \\
         &   JKT      &  2001 Jan 29  &  2000 &  $R$ & standards & 2 \\
NGC 2859 &   SDSS     &  \nodata      &    54 &  $r$ & SDSS      & \\
         &   WIYN     &  1995 Dec 26  &   180 &  $R$ & \nodata   & \\
NGC 2880 &   WIYN     &  1997 Mar 2   &   300 &  $R$ & INT-WFC   & \\
         &   INT-WFC  &  2004 Mar 16  &  1200 &  $r$ & standards & \\
NGC 2950 &   SDSS     &  \nodata      &    54 &  $r$ & SDSS      & \\
NGC 2962 &   WIYN     &  1997 Mar 2   &   300 &  $R$ & PH98      & \\
         &   SDSS     &  \nodata      &    54 &  $r$ & SDSS      & \\
NGC 3049 &   SDSS     &  \nodata      &    54 &  $r$ & SDSS      & \\
         &   NOT      &  2002 Apr 18  &   600 &  $R$ & \nodata   & \\
NGC 3185 &   SDSS     &  \nodata      &    54 &  $r$ & SDSS      & \\
NGC 3351 &   SDSS     &  \nodata      &    54 &  $r$ & SDSS      & \\
NGC 3368 &   SDSS     &  \nodata      &    54 &  $r$ & SDSS      & \\
NGC 3412 &   INT-WFC  &  2004 Mar 15  &  2400 &  $r$ & standards & \\
         &   SDSS     &  \nodata      &    54 &  $r$ & SDSS      & \\
NGC 3485 &   NOT      &  2002 Apr 18  &   600 &  $R$ & PH98      & \\
NGC 3489 &   INT-WFC  &  2004 Mar 14  &  1200 &  $r$ & standards & \\
         &   SDSS     &  \nodata      &    54 &  $r$ & SDSS      & \\
NGC 3504 &   JKT      &  1998 Mar  2  &   600 &  $R$ & PH98      & 2 \\
NGC 3507 &   SDSS     &  \nodata      &    54 &  $r$ & SDSS      & \\
         &   NOT      &  2002 Apr 18  &   600 &  $R$ & \nodata   & \\
NGC 3729 &   INT-WFC  &  2004 Mar 16  &   120 &  $r$ & standards & \\
         &   SDSS     &  \nodata      &    54 &  $r$ & SDSS      & \\
NGC 3941 &   INT-WFC  &  2004 Mar 15  &   120 &  $r$ & standards & \\
         &   SDSS     &  \nodata      &    54 &  $r$ & SDSS      & \\
NGC 3945 &   INT-WFC  &  2004 Mar 15  &   120 &  $r$ & standards & \\
         &   SDSS     &  \nodata      &    54 &  $r$ & SDSS      & \\
NGC 3982 &   SDSS     &  \nodata      &    54 &  $r$ & SDSS      & \\
NGC 4037 &   INT-WFC  &  2004 Mar 16  &   120 &  $r$ & standards & \\
         &   SDSS     &  \nodata      &    54 &  $r$ & SDSS      & \\
NGC 4045 &   SDSS     &  \nodata      &    54 &  $r$ & SDSS      & \\
         &   INT-WFC  &  2004 Mar 16  &   120 &  $r$ & standards & \\
         &   WIYN     &  1998 Mar 21  &   300 &  $R$ & \nodata   & \\
NGC 4102 &   INT-WFC  &  2004 Mar 17  &  1200 &  $r$ & standards & \\
         &   SDSS     &  \nodata      &    54 &  $r$ & SDSS      & \\
NGC 4143 &   INT-WFC  &  2004 Mar 16  &   120 &  $r$ & standards & \\
         &   SDSS     &  \nodata      &    54 &  $r$ & SDSS      & \\
NGC 4151 &   INT-PFCU &  1996 Feb 11  &   600 &  $R$ & SDSS      & 3 \\
         &   SDSS     &  \nodata      &    54 &  $r$ & SDSS      & \\
NGC 4203 &   SDSS     &  \nodata      &    54 &  $r$ & SDSS      & \\
         &   INT-WFC  &  2004 Mar 16  &    30 &  $r$ & standards & \\
NGC 4245 &   INT-WFC  &  2004 Mar 14  &  1200 &  $r$ & standards & \\
         &   SDSS     &  \nodata      &    54 &  $r$ & SDSS      & \\
NGC 4267 &   SDSS     &  \nodata      &    54 &  $r$ & SDSS      & \\
NGC 4314 &   INT-WFC  &  2004 Mar 17  &    60 &  $r$ & standards & \\
NGC 4319 &   JKT      &  1998 Dec 24  &  3000 &  $R$ & \nodata   & 1,2 \\
NGC 4340 &   SDSS     &  \nodata      &    54 &  $r$ & SDSS      & \\
NGC 4371 &   INT-WFC  &  2004 Mar 16  &  1200 &  $r$ & standards & \\
         &   SDSS     &  \nodata      &    54 &  $r$ & SDSS      & \\
NGC 4386 &   INT-WFC  &  2004 Mar 17  &    60 &  $r$ & standards & \\
NGC 4477 &   SDSS     &  \nodata      &    54 &  $r$ & SDSS      & \\
NGC 4531 &   SDSS     &  \nodata      &    54 &  $r$ & SDSS      & \\
NGC 4596 &   INT-PFCU &  1996 Feb 11  &   600 &  $R$ & SDSS      & 3 \\
         &   SDSS     &  \nodata      &    54 &  $r$ & SDSS      & \\
NGC 4608 &   SDSS     &  \nodata      &    54 &  $r$ & SDSS      & \\
NGC 4612 &   MDM      &  1996 Mar 13  &   300 &  $R$ & SDSS      & \\
         &   SDSS     &  \nodata      &    54 &  $r$ & SDSS      & \\
NGC 4643 &   INT-WFC  &  2004 Mar 17  &    60 &  $r$ & standards & \\
NGC 4665 &   SDSS     &  \nodata      &    54 &  $r$ & SDSS      & \\
NGC 4691 &   INT-WFC  &  2004 Mar 17  &    60 &  $r$ & standards & \\
NGC 4699 &   NOT      &  2001 Apr 10  &   600 &  $R$ & PH98      & 4 \\
\enddata
\end{deluxetable}

\addtocounter{table}{-1}

\begin{deluxetable}{lllrrrr}
\tablewidth{0pt}
\tablecaption{continued} 
\tablecolumns{7}
\tablehead{
\colhead{Galaxy} & \colhead{Telescope/Instrument} & \colhead{Date} & \colhead{$t_{exp}$ (s)} &
\colhead{Filter} & \colhead{Calibration} & \colhead{Notes}}

\startdata
NGC 4725 &   SDSS     &  \nodata      &    54 &  $r$ & SDSS      & \\
NGC 4754 &   SDSS     &  \nodata      &    54 &  $r$ & SDSS      & \\
NGC 4995 &   INT-WFC  &  2004 Mar 17  &   600 &  $r$ & standards & 5 \\
         &   NOT      &  2001 Apr 10  &   600 &  $R$ & \nodata   & \\
NGC 5338 &   INT-WFC  &  2004 Mar 15  &   600 &  $r$ & standards & \\
         &   SDSS     &  \nodata      &    54 &  $r$ & SDSS      & \\
NGC 5377 &   SDSS     &  \nodata      &    54 &  $r$ & SDSS      & \\
NGC 5701 &   Lowell   &  1989 Apr 1   &   600 &  $R$ & PH98      & 6 \\
NGC 5740 &   INT-WFC  &  2004 Mar 16  &   600 &  $r$ & standards & \\
         &   SDSS     &  \nodata      &    54 &  $r$ & SDSS      & \\
NGC 5750 &   INT-WFC  &  2004 Mar 16  &   120 &  $r$ & standards & \\
NGC 5806 &   SDSS     &  \nodata      &    54 &  $r$ & SDSS      & \\
         &   INT-WFC  &  2004 Mar 17  &   600 &  $r$ & standards & \\
NGC 5832 &   INT-WFC  &  2004 Mar 16  &   600 &  $r$ & standards & \\
NGC 5957 &   SDSS     &  \nodata      &    54 &  $r$ & SDSS      & \\
         &   NOT      &  2002 Apr 18  &   600 &  $R$ & \nodata   & \\
NGC 6012 &   INT-WFC  &  2004 Mar 15  &   600 &  $r$ & standards & \\
NGC 6654 &   INT-WFC  &  2003 Sep 21  &  1200 &  $r$ & INT-WFC   & 7 \\
         &   INT-WFC  &  2004 Mar 16  &   600 &  $r$ & standards & \\
NGC 7177 &   INT-WFC  &  2000 Jul 31  &  1200 &  $R$ & PH98      & 2 \\
NGC 7280 &   WIYN     &  1996 Oct 10  &   300 &  $R$ & PH98      & \\
         &   INT-WFC  &  2003 Sep 20  & 10200 &  $r$ & PH98      & \\
NGC 7743 &   WIYN     &  1996 Oct 9   &   300 &  $R$ & PH98      & 4 \\
IC 676   &   WIYN     &  1997 Mar 2   &   300 &  $R$ & INT-WFC   & \\
         &   INT-WFC  &  2004 Mar 15  &   120 &  $r$ & standards & \\
IC 1067  &   SDSS     &  \nodata      &    54 &  $r$ & SDSS      & \\
UGC 3685 &   INT-WFC  &  2004 Mar 15  &   600 &  $r$ & standards & \\
UGC 11920 &   WIYN     &  1996 Jun 16  &   300 &  $R$ & \nodata   & 1 \\
\enddata

\tablecomments{For each galaxy, the \textit{first} line lists the main data
source, used for generating the surface-brightness profile; subsequent lines
list any additional observations/sources used for comparison, calibration,
etc.  We do not list the observation date for SDSS images.  $t_{exp} =$
cumulative exposure time in seconds.  Filter = filter used for observations
(note that all calibrations are in Cousins $R$).  Calibration sources: PH98 =
aperture photometry from the compilation of \citet{ph98}; SDSS = standard SDSS
photometric calibration, converted to Cousins $R$ as described in the text.
Notes: 1.\ No photometric calibration possible.  2.\ Archival data.  3.\ Data
from BARS Project; profile combines 20s and 600s exposures.  4.  Uncertain
calibration.  5.\ Profile at $r < 14\arcsec$ is from higher-resolution,
unsaturated NOT image, scaled to match INT-WFC image.  6.\ Image from
\citet{frei96}.  7.\ Profile from 2003 observations is better quality;
calibrated using 2004 observation.}

\end{deluxetable}

\clearpage

\begin{deluxetable}{llrrrrrrl}
\tablewidth{0pt}
\tablecaption{Outer Disk Classifications and Measurements\label{tab:results}} 
\tablecolumns{9}
\tablehead{
\colhead{Galaxy} & \colhead{Profile Type} & \colhead{$h_{i}$} & \colhead{$h_{o}$} &
\colhead{$R_{\rm brk}$} & \colhead{$\mu_{0,i}$} & \colhead{$\mu_{0,o}$} &
\colhead{$\mu_{\rm brk}$} & \colhead{Notes} \\
 & & ($\arcsec$) & ($\arcsec$) & ($\arcsec$) & & & & }

\startdata
  NGC 718 &             II.o-OLR &  39.9 &  17.1 &    40.5 & 20.88 & 19.34 &  22.0 & \\
  NGC 936 &          II.o-OLR(?) &  51.5 &  27.4 &    95.0 & 20.04 & 18.29 &  22.1 & \\
 NGC 1022 &                    I &  23.8 & \nodata &  $>$150 & \nodata & \nodata & \nodata & 1 \\
 NGC 2273 &              II.o-CT &  30.0 &  14.0 &    98.4 & 20.58 & 16.50 &  24.2 & \\
 NGC 2681 &                    I &  27.5 & \nodata &  $>$180 & 19.70 & \nodata & $>$ 26.6 & \\
 NGC 2712 &                    I &  19.7 & \nodata &  $>$120 & 19.95 & \nodata & $>$ 27.1 & \\
 NGC 2787 &                    I &  26.7 & \nodata &  $>$170 & 19.05 & \nodata & $>$ 24.1 & \\
 NGC 2859 &             II.o-OLR & \nodata &  31.9 & $\sim$105 & \nodata & 19.08 &  23.1 & 2,3 \\
 NGC 2880 &                III-s &  20.0 & \nodata & $\sim$67 & 19.77 & \nodata & \nodata & \\
 NGC 2950 &              II.o-CT &  31.8 &  21.7 &    92.1 & 20.41 & 19.00 &  23.7 & \\
 NGC 2962 &             II.o-OLR & 135.8 &  20.9 &    68.4 & 22.33 & 19.32 &  22.9 & \\
 NGC 3049 &          II.o-OLR(?) &  29.5 &  12.7 &    53.7 & 21.04 & 18.42 &  23.0 & 4 \\
 NGC 3185 &             II.o-OLR &  46.3 &  12.5 &    80.9 & 22.39 & 17.37 &  24.4 & \\
 NGC 3351 &             II.o-OLR &  96.5 &  46.6 &   141.1 & 20.69 & 18.94 &  22.3 & \\
 NGC 3368 &             II.o-OLR & 146.8 &  60.6 &   172.3 & 21.14 & 19.28 &  22.4 & \\
 NGC 3412 &  II.o-OLR(?) + III-s &  31.5 &  20.7 &    56.4 & 18.25 & 19.27 &  21.3 & 5,6 \\
 NGC 3485 &                    I &  21.9 & \nodata &  $>$135 & 20.67 & \nodata & $>$ 27.2 & \\
 NGC 3489 &                III-d &  17.2 &  58.4 &    85.0 & 17.95 & 22.03 &  23.3 & \\
 NGC 3504 &             II.o-OLR &  52.3 &  19.4 &    60.9 & 21.33 & 19.19 &  22.6 & \\
 NGC 3507 &          II.o-OLR(?) &  36.7 &  18.9 &    75.1 & 20.31 & 18.22 &  22.6 & 4 \\
 NGC 3729 &          II.o-OLR(?) &  24.3 &  17.2 &    78.5 & 19.52 & 18.02 &  23.0 & \\
 NGC 3941 &                III-s &  24.2 &  71.9 & $\sim$115 & 19.12 & \nodata & \nodata & \\
 NGC 3945 &             II.o-OLR & 145.0 &  35.9 &   119.9 & 22.19 & 19.42 &  23.2 & \\
 NGC 3982 &     II.o-OLR + III-d & \nodata &  10.3 &    12.2 & \nodata & 18.05 &  19.3 & 2,5 \\
          &                    &  10.3 &  15.3 &    52.9 & 18.06 & 19.83 &  23.6 & \\
 NGC 4037 &                 II.o &  34.2 &  22.2 &    84.7 & 21.24 & 19.85 &  24.0 & \\
 NGC 4045 &                III-d &  21.9 &  34.2 &    70.0 & 19.82 & 21.35 &  23.5 & \\
 NGC 4102 &  II.o-OLR + III-s(?) & 344.1 &  15.4 &    42.9 & 20.58 & 17.59 &  20.9 & 5,7 \\
 NGC 4143 &                III-s &  14.3 & \nodata & $\sim$70 & 18.06 & \nodata & \nodata & \\
 NGC 4151 &                    I &  78.6 & \nodata &  $>$310 & 22.12 & \nodata & $>$ 26.5 & \\
 NGC 4203 & II.o-OLR(?) + III-s(?) &  38.0 &  24.4 &    58.1 & 19.96 & 18.95 &  21.7 & 5,8 \\
 NGC 4245 &                    I &  29.9 & \nodata &  $>$220 & 20.49 & \nodata & $>$ 27.8 & \\
 NGC 4267 &                    I &  26.1 & \nodata &  $>$180 & 19.68 & \nodata & $>$ 26.7 & \\
 NGC 4314 &                 II.i &  27.7 & \nodata &  $>$210 & 19.50 & \nodata &  28.9 & \\
 NGC 4319 &                III-d &  12.0 &  55.8 &    55.0 & \nodata & \nodata & \nodata & 1 \\
 NGC 4340 &                    I &  48.4 & \nodata &  $>$220 & 21.47 & \nodata & $>$ 26.5 & \\
 NGC 4371 &                III-d &  37.1 &  55.4 &   190.0 & 19.96 & 21.81 &  25.6 & \\
 NGC 4386 &                III-s &  20.5 &  43.0 & $\sim$75 & 19.88 & \nodata & \nodata & \\
 NGC 4477 &                    I &  35.7 & \nodata &  $>$200 & 19.87 & \nodata & $>$ 26.0 & 4 \\
 NGC 4531 &                III-s &  26.6 & \nodata & $\sim$125 & 19.81 & \nodata & \nodata & \\
 NGC 4596 &                    I &  39.5 & \nodata &  $>$250 & 19.85 & \nodata & $>$ 26.6 & \\
 NGC 4608 &                 II.i &  29.0 & \nodata &  $>$175 & 19.85 & \nodata & $>$ 26.6 & \\
 NGC 4612 &                III-d &  14.2 &  39.3 &    50.0 & 18.67 & 21.26 &  22.7 & \\
 NGC 4643 &                    I &  53.8 & \nodata &  $>$285 & 21.42 & \nodata & $>$ 26.2 & \\
 NGC 4665 &                    I &  36.2 & \nodata &  $>$200 & 19.86 & \nodata & $>$ 25.4 & \\
 NGC 4691 &                III-d &  29.5 &  47.8 &   125.0 & 19.96 & 21.71 &  24.5 & \\
 NGC 4699 &                III-d &  12.8 &  35.3 &    41.6 & 16.57 & 18.86 &  20.0 & \\
 NGC 4725 &             II.o-OLR & 184.1 &  55.2 &   268.4 & 22.14 & 18.46 &  24.2 & \\
 NGC 4754 &                    I &  37.6 & \nodata &  $>$260 & 19.59 & \nodata & $>$ 26.5 & \\
 NGC 4995 &                 II.i &  17.4 & \nodata &   $>$90 & 18.55 & \nodata & $>$ 24.5 & \\
 NGC 5338 &           II.o-CT(?) &  25.2 &  17.1 &    67.3 & 21.03 & 19.65 &  23.9 & \\
 NGC 5377 &             II.o-OLR & 153.6 &  29.8 &   115.7 & 22.70 & 19.01 &  23.5 & \\
 NGC 5701 &             II.o-OLR & 110.0 &  18.3 &   120.0 & 21.74 & 15.70 &  23.1 & 3 \\
 NGC 5740 &                III-d &  17.4 &  27.7 &   104.0 & 19.04 & 21.34 &  25.5 & \\
 NGC 5750 &                    I &  21.5 & \nodata &  $>$160 & 19.26 & \nodata & $>$ 26.8 & \\
 NGC 5806 &                III-d &  29.6 &  50.6 &   114.0 & 20.05 & 21.76 &  24.2 & \\
 NGC 5832 &                 II.i &  20.9 & \nodata &  $>$165 & 19.56 & \nodata & $>$ 28.2 & \\
 NGC 5957 &                 II.o &  33.3 &  17.9 &    81.8 & 21.69 & 19.53 &  24.5 & \\
 NGC 6012 &                III-d &  23.4 &  58.7 &   118.0 & 20.62 & 23.76 &  25.9 & \\
 NGC 6654 &             II.o-OLR &  58.4 &  13.0 &    62.7 & 21.05 & 16.97 &  22.3 & \\
 NGC 7177 &                III-d &  16.7 &  83.8 &    89.0 & 18.81 & 23.47 &  24.1 & \\
 NGC 7280 &          II.o-OLR(?) &  25.4 &  11.2 &    42.5 & 20.20 & 17.90 &  22.1 & 4 \\
 NGC 7743 &                    I &  39.9 & \nodata &  $>$145 & 21.08 & \nodata & $>$ 25.1 & \\
   IC 676 &                    I &  15.6 & \nodata &  $>$105 & 20.25 & \nodata & $>$ 27.5 & \\
  IC 1067 &          II.o-OLR(?) &  32.6 &  13.7 &    39.9 & 21.34 & 19.27 &  22.7 & \\
 UGC 3685 &                    I &  44.4 & \nodata &  $>$195 & 22.34 & \nodata & $>$ 27.1 & \\
UGC 11920 &          II.o-OLR(?) &  87.6 &  24.5 &    60.0 & \nodata & \nodata & \nodata & 1 \\
\enddata 

\tablecomments{Classifications and measurements of surface brightness
profiles.  For each galaxy, we list the profile type
(Section~\ref{sec:profile-types}), the exponential scale lengths of the inner
and outer parts of the profile, the break radius of Type~II and III profiles,
the central surface brightnesses of the fitted exponentials, and the surface
brightness at the break radius.  Surface brightnesses are observed values, and
have not been corrected for Galactic extinction, inclination, or redshift.
Note that Type~I profiles by definition have no ``outer'' part and have only
an upper limit for the break radius; some Type~II profiles have
non-exponential inner parts.  For NGC~3982, we list values for both the inner
zone (Type~II.o-OLR) and the outer zone (Type~III-d).  Notes: 1 = no
photometric calibration; 2 = non-exponential inner profile; 3 = very gradual,
smooth break; 4 = tentative evidence for Type~III profile at large radii; 5 =
profile has multiple breaks or transitions; 6 = $R_{\rm brk}$ for III-s part of 
profile $\sim 115\arcsec$;  7 = $R_{\rm brk}$ for III-s part of 
profile $\sim 110\arcsec$;  8 = $R_{\rm brk}$ for III-s part of 
profile $\sim 130\arcsec$.}

\end{deluxetable}

\begin{deluxetable}{lrrr}
\tablewidth{0pt}
\tablecaption{Inner Break Radii for Type II.i Profiles\label{tab:2i}} 
\tablecolumns{4}
\tablehead{
\colhead{Galaxy} & \colhead{$R_{\rm brk}$} & \colhead{Bar \amax} &
\colhead{Bar \lbar} \\
 & ($\arcsec$) & ($\arcsec$) & ($\arcsec$) }

\startdata
NGC 4314 &    73.5 &  73.8  & 88.0  \\
NGC 4608 &    $\sim 47$ &  53.8  & 59.9  \\
NGC 4995 &    24.5 &  22.5  & 26.7 \\
NGC 5832 &    $\sim 45$ &  42.7  & 49.3 \\
\enddata 

\tablecomments{Break radius measurements for Type II.i profiles.  These are
the inner breaks, located near the end of the bar; also listed are the
(deprojected) bar-radius measurements \amax{} and \lbar{} from Table~\ref{tab:basic}.  Lower limits on
possible breaks at larger radii (e.g., disk truncations) are given in
Table~\ref{tab:results}.}

\end{deluxetable}


\clearpage

\begin{figure}
\begin{center}
\includegraphics[scale=0.9]{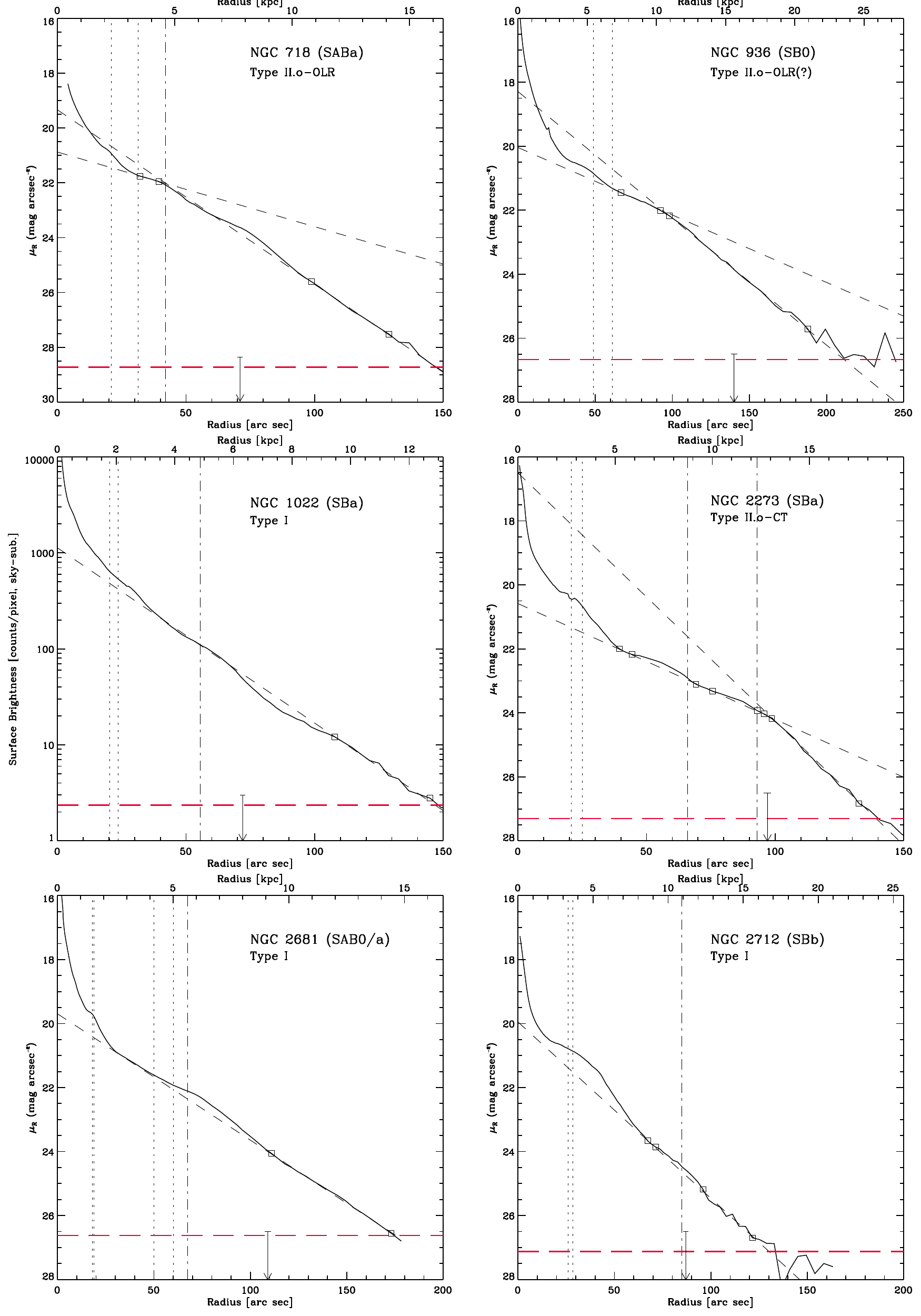}
\end{center}

\caption{Azimuthally averaged surface brightness profiles for all the galaxies
in our sample.  For each galaxy, we also show lower and upper limits to the
bar size (vertical dotted lines), the radius of any outer rings (vertical
dot-dashed lines), and exponential fits to different parts of the profile
(diagonal dashed lines; small boxes mark regions used for fits).  Horizontal
dashed red lines mark the sky-uncertainty limit $\mulim$; small arrows
indicate $R_{25}$.\label{fig:profiles}}

\end{figure}

\clearpage

\begin{figure}
\begin{center}
\includegraphics[scale=0.9]{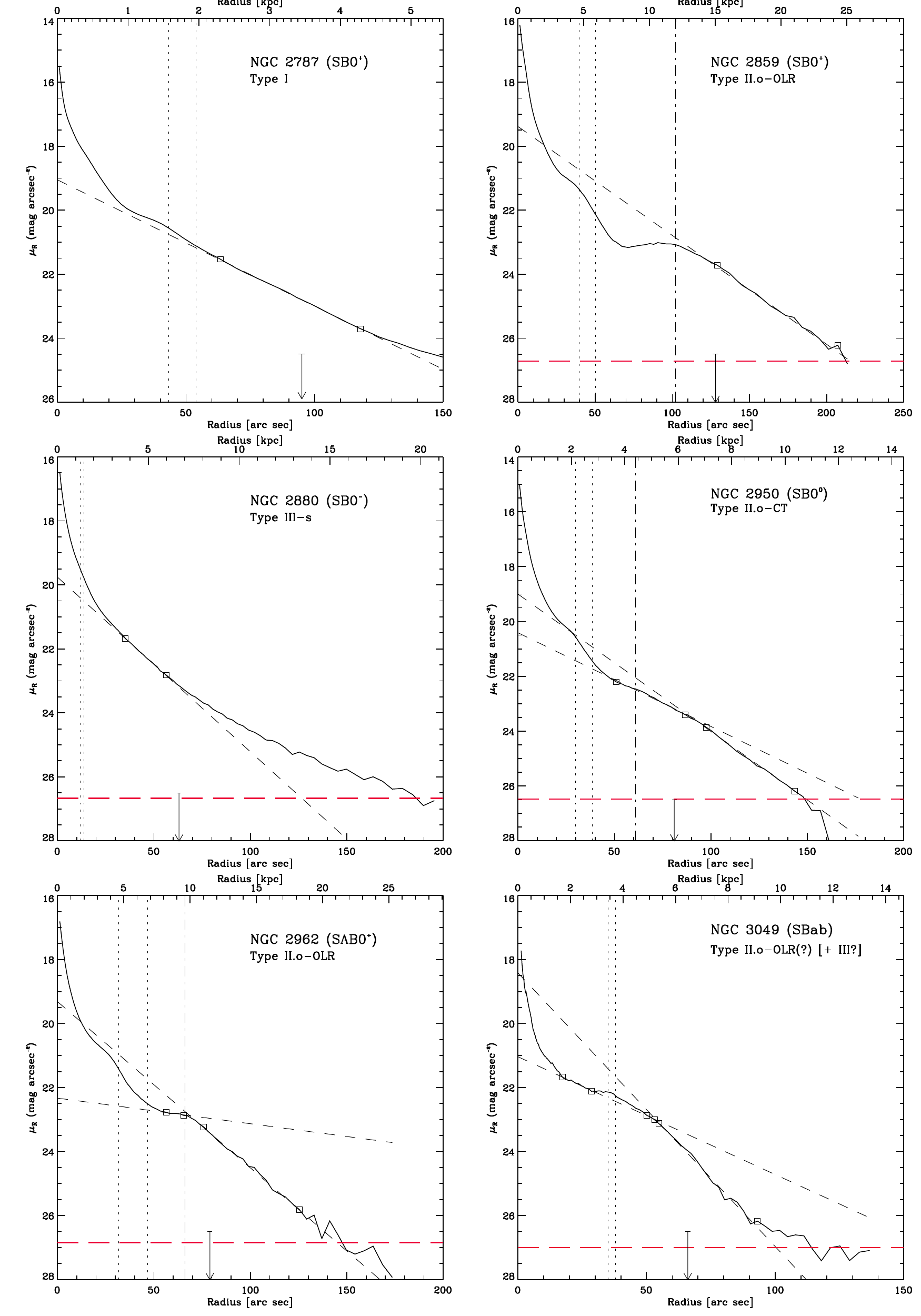}
\end{center}

\addtocounter{figure}{-1}
\caption{continued.}

\end{figure}

\clearpage

\begin{figure}
\begin{center}
\includegraphics[scale=0.9]{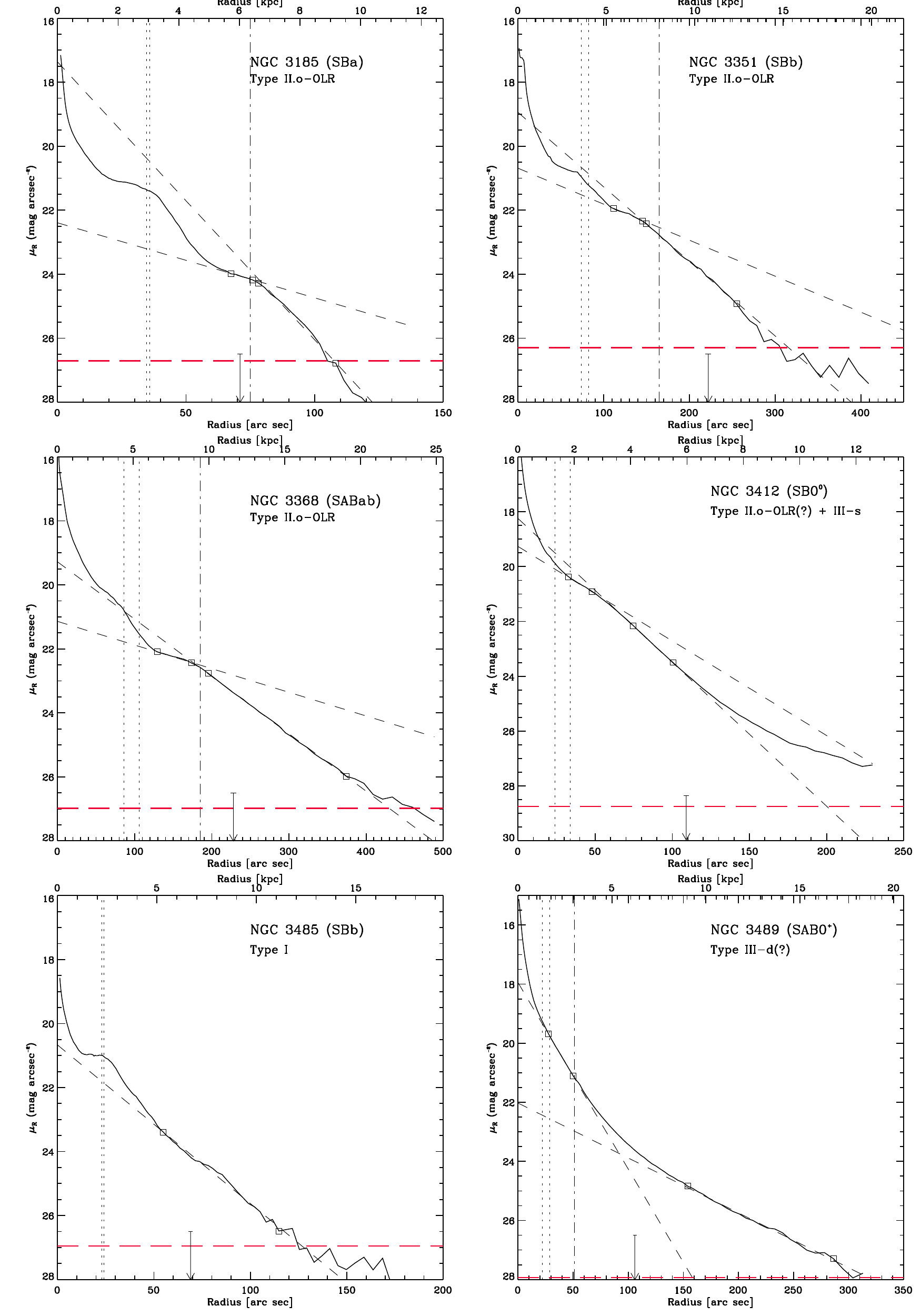}
\end{center}

\addtocounter{figure}{-1}
\caption{continued.}

\end{figure}

\clearpage

\begin{figure}
\begin{center}
\includegraphics[scale=0.9]{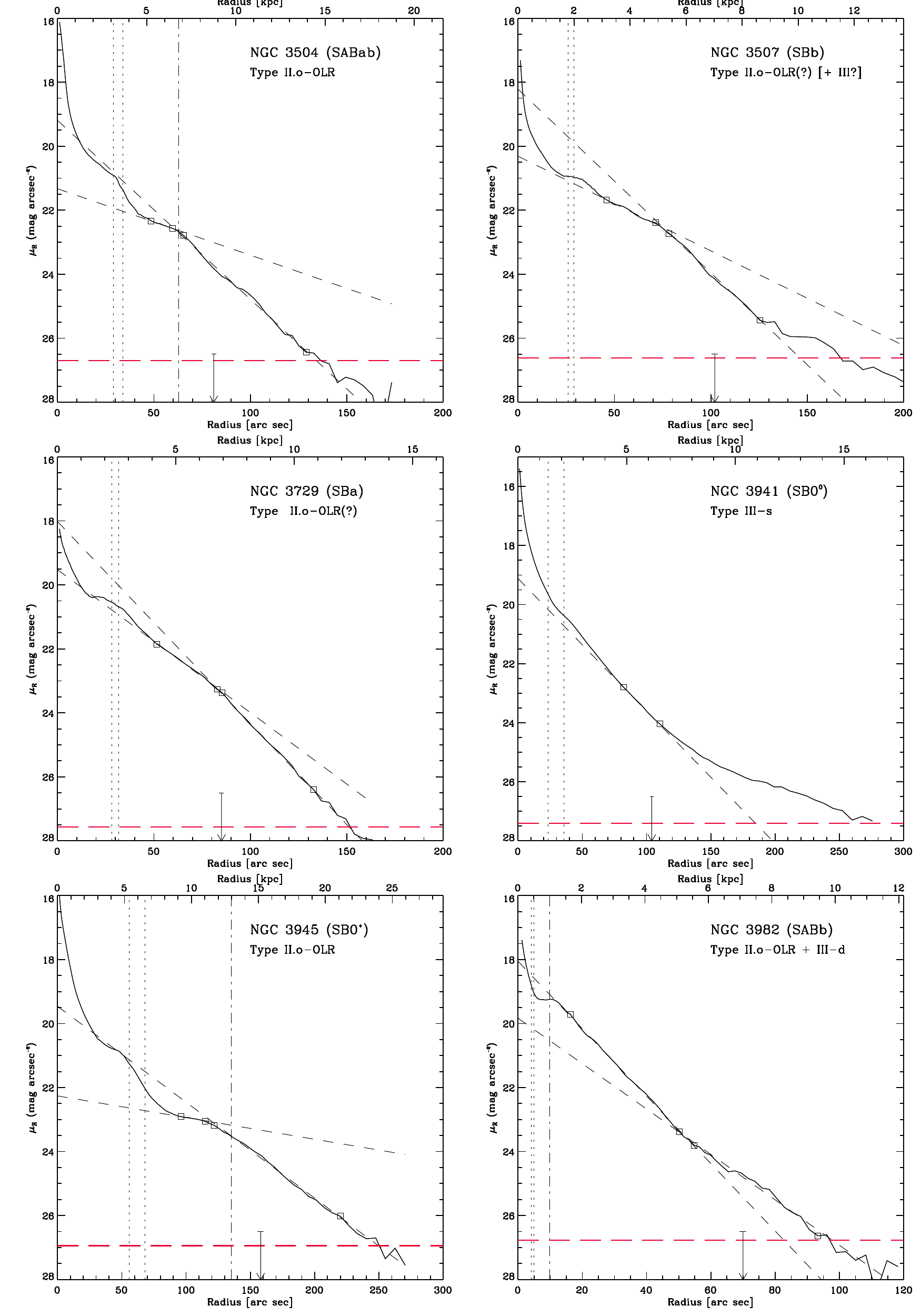}
\end{center}

\addtocounter{figure}{-1}
\caption{continued.}

\end{figure}

\clearpage

\begin{figure}
\begin{center}
\includegraphics[scale=0.9]{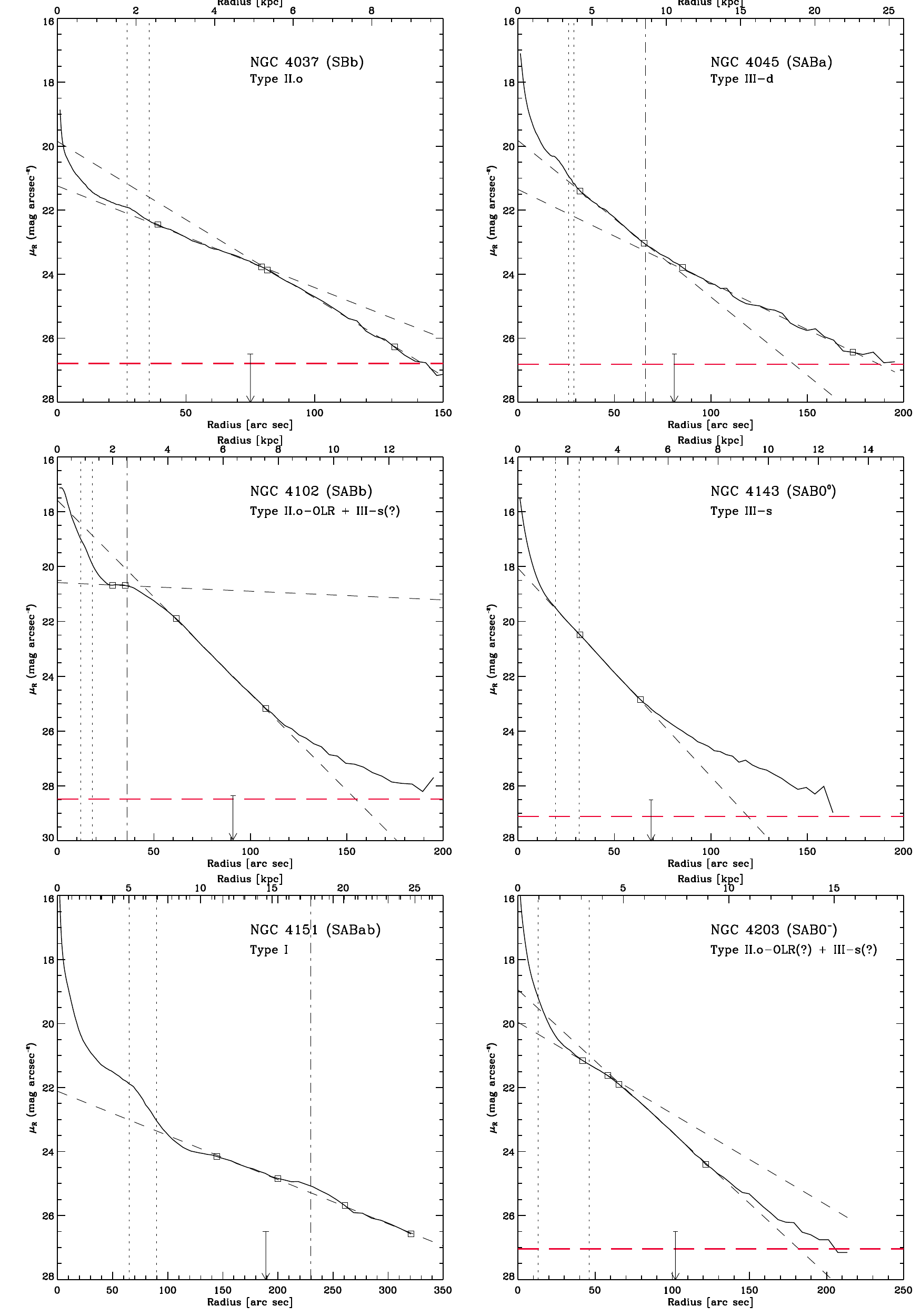}
\end{center}

\addtocounter{figure}{-1}
\caption{continued.}

\end{figure}

\clearpage

\begin{figure}
\begin{center}
\includegraphics[scale=0.9]{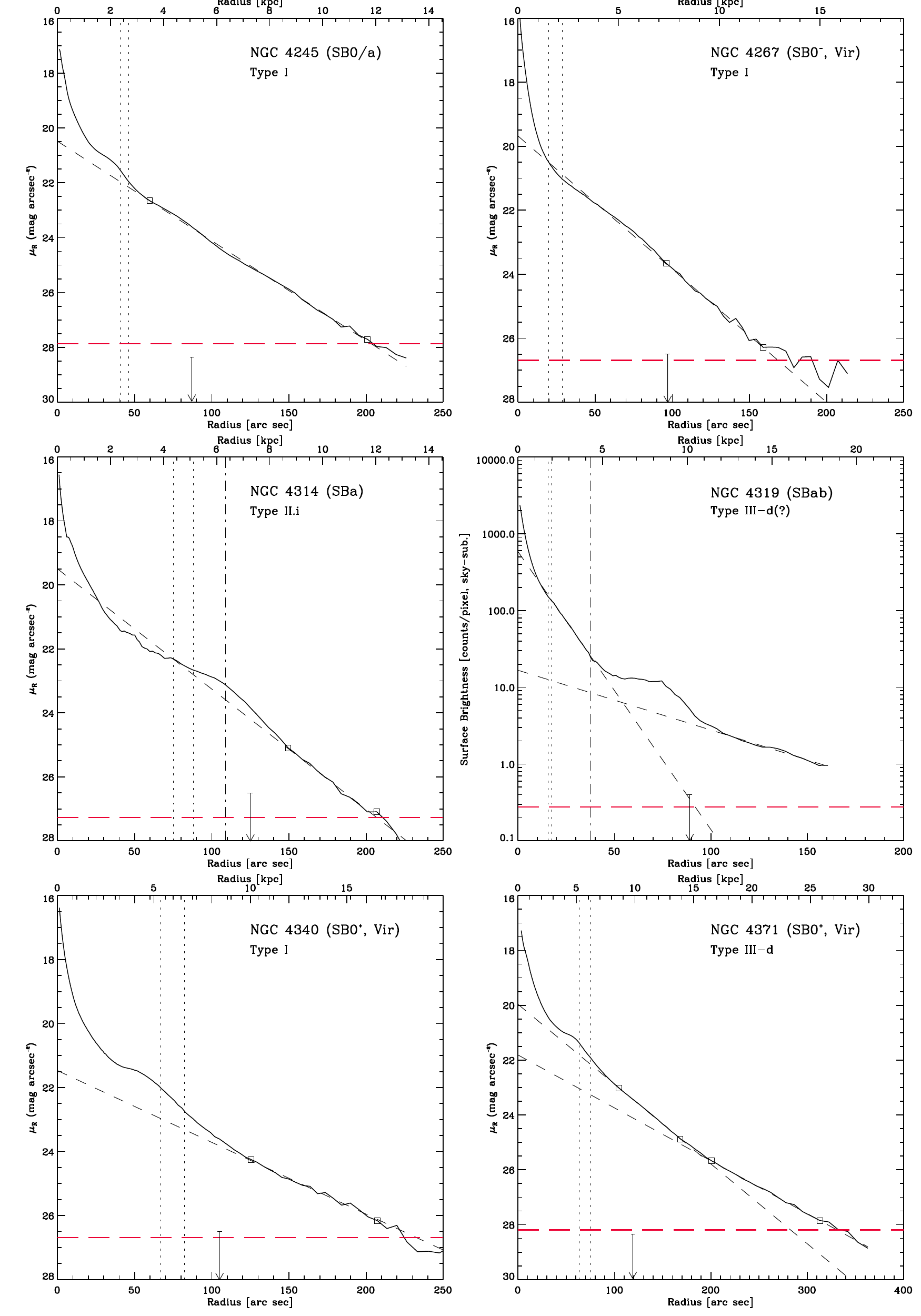}
\end{center}

\addtocounter{figure}{-1}
\caption{continued.}

\end{figure}

\clearpage

\begin{figure}
\begin{center}
\includegraphics[scale=0.9]{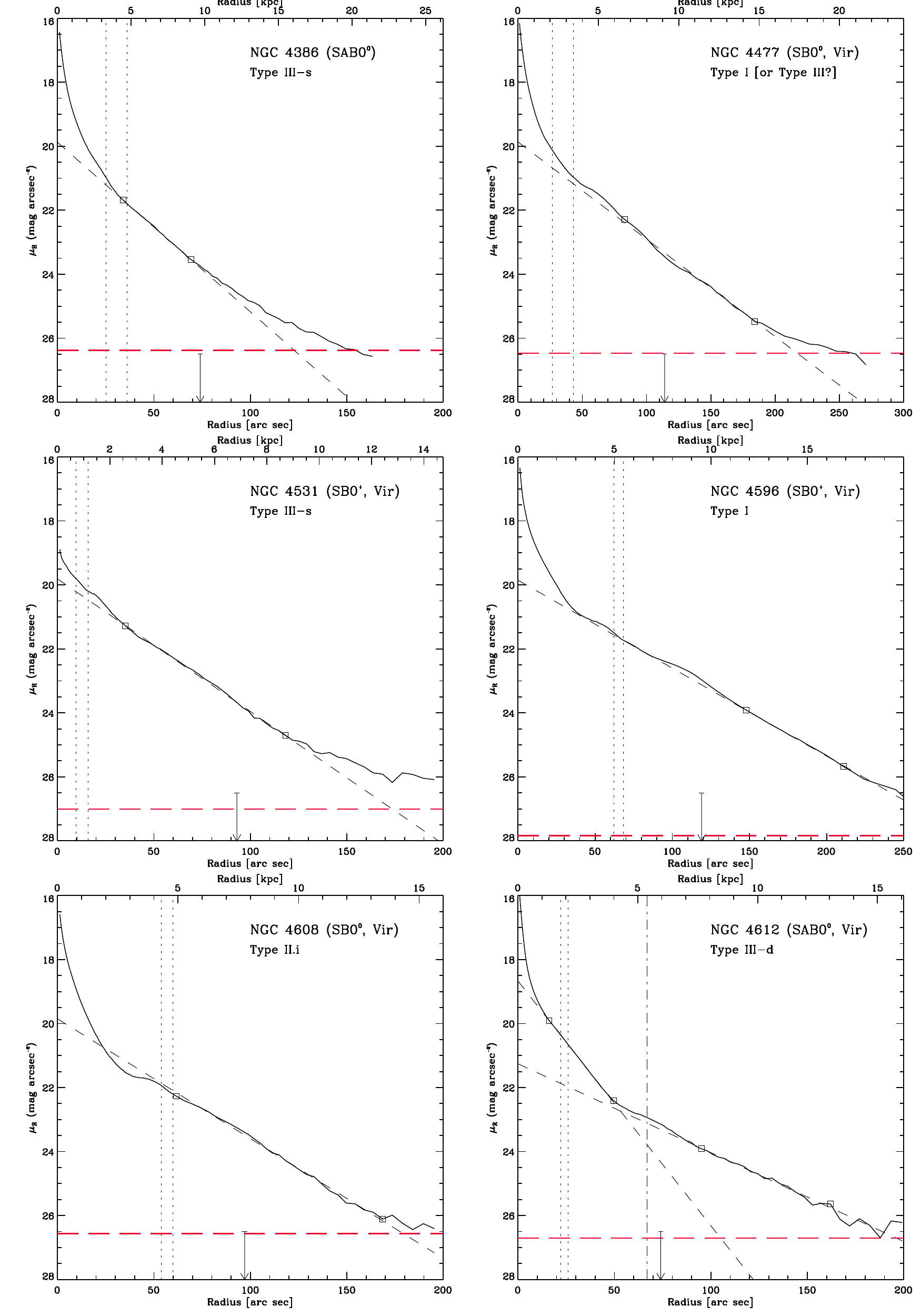}
\end{center}

\addtocounter{figure}{-1}
\caption{continued.}

\end{figure}

\clearpage

\begin{figure}
\begin{center}
\includegraphics[scale=0.9]{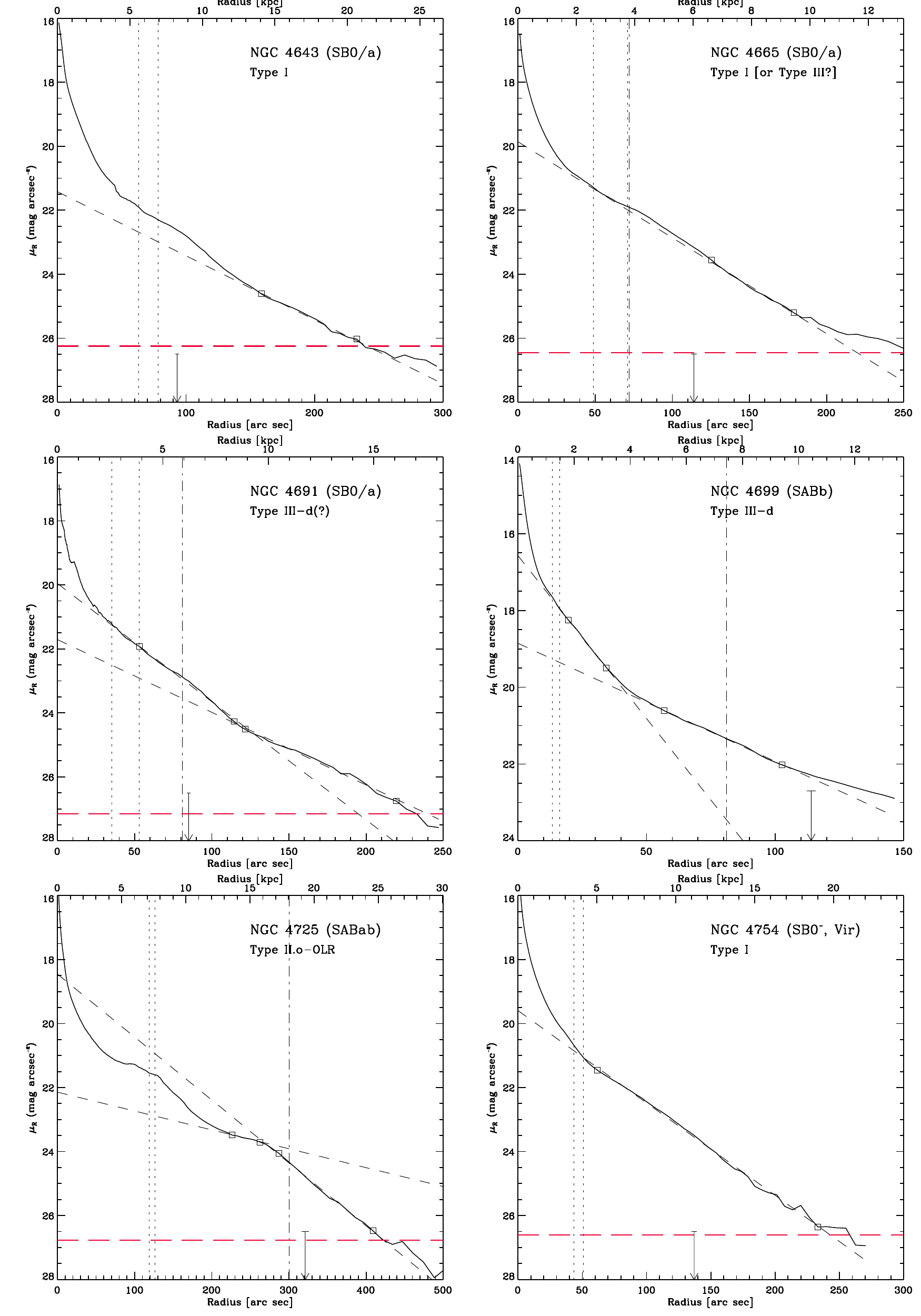}
\end{center}

\addtocounter{figure}{-1}
\caption{continued.}

\end{figure}

\clearpage

\begin{figure}
\begin{center}
\includegraphics[scale=0.9]{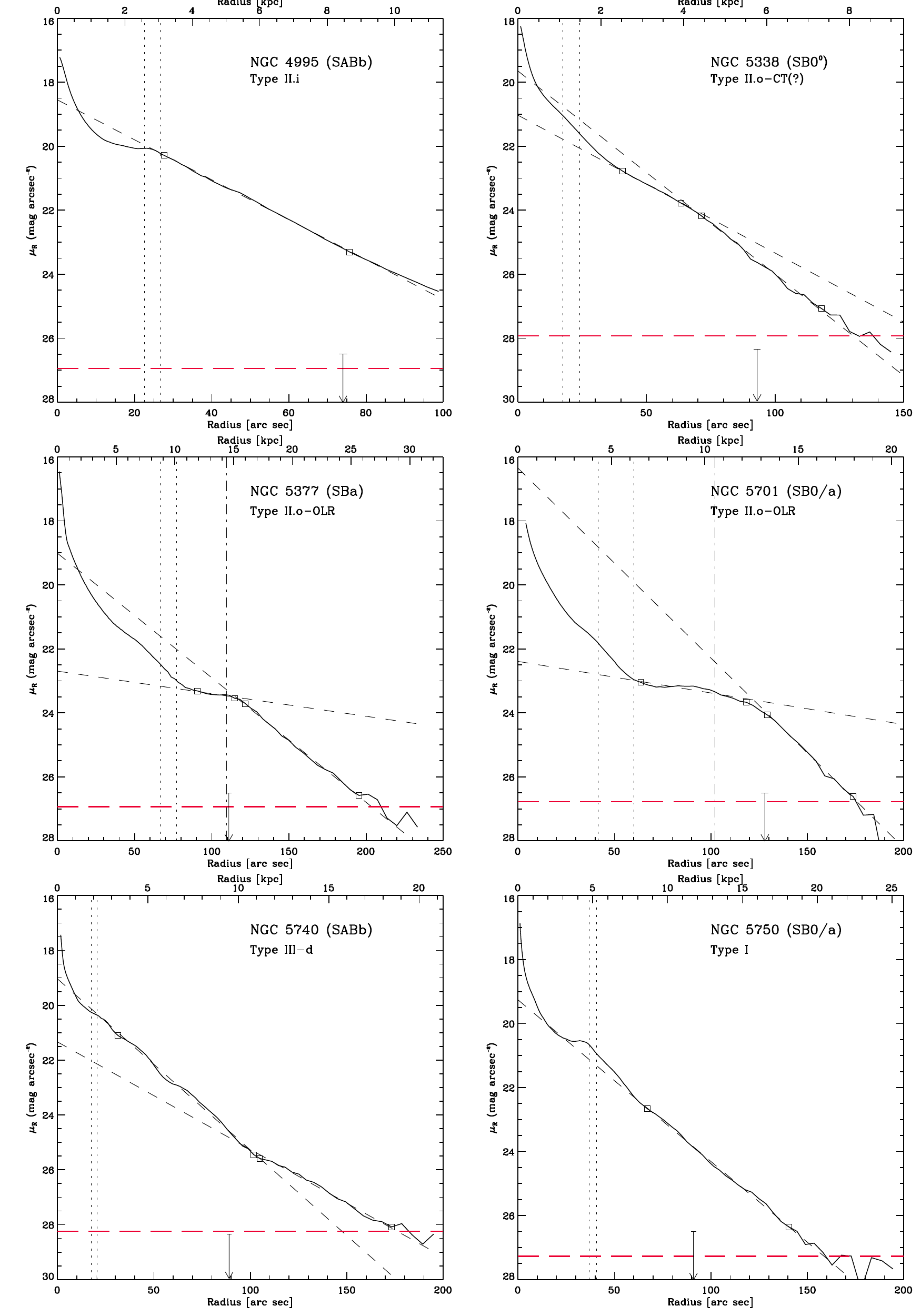}
\end{center}

\addtocounter{figure}{-1}
\caption{continued.}

\end{figure}

\clearpage

\begin{figure}
\begin{center}
\includegraphics[scale=0.9]{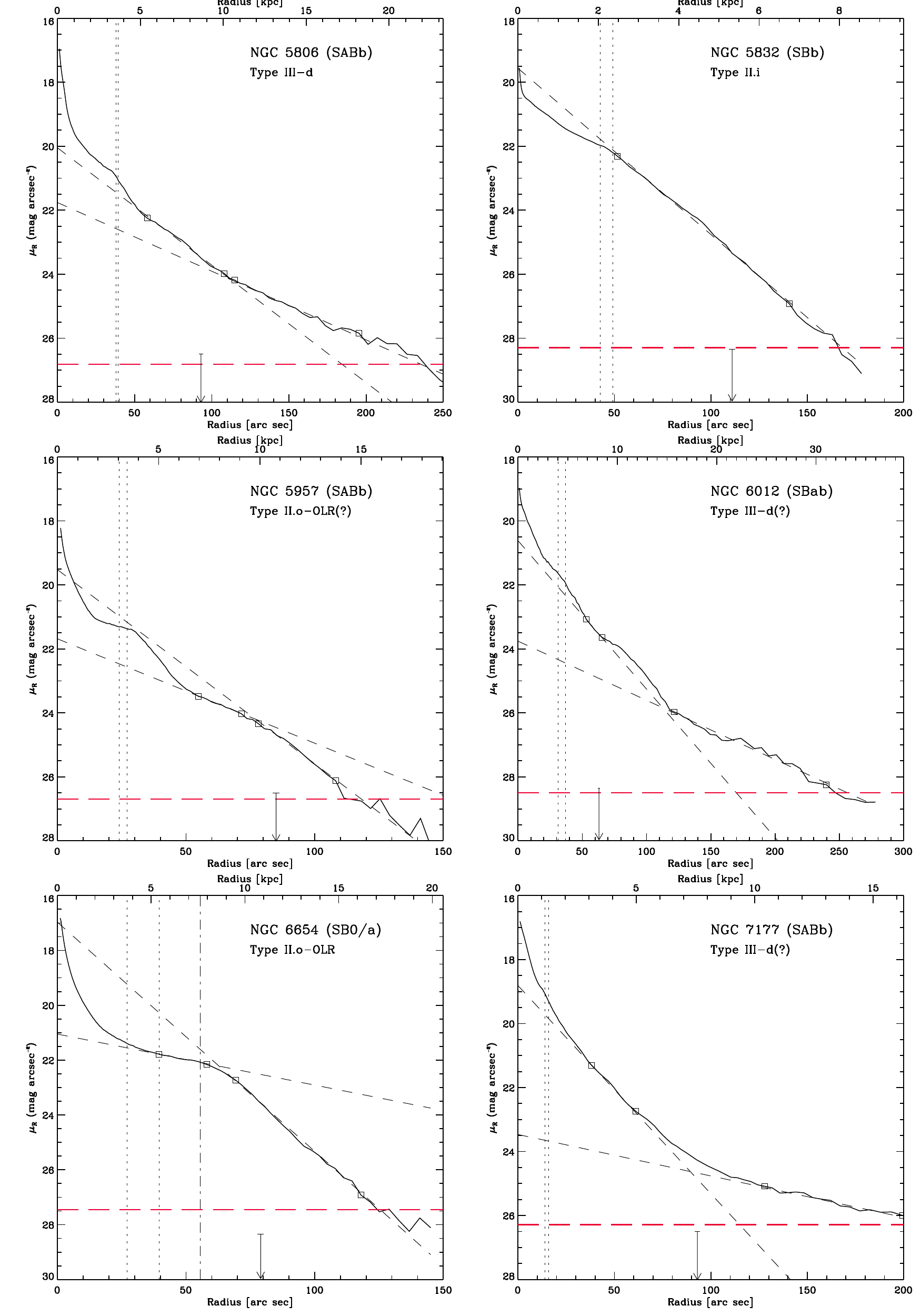}
\end{center}

\addtocounter{figure}{-1}
\caption{continued.}

\end{figure}

\clearpage

\begin{figure}
\begin{center}
\includegraphics[scale=0.9]{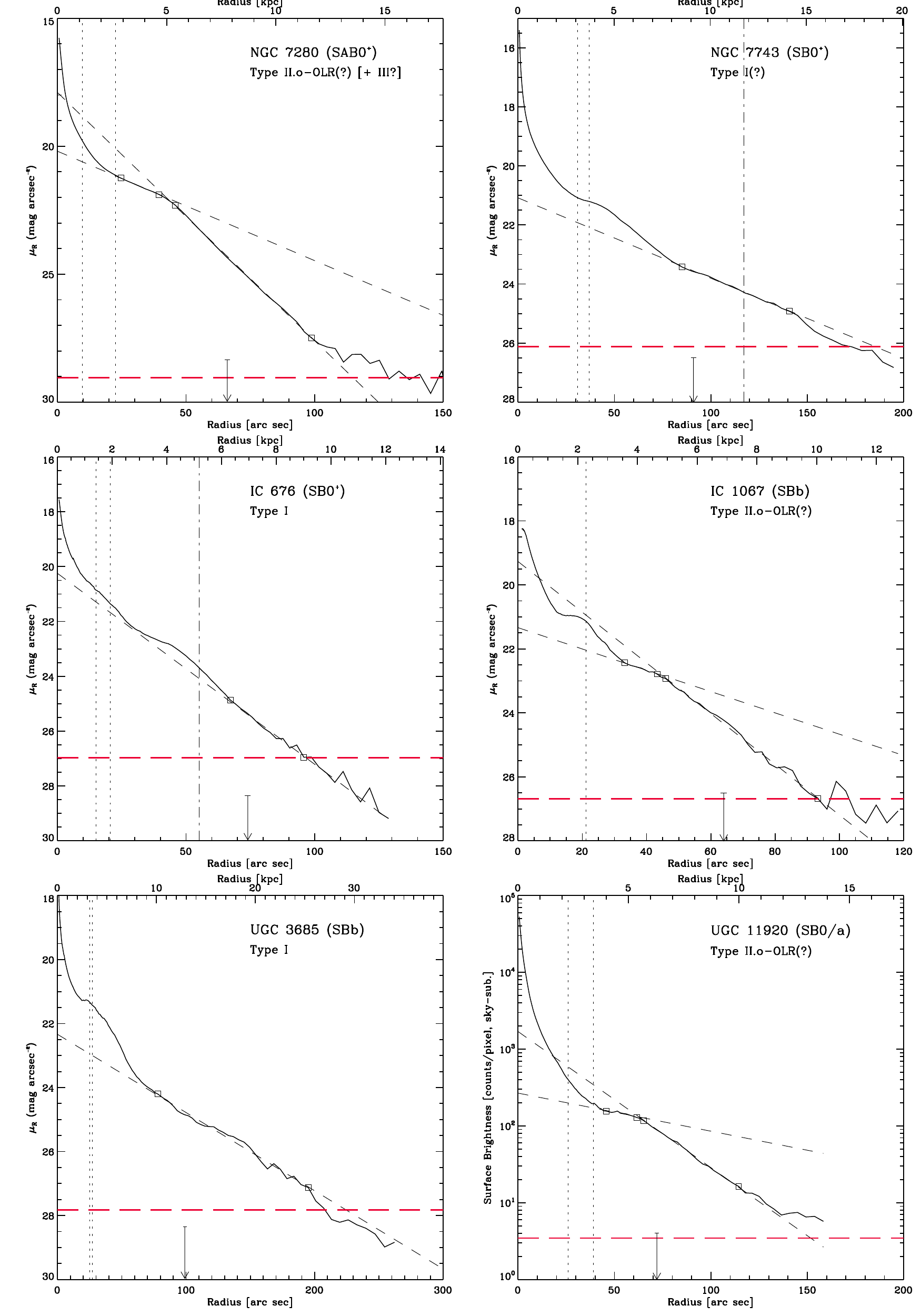}
\end{center}

\addtocounter{figure}{-1}
\caption{continued.}

\end{figure}


\begin{thebibliography}{}

\bibitem[Adamson et al.(1987)]{adamson87} Adamson, A. J., Adams, D. 
J., \& Warwick, R. S. 1987, \mnras, 224, 367

\bibitem[Adelman-McCarthy et al.(2007)]{sdss-dr5} Adelman-McCarthy, 
J. K., et al.\ 2007, \apjs, in press (arXiv:0707.3380)

\bibitem[Ajhar et al.(2001)]{ajhar01} Ajhar, E. A., Tonry, J. L.,
Blakeslee, J. P., Riess, A. G., \& Schmidt, B. P. 2001, \apj, 559, 584

\bibitem[Aladro et al.(2007)]{aladro07} Aladro, R., Guti\'{e}rrez, L., Erwin, P., 
\& Beckman, J. E. 2007, in Pathways through an Eclectic
Universe, eds.\ J. H. Knapen, T. J. Mahoney, \& A. Vazdekis (San Francisco:
ASP Conference Series), in press.

\bibitem[Athanassoula \& Misiriotis(2002)]{athan02} Athanassoula, E., \& 
Misiriotis, A. 2002, \mnras, 330, 35

\bibitem[Barton \& Thompson(1997)]{barton97} Barton, I. J., \& 
Thompson, L. A. 1997, \aj, 114, 655

\bibitem[Battaner, Florido, \& Jim\'{e}nez-Vicente(2002)]{battaner02} 
Battaner, E., Florido, E., \& Jim\'{e}nez-Vicente, J. 2002, \aap, 
388, 213

\bibitem[Baggett, Baggett, \& Anderson(1998)]{bba98} Baggett, W. E.,
Baggett, S. M.,\& Anderson, K. S. J. 1998, \aj, 116, 1626

\bibitem[Bland-Hawthorn et al.(2005)]{bland-hawthorn05}
Bland-Hawthorn, J., Vlaji\'{c}, M., Freeman, K. C., \& Draine, B. T. 
2005, \apj, 629, 239

\bibitem[Buta(1986)]{buta86} Buta, R. 1986, \apjs, 61, 609
\bibitem[Buta(1995)]{buta95} Buta, R. 1995, \apjs, 96, 39
\bibitem[Buta \& Crocker(1993)]{bc93} Buta, R., \& Crocker,
D. A. 1993, \aj, 105, 1344
\bibitem[Buta \& Combes(1996)]{buta96} Buta, R., \& Combes, 
F.\ 1996, Fundamentals of Cosmic Physics, 17, 95 

\bibitem[Byun et al.(1996)]{byun96} Byun, Y.-I., et al.\ 1996, \aj,
111, 1889

\bibitem[Debattista et al.(2006)]{debattista06} Debattista, V. P., Mayer, L.,
Carollo, C. M., Moore, B., Wadsley, J., \& Quinn, T. 2006, \apj, 645, 209

\bibitem[Eskridge et al.(2000)]{eskridge00} Eskridge, P. B., et al.\ 2000,
\aj, 119, 536

\bibitem[de Grijs, Kregel, \& Wesson(2001)]{degrijs01} de Grijs, 
R., Kregel, M., \& Wesson, K. H. 2001, \mnras, 324, 1074

\bibitem[de Jong \& van der Kruit(1994)]{dejong94} de Jong, R. S., \&
van der Kruit, P. C. 1994, \aaps, 106, 451

\bibitem[de Vaucouleurs(1959)]{deV59} de Vaucouleurs, G. 1959,
Handbuch der Physik, 53, 311
\bibitem[de Vaucouleurs \& de Vaucouleurs(1964)]{rc1} de Vaucouleurs, 
G., \& de Vaucouleurs, A. 1964, Reference Catalogue of Bright Galaxies
(Austin: University of Texas)
\bibitem[de Vaucouleurs et al.(1991)]{rc3} de Vaucouleurs, G.,
de Vaucouleurs, A., Corwin, H. G., Buta, R.~J., Paturel, G., \&
Fouqu\'{e}, P. 1991, Third Reference Catalogue of Bright Galaxies
(New York: Springer-Verlag) (RC3)

\bibitem[Elmegreen \& Hunter(2006)]{elmegreen06} Elmegreen, B. G., \& 
Hunter, D. A. 2006, \apj, 636, 712

\bibitem[Erwin(2005)]{erwin05} Erwin, P. 2005, \mnras, 364, 283
\bibitem[Erwin(2007a)]{erwin07-rc3} Erwin, P. 2007a, in prep
\bibitem[Erwin(2007b)]{erwin07-n1543} Erwin, P. 2007b, in prep
\bibitem[Erwin \& Sparke(1999)]{erwin99} Erwin, P., \& Sparke, L. S. 
1999, \apjl, 512, L37
\bibitem[Erwin \& Sparke(2003)]{erwin03} Erwin, P., \& Sparke, L. S. 
2003, \apjs, 146, 299
\bibitem[Erwin et al.(2003)]{erwin03-id} Erwin, P., Vega Belt\'{a}n, J. 
C., Graham, A. W., \& Beckman, J. E. 2003, \apj, 597, 929
2003, \apjs, 146, 299
\bibitem[Erwin et al.(2005)]{erwin05-type3} Erwin, P., Beckman, J. E., 
\& Pohlen, M. 2005, \apjl, 626, L81
\bibitem[Erwin et al.(2007)]{erwin07} Erwin, P., Pohlen, M., \& 
Beckman, J. E. 2007, in prep

\bibitem[Eskridge et al.(2002)]{eskridge02} Eskridge, P. B., et al.\ 2002,
\apjs, 143, 73

\bibitem[Ferguson \& Clarke(2001)]{ferguson01} Ferguson, A. M. N., \&
Clarke, C. J. 2001, \mnras, 325, 781

\bibitem[Forbes(1996)]{forbes96} Forbes, D. A. 1996, \aj, 112, 1409

\bibitem[Freedman(2001)]{freedman01} Freedman, W. L., et al.\ 2001, \apj, 553,
47

\bibitem[Freeman(1970)]{freeman70} Freeman, K. C. 1970, \apj, 160, 811

\bibitem[Frei et al.(1996)]{frei96} Frei, Z., Guhathakurta, P., Gunn, J. E., 
\& Tyson, J. A. 1996, \aj, 111, 174

\bibitem[Garrido et al.(2005)]{garrido05} Garrido, O., Marcelin, M., Amram, 
P., Balkowski, C., Gach, J. L., \& Boulesix, J. 2005, \mnras, 362, 127

\bibitem[Gavazzi et al.(2003)]{gavazzi03} Gavazzi, G., Boselli, A., 
Donati, A., Franzetti, P., \& Scodeggio, M. 2003, \aap, 400, 451

\bibitem[Governato et al.(2007)]{governato07} Governato, F., Willman, B.,
Mayer, L., Brooks, A., Stinson, G. and Valenzuela, O., Wadsley, J., \&
Quinn, T. 2007, \mnras, 374, 1479

\bibitem[Gutierrez et al.(2007)]{gutierrez07} Guti\'{e}rrez, L., Erwin, P., 
Aladro, R., Beckman, J. E., \& Pohlen, M. 2007, in prep

\bibitem[Gunn et al.(1998)]{gunn98} Gunn, J. E., et al.\ 1998, \aj, 
116, 3040
\bibitem[Gunn et al.(2006)]{gunn06} Gunn, J. E., et al.\ 2006, \aj, 
131, 2332

\bibitem[Holtzman et al.(1995)]{holtzman95} Holtzman, J. A., Burrows, 
C. J., Casertano, S., Hester, J. J., Trauger, J. T., Watson, A. M., 
\& Worthey, G. 1995, \pasp, 107, 1065

\bibitem[Hunter \& Elmegreen(2006)]{hunter06} Hunter, D. A., \&  
Elmegreen, B. G. 2005, \apjs, 162, 49

\bibitem[Ibata et al.(2005)]{ibata05} Ibata, R., Chapman, S., Ferguson, A. M. 
N., Lewis, G., Irwin, M., \& Tanvir, N. 2005, \apj, 634, 287

\bibitem[Kent \& Glaudell(1989)]{kent89} Kent, S. M., \& Glaudell, 
G. 1989, \aj, 98, 1588

\bibitem[Kennicutt(1989)]{kennicutt89} Kennicutt, R. C. 1989, \apj, 344, 685
\bibitem[Kennicutt \& Edgar(1986)]{kennicutt86} Kennicutt, R. C., \&
Edgar, B. K. 1986, \apj, 300, 132
\bibitem[Kennicutt et al.(2003)]{kennicutt03} Kennicutt, R. C. et al.\ 2003,
\pasp, 115, 928

\bibitem[Koopmann \& Kenney(1998)]{koopmann98} Koopmann, R. A., \&
Kenney, J. D. P. 1998, \apj, 497, L75

\bibitem[Kormendy(1977)]{kormendy77} Kormendy, J. 1977, \apj, 214, 359
\bibitem[Kormendy(1979)]{kormendy79} Kormendy, J. 1979, \apj, 227, 714
\bibitem[Kormendy(1984)]{kormendy84} Kormendy, J. 1984, \apj, 286, 132

\bibitem[Krumm \& Shane(1982)]{krumm82} Krumm, N., \& Shane, W. W. 
1982, \aap, 116, 237

\bibitem[Landolt(1992)]{landolt92} Landolt, A. U., 1992, \aj, 104, 340

\bibitem[Lauer et al.(1995)]{lauer95} Lauer, T. R., et al.\ 1995, 
\aj, 110, 2622

\bibitem[Laurikainen et al.(2004)]{laurikainen04} Laurikainen, E., 
Salo, H., Buta, R., \& Vasylyev, S. 2004, \mnras, 355, 1251
\bibitem[Laurikainen et al.(2005)]{laurikainen05} Laurikainen, E., 
Salo, H., \& Buta, R. 2005, \mnras, 362, 1319

\bibitem[Lourenso \& Beckman(2001)]{lourenso01} Lourenso, S., \& Beckman, J.
E. 2001, Ap\&SS, 276, 1161

\bibitem[MacArthur, Courteau, \& Holtzman(2003)]{macarthur03} 
MacArthur, L. A., Courteau. S., \& Holtzman, J. A. 2003, \apj, 582, 
689

\bibitem[Matthews \& Gallagher(1997)]{matthews97} Matthews, L. D., \&
Gallagher, J. S. 1997, \aj, 114, 1899

\bibitem[Mei et al.(2007)]{mei07} Mei, S., Blakeslee, J. P., C\^{o}t\'{e}, P.,
Tonry, J. L., West, M. J., Ferrarese, L., Jord\'{a}n, A., Peng, E. W.,
Anthony, A., \& Merritt, D. 2007, \apj, 655, 144

\bibitem[Men\'{e}ndez-Delmestre et al.(2007)]{menendez-delmestre07} 
Men\'{e}ndez-Delmestre, K., Sheth, K., Schinnerer, E., Jarrett, T. H., \& 
Scoville, N. Z. 2007, \apj, 657, 790

\bibitem[Navarro \& White(1994)]{navarro94} Navarro, J.~F. \& White,
S.~D.~M. 1994, \mnras, 267, 401

\bibitem[Nilson(1973)]{ugc} Nilson, P. 1973, Uppsala General Catalog
of Galaxies, Uppsala Astron.\ Obs.\ Annals, 5, 1

\bibitem[Phillips et al.(1991)]{phillips91} Phillips, S., Evans, R., Davies, 
J.~I., \& Disney, M.~J. 1991, \mnras, 253, 496

\bibitem[Pohlen \& Trujillo(2006)]{pohlen-trujillo} Pohlen, M., \& Trujillo,
I.\ 2006, \aap, 454, 759 
\bibitem[Pohlen et al.(2002)]{pohlen02} Pohlen, M., Dettmar, R.-J.,
L\"utticke, R., \& Aronica, G. 2002, \aap, 392, 807
\bibitem[Pohlen et al.(2004)]{pohlen04} Pohlen, M., Beckman, J. E., S.
H\"{u}ttemeister, S., Knapen, J. H., Erwin, P., \& Dettmar, R.-J. 2004,
Penetrating Bars through Masks of Cosmic Dust: The Hubble Tuning Fork Strikes
a New Note, ed.\ D. L. Block, I. Puerari, K. C. Freeman, R. Groess, \& E. K.
Block (Dordrecht: Springer), 731

\bibitem[Prugniel \& Heraudeau(1998)]{ph98} Prugniel, P., \& Heraudeau, P.
1998, \aaps, 128, 299

\bibitem[Riess et al.(2005)]{riess05} Riess, A. G., et al. 2005, 
\apj, 627, 579

\bibitem[Robertson et al.(2004)]{robertson04} Robertson, B., Yoshida, N.,
Springel, V., \& Hernquist, L. 2004, \apj, 606, 32

\bibitem[Saha et al.(2001)]{saha01} Saha, A., et al. 2001, 
\apj, 562, 314

\bibitem[Sandage(1961)]{hubble-atlas} Sandage, A. 1961, The Hubble 
Atlas of Galaxies (Washington, D.C.: Carnegie Institution)
\bibitem[Sandage \& Visvanathan(1978)]{sv78} Sandage, A., \& Visvanathan, N. 
1978, \apj, 223, 707

\bibitem[Schaye(2004)]{schaye04} Schaye, J. 2004, \apj, 609, 667

\bibitem[Slyz et al.(2002)]{slyz02} Slyz, A. D., Devriendt, J. E. G., 
Silk, J., \& Burkert, A. 2002, \mnras, 333, 894

\bibitem[Smith et al.(2002)]{smith02} Smith, J. A., et al.\ 2002, \aj, 123,
2121

\bibitem[Stetson \& Gibson(2001)]{stetson01} Stetson, P. B., \& Gibson,
B. K. 2001, \mnras, 328, L1

\bibitem[Tonry et al.(2001)]{tonry01} Tonry, J. L., Dressler, A., 
Blakeslee, J. P., Ajhar, E. A., Fletcher, A. B., Luppino, G. A., 
Metzger, M. R., \& Moore, C. B. 2001, \apj, 546, 681

\bibitem[Valenzuela \& Klypin(2003)]{valenzuela03} Valenzuela, O., \& Klypin, 
A. 2003, \mnras, 345, 406

\bibitem[van den Bergh(1976)]{vandb76} van den Bergh, S. 1976, \apj,
206, 883

\bibitem[van der Kruit(1979)]{vdk79} van der Kruit, P.~C., 1979, \aaps, 38, 15
\bibitem[van der Kruit(1987)]{vdk87} van der Kruit, P.~C., 1987, \aap, 173, 59
\bibitem[van der Kruit \& Searle(1981a)]{vdk81a} van der Kruit, P.~C., \& 
Searle, L., 1981a, \aap, 95, 105
\bibitem[van der Kruit \& Searle(1981b)]{vdk81b} van der Kruit, P.~C., \& 
Searle, L., 1981b, \aap, 95, 116

\bibitem[van Driel \& Buta(1991)]{vandriel91} van Driel, W., \& Buta, 
R. J. 1991, \aap, 245, 7

\bibitem[Weiner et al.(2001)]{weiner01} Weiner, B. J., Williams, T.
B., van Gorkom, J. H., \& Sellwood, J. A. 2001, \apj, 546, 916

\bibitem[Wozniak \& Pierce(1991)]{wozniak91} Wozniak, H., \& Pierce, 
M. J. 1991, \aaps, 88, 325

\bibitem[Younger et al.(2007)]{younger07} Younger, J. D., Cox, T. J., Seth, A.
C., \& Hernquist, L. 2007, \apj, in press (arXiv:0707.4481)

\bibitem[Yoshii \& Sommer-Larsen(1989)]{yoshii89} Yoshii, Y., \& 
Sommer-Larsen, J. 1989, \mnras, 236, 779

\bibitem[York et al.(2000)]{york00} York, D. G., et al.\ 2000, \aj,
120, 1579

\bibitem[Zhang \& Wyse(2000)]{zhang00} Zhang, B., \& Wyse, R. F. G. 2000, 
\mnras, 313, 310

\end{thebibliography}
\end{document}